\documentclass[3p,11pt]{elsarticle}
\journal{...}
\emergencystretch=\maxdimen
\biboptions{sort&compress}
\usepackage{geometry}
\usepackage{hyperref}
 \hypersetup{
 	plainpages = false, 
	bookmarksopen = true,
	bookmarksnumbered = true,
	breaklinks = true,
	linktocpage,
	colorlinks = true,
	linkcolor = purple,
	urlcolor  = black,
	citecolor = purple,
	}
\usepackage{multicol}
\usepackage{multirow}
\usepackage{imakeidx}
\usepackage{nomencl}
\makenomenclature
\usepackage[xindy]{glossaries}
\makeglossaries
\nomenclature[g0]{\textit{Greek letters}}{}
\nomenclature[s0]{\textit{Scaling factors}}{}
\nomenclature[sb0]{\textit{Subscripts}}{}
\usepackage[titletoc,title]{appendix}
\usepackage[mathscr]{euscript}
\usepackage{graphics}
\usepackage{graphicx}
\usepackage{epsf,psfrag}
\usepackage{epstopdf}
\usepackage[para]{threeparttable}
\usepackage{subfig}
\usepackage{epsfig}
\usepackage{pdflscape}
\usepackage{rotating}
\usepackage{lscape}
\usepackage{float}
\usepackage{stfloats}
\usepackage{textgreek}
\usepackage{longtable}
\usepackage{lscape}
\usepackage{capt-of}
\usepackage{comment}
\usepackage{tablefootnote}
\usepackage{footnote}
\usepackage{caption}
\usepackage[autopunct=true]{csquotes}
\usepackage{rotating}
\usepackage{txfonts}
\usepackage[normalem]{ulem}
\usepackage{nccmath}
\usepackage{setspace}
\usepackage{color,xcolor,soul}
\usepackage{amsfonts,amsmath,amssymb}
\usepackage{latexsym,array}
\usepackage{textcomp}
\usepackage{mathtools}

\newcommand{\RomanNumeralCaps}[1]
\linenumbers
\makesavenoteenv{tabular}
\makesavenoteenv{table}
\emergencystretch=\maxdimen
\usepackage[left,displaymath, mathlines]{lineno}

\def\fig{Figure~}
\def\figs{Figures~}
\def\eqn{Eq.~}
\def\eqns{Eqs.~}
\def\tab{Table~}

\def\micro{\textmu}
\begin{document}

\setcounter{page}{1}
%
%
\begin{frontmatter} 
\title{\textcolor{blue}{Piezoelectricity and flexoelectricity in biological cells: \\ The role of cell structure and organelles}}
\author[labela]{Akepogu Venkateshwarlu\corref{coradd}}\ead{avenkateshwarlu@wlu.ca}
\author[labela]{Akshayveer}\ead{aakshayveer@wlu.ca}
\author[labelb]{Sundeep Singh}\ead{sunsingh@upei.ca}
\author[labela]{Roderick Melnik}\ead{rmelnik@wlu.ca}
\address[labela]{MS2Discovery Interdisciplinary Research Institute, Wilfrid Laurier University, 75 University Avenue West, Waterloo, Ontario, N2L 3C5, Canada}
\address[labelb]{Faculty of Sustainable Design Engineering, University of Prince Edward Island, Charlottetown, PE C1A 4P3, Canada}
\cortext[coradd]{\textit{Corresponding author. }}
%
\begin{abstract}
Living tissues experience various external forces on cells, influencing their behaviour, physiology, shape, gene expression, and destiny through interactions with their environment. Despite much research done in this area, challenges remain in our better understanding of the behaviour of the cell in response to external stimuli, including the arrangement, quantity, and shape of organelles within the cell. This study explores the electromechanical behaviour of biological cells, including organelles like microtubules, mitochondria, nuclei, and cell membranes. Two distinct cell structures have been developed to explore the cell responses to mechanical displacement, resembling actual cell shapes. The finite element method has been utilized to integrate the linear piezoelectric and non-local flexoelectric effects accurately. It is found that the longitudinal stress is absent and only the transverse stress plays a crucial role when the mechanical load is imposed on the top side of the cell through compressive displacement. The impact of flexoelectricity is elucidated by introducing a new parameter called \textit{the maximum electric potential ratio} ($V_{\text{R,max}}$). It has been found that $V_{\text{R,max}}$ depends upon the orientation angle and shape of the microtubules. 
Further, the study reveals that the number of microtubules significantly impacts effective elastic and piezoelectric coefficients, affecting cell behavior based on structure, microtubule orientation, and mechanical stress direction.
The insight obtained from the current study can assist in advancements in medical therapies such as tissue engineering and regenerative medicine.
\end{abstract}
\begin{keyword}
Electromechanical coupling \sep Shape of the organelles \sep Electric potential ratio \sep Microtubule's orientation \sep Medical therapies \sep Cell structure \sep Flexoelectricity 
\end{keyword}
\end{frontmatter}
\section{Introduction}
\noindent 
Cell mechanics plays a vital role in the development, functioning, metabolism, and regulation of nuclear responses in cells. 
The primary factors that determine the mechanical properties of the cell are the elasticity of the cell, adhesion between the cell and its substrate, and membrane tension \cite{WANG2015,Tang2021}. The biological cell exhibits a highly intricate internal structure. The cell comprises two primary constituents: the cytoplasm and the nucleus. The cytoplasm is a fluidic matrix that commonly surrounds the nucleus and is enclosed by the cellular plasma membrane \citep{Holy2003,Link2023}.

Organelles are subcellular structures that are microscopic and are located within the cytoplasm. These structures are responsible for carrying out various functions that are essential for maintaining the stability of the cell's internal environment, also known as homeostasis \citep{Basoli2018}. These functions include energy generation, protein synthesis and release, toxin removal, and the reception of external stimuli.
The organelles undergo alterations by changing their shape, quantity, and spatial arrangement inside the cell, which affects their structure and organization \citep{Zhang2007,Lin2012,HoyerFender2013}. Cellular dynamics are directly impacted by processes such as division, merging, morphological changes, transportation, and external stimuli \citep{Fenton2021}. 
For instance, mitochondria are attached to the cytoskeleton at precise sites within the cell. Neurons rely on mitochondrial anchoring to ensure that mitochondria are solidly attached to presynaptic locations in axons, enabling them to provide the necessary energy for neurotransmission \citep{Chada2004,Gutnick2019,Cason2022}.

The regulation of microtubule dynamics is crucial for the proper functioning and division of all eukaryotic cells. The wide range of actions is primarily facilitated by the dynamic ends of microtubules, which can attach to different targets within the cell, produce mechanical forces, and interact with actin microfilaments \citep{RodriguezGarcia2020}. Microtubules are cellular structures that play an essential part in understanding how related proteins, mechanical forces, and microtubule-targeting medicines, such as cancer chemotherapeutics, regulate the mechanisms of microtubule function \citep{Gudimchuk2021, Gudimchuk2023}. Microtubules also play a crucial role in facilitating the transportation of diverse cellular structures, such as lysosomes, mitochondria, and other organelles \citep{Tuszynski2020,Kummer2021}. The structure of microtubules is based on dimers composed of α- and β-tubulin, which exhibit an elongated, non-branching polymer pattern \citep{Li2017}. Each microtubule consists of approximately thirteen dimeric tubulin molecules arranged in a circular configuration \citep{SETAYANDEH2016}. The length of the microtubule can be modified through the addition or removal of dimers. The microtubules, which have rigid structures found within cells, can cause significant strain fluctuations due to their small size and interaction with a flexible cellular matrix. Further, they provide stiffness to the structures, and their number depends on cell type, structural characteristics, and the primary function \citep{Ishikawa2017}. 
Among the many cell types that may be found in a wide variety of organisms, ranging from protists to mammals, cilia and flagella are two structures that are particularly notable and widely recognized. Protists have the ability to arrange microtubules in various diverse formations, which results in the construction of structures that are both unique and comprehensive \cite{Wanderley2010}.
Some motile cilia contain different numbers of microtubules, such as singlets in apicomplexans, fungi, and Caenorhabditis elegans \citep{Mollaret1997,Bezler2022}, symmetrical basal bodies with nine microtubules, and complex organelles like motile cilium with 9+2 structures \citep{Zhang2007,Lin2012,HoyerFender2013}. Axonemes in protozoans, insects, arthropods, and nematodes have varying numbers of microtubule doublets, precisely 10, 12, 14, 16, or 18 \citep{Roggen1966,Deurs1973,Dallai1992,Spreng2019,Ferreira2023}. 
External stimuli increase actin stress fibres in cells exposed to shear stress or geometric patterns, resulting in an increase in the cytoskeletal network \citep{Guolla2012}.

Many computational studies \citep{Liu2012,Yang2015,Joshan2022,Thai2022,WANG2023} have focused on the mechanical coupling between strain and stress in nanowires and plates. It has been reported that a significant proportion of biological components, such as microtubules, bones, and collagen, exhibit piezoelectric properties. For instance, during walking, the human tibia produces a piezoelectric potential of approximately 300 \micro V \citep{Iwao1977}. Therefore, it is important to conduct a thorough investigation into the mechanics of the cell and its organelles in order to gain a better understanding of their piezoelectric properties. 
The discovery of piezoelectricity has a long history, starting with the Curie brothers, who, in 1880, observed that some crystals display both positive and negative charges under pressure, which disappear once the pressure is released. This resulted in their discovery of piezoelectricity. Later, scientists used the term ``piezoelectricity" to describe the direct relationship between mechanical stress and electricity \citep{Fukada1955,Fukada1957,Fukada1964,Fukada1968,Fukada1983,Fukada2000, Chae2018}. Since then, piezoelectricity,understood as the direct relationship between electrical and mechanical energy, is widely used in various applications, including artificial muscles, wearable sensors, advanced microscopes, flexible actuators, energy harvesting devices, and minimally invasive surgery  \citep{SHAMOS1967,ANDERSON1970,Fukada1983,Fletcher2010,He2017,LUN2020,ATEF2022,Rout2023}.
On the other hand, flexoelectricity is a bidirectional linear interaction between the electric field and strain gradients, unlike piezoelectricity, which occurs due to electric polarization and strain. It is characterized by strain gradients or inhomogeneous strains, while uniform strains cause piezoelectricity \citep{DENG2014,LUN2020,ATEF2022}. While non-centrosymmetric materials in physical and biological systems exhibit linear piezoelectricity, which is the coupling of a single strain component with a single electric field component \citep{He2017,Rout2023}, flexoelectricity can occur even in materials with centrosymmetry, especially when inhomogeneities at micro length scales lead to significant strain gradients and electric field formation \citep{Deng2017,Abdollahi2019}. Despite their minimal effects, often at larger scales, flexoelectric contributions to electromechanical coupling are likely to be substantial at smaller scales \citep{Nguyen2013,Thai2022}.

Given these coupled effects, it becomes apparent that further research is necessary to fully understand how living cells behave, especially in the context of their variable organelle shapes. The structure and orientation of the organelles are essential for comprehending the dynamics of the cell. Moreover, the quantity of microtubules within a cell plays a crucial role in cell dynamics. These aspects have not yet been thoroughly investigated \citep{Spreng2019,Ferreira2023}. Recent experimental studies \citep{Zhang2023,Kobori2024} show that microtubules are mostly composed of $\alpha$- and $\beta$-tubulin dimerization, resulting in a polarized tubular structure. Cells can extend and contract at the plus end. Cell morphology may depend on microtubule reorientation. Non-centrosomal microtubules (NC-MTs) may affect cell adhesion and division, depending on cell type. However, these implications have not been explored in the context of piezoelectric and flexoelectric effects. Despite notable progress, there is still a considerable dearth of understanding the functioning of individual organelles and their integrated behaviour within the living cells when they are exposed to external stimuli such as piezoelectricity and flexoelectricity \citep{Fletcher2010,He2017,LUN2020,ATEF2022,Thai2022,Rout2023}. 
The significance of mechanical stimuli in regulating cell functions is becoming even more apparent, given the fact that the understanding of the immediate structural reaction of cells to mechanical stresses is still inadequate \citep{Guolla2012}. 
 
Therefore, the primary aim of this work is to investigate the impact of externally exerted mechanical displacement on the characteristics of individual cells, with a specific emphasis on the variable shapes and sizes of organelles. In what follows, we carry out an extensive parametric analysis to examine the effects of piezoelectricity and flexoelectricity by varying the number of microtubules from 2 to 18 along with their shapes when the cells are subjected to the mechanical displacements. The mechanical response of the cell is elucidated by determining the effective elastic coefficients, and the contribution of piezoelectricity resulting from mechanical displacement is quantified through the effective piezoelectric coefficients.
 
The rest of the paper has been structured as follows: Section \ref{Sec:PoM} describes the model that governs the coupled electro-elastic behaviour of the cells, including the geometry of the cell structure and the boundary conditions. 
The results obtained from the present study, which include the effects of piezoelectricity and flexoelectricity on the distribution of electric potential, strain, effective elastic coefficients, and effective piezoelectric coefficients, are delineated in Section \ref{Section:RandD}. The findings are summarized in Section \ref{Conlusions}.
\section{Physics-based model of the biological cell}\label{Sec:PoM}
\begin{figure}[htbp]
	\centering
\subfloat[Cell structure-1 (CS-1)]{\includegraphics[width=0.7\linewidth]{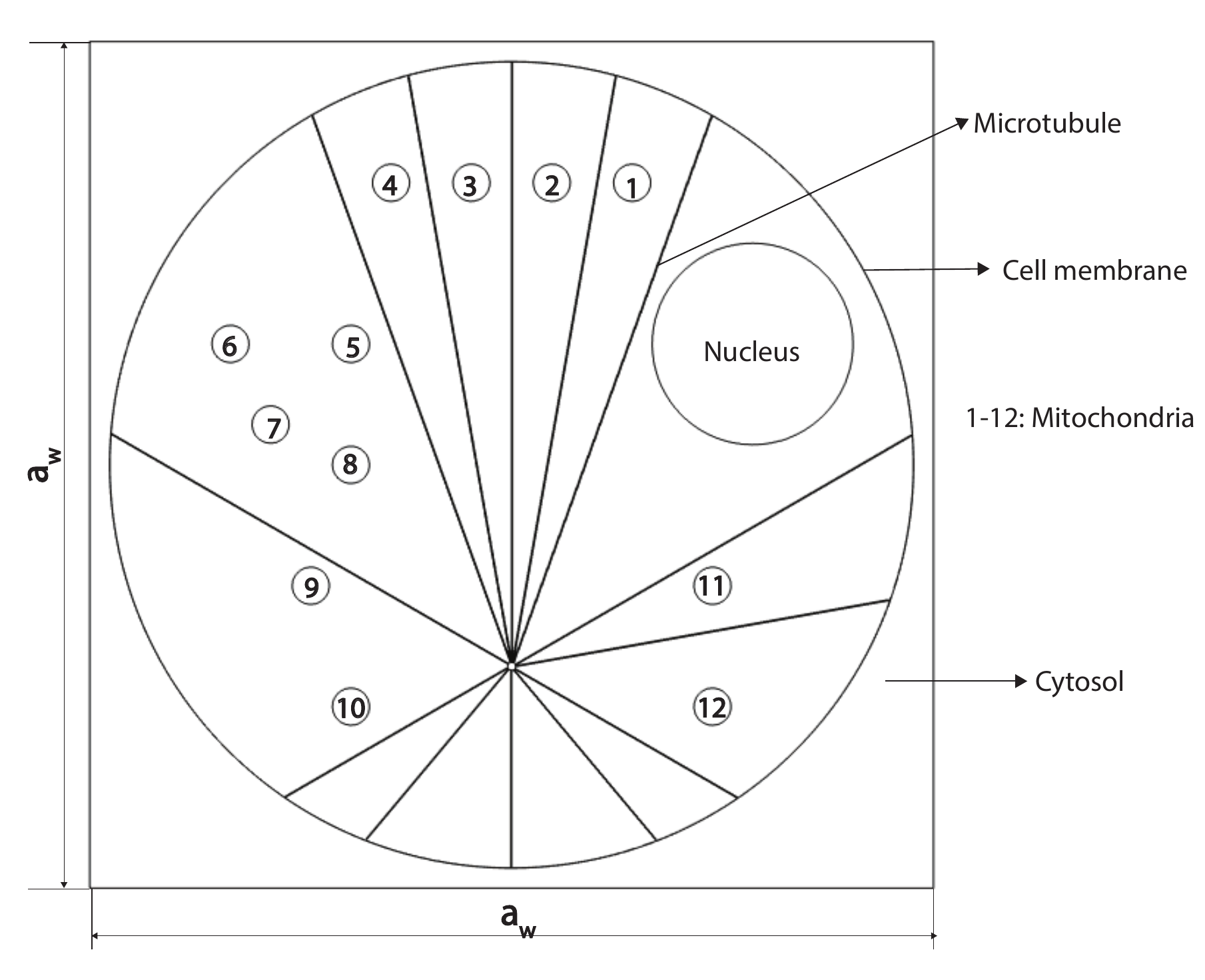}}\\
\subfloat[Cell structure-2 (CS-2)]{\includegraphics[width=0.7\linewidth]{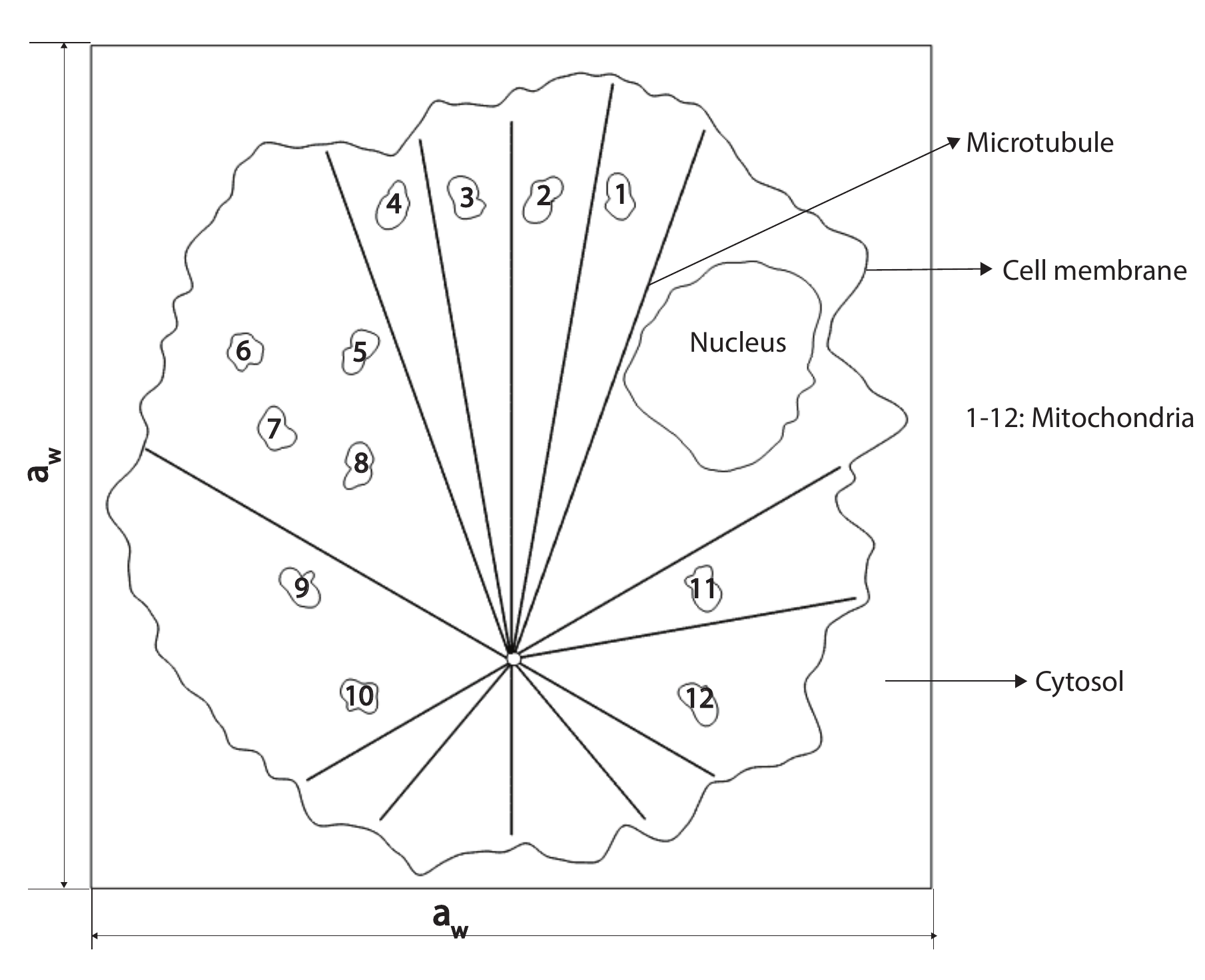}}
	\caption{Schematic representation of the structure of the two-dimensional biological cell, including various organelles such as the cell membrane, nucleus, mitochondria, and microtubules, with (a) idealized shapes and (b) shapes that closely resemble the experimental studies.}
	\label{fig:1}
\end{figure}

\noindent This section details the developed model of the biological cell, including material properties, mathematical framework, numerical setup, and boundary conditions necessary for problem solving.
The cell has a width of $a_{w}=21$ \micro m \citep{Connie2016b,Natasha2016a,SUNDEEPSINGH2020}. It contains distinct organelles with well-defined shapes in cell structure-1 in the $x_{1}-x_{3}$ plane. However, the realistic organelles lack a specific and well-defined shape. Therefore, we have also selected arbitrary shapes for intracellular organelles, such as the nucleus, mitochondria, cytoplasm, microtubules, cell membrane, and centrosome, for cell structure-2. The arbitrary shapes of the organelles have been created using an in-house developed MATLAB code. The organelles possess unique structures that exhibit a diverse range of sizes and shapes within cellular compartments. For instance, mitochondria are portrayed as rigid, long, cylinder-shaped structures that resemble bacteria. They are highly elastic and dynamic organelles that are always undergoing shape changes \citep{Uzman2003,Parker2017,Fenton2021}. 
The morphology of the nucleus \citep{Tange2002,Norppa2003,Robert2004,Webster2009,Khatau2012}, cell membrane \citep{Norppa2003,MANNELLA2006,BOZELLI2020}, and mitochondria \citep{Kalt1975,Paumard2002,MANNELLA2006,VELOURS2009} exhibits distinct characteristics in the cell structures, as shown in \fig\ref{fig:1}. 
The shape of the cell model has been created based on the experimental studies \citep{Vaziri2008,Connie2016a,Natasha2016a,Moujaber2020,Parker2017} as depicted in \figs\ref{fig:1}(a) and \ref{fig:1}(b). Similarly, the shapes of organelles such as the nucleus \citep{Tange2002,Norppa2003,Robert2004,Webster2009,Khatau2012}, cell membrane \citep{Norppa2003,MANNELLA2006,BOZELLI2020}, and mitochondria \citep{Kalt1975,Paumard2002,MANNELLA2006,VELOURS2009}, are modelled based on experimental and other computational studies. 
The dimensions of the organelles have been taken as follows \citep{Ofek2009}: nucleus, 5 \micro m \citep{Connie2016b}; mitochondria, 0.92 \micro m \citep{Connie2016b}; microtubule, 25 nm (diameter), length 5-15 \micro m \citep{Cooper2000,Moujaber2020,Parker2017}; cell membrane, 20 \micro m (diameter), 7 \micro m \citep{Connie2016b,Natasha2016a}. 

Both cell structures are presumed to possess an equivalent quantity of organelles, comprising 12 mitochondria, 18 microtubules, 1 nucleus, 1 cytosol, and 1 cell membrane, for the purpose of conducting a comparison analysis. 
Furthermore, it has been considered that all intercellular organelles have identical surface areas compared to their counterparts in the cell structures.
The lengths of the microtubules are varied, and their inward ends have been connected to the centrosome. The cell membrane has been conceptualized as a structural barrier that delineates the intracellular compartment of the cell from the extracellular matrix environment \citep{Gudimchuk2021}.

The material properties pertaining to the organelles that have been utilized in the current study are outlined in \tab\ref{tab:2b}. This study focuses on the piezoelectric and flexoelectric coefficients that relate to microtubules, while neglecting the consideration of these coefficients for other organelles; they are assumed to be zero \citep{SUNDEEPSINGH2020}. 

\begin{table}[!t]
	\begin{center}
		\caption{Physical properties of the organelles considered in the current study.}\label{tab:2b}
		\scalebox{0.9}
		{
			\begin{tabular}{p{0.4in}p{1.0in}p{1.2in}p{1.2in}p{1.4in}} \hline
				S.No. &Parameter &   Elastic modulus (E)  & Poisson's ratio ($\nu$) &Relative permittivity (Dielectric constant)   \\ \hline
				1 & Nucleus & 1 kPa \citep{KATTI2017} & 0.3 \citep{KATTI2017}& 80 \citep{GOWRISHANKAR2006,Tiwari2009} \\
				2 & Mitochondria & 50.3 kPa \citep{Yongbo2015} & 0.3 \citep{SUNDEEPSINGH2020} & 80 \citep{GOWRISHANKAR2006,Tiwari2009} \\
				3 & Microtubule & 2 GPa \citep{BARRETO2013}  & 0.3  &40 \citep{Yongbo2015}  \\ 
				 &  & 1.9 GPa \citep{KATTI2017}  & 0.3  \citep{PAMPALONI2008,BARRETO2013,KATTI2017}  &\\  
				4 & Cell membrane & 1.8 kPa \citep{KATTI2017}  & 0.3 \citep{GOWRISHANKAR2006,Tiwari2009} & 3 \citep{GOWRISHANKAR2006,Tiwari2009}\\ 
				5 & Cytoplasm & 0.25 kPa \citep{Ingber2000,BARRETO2013,KATTI2017} & 0.49 \citep{Ingber2000,GOWRISHANKAR2006,Tiwari2009,BARRETO2013} & 80 \citep{GOWRISHANKAR2006,Tiwari2009}\\
				6 & Actin cortex & 2 kPa \citep{STRICKER2010,BARRETO2013} &0.3 \citep{STRICKER2010,BARRETO2013} \\ \hline
			\end{tabular}
		}
	\end{center}
\end{table}
The fully coupled electromechanical model, based on the extended linear theory of continuum mechanics for dielectric continua, has been used to analyze the two-dimensional biological cell domain in a steady-state regime. 
Under equilibrium conditions and based on Gauss's law, the governing equations for the current two-dimensional computational domain are given as:
\begin{gather}
	\left(\sigma_{{ij}}-\hat{\sigma}_{ij,k} \right)_{,j}=0, \\
	D_{i,i}=0,
	\label{eqn:06}
\end{gather}
where $\sigma_{{ij}}$ are the components of the mechanical stress tensor, $D_{{i,i}}$ are the components of the electric displacement vector, and $\hat{\sigma}_{ijk}$ is the higher-order flexoelectric tensor. $\sigma_{{ij}}$, $D_{{i,i}}$, and $\hat{\sigma}_{ijk}$ have been calculated by their fundamental constitutive relationships based on the Gibbs free energy equation (\textit{G}) as follows:
\begin{gather}
	G=\frac{1}{2}c_{{ijkl}}\varepsilon_{{ij}}\varepsilon_{{kl}}-e_{{kij}}E_{k}\varepsilon_{{ij}}-\frac{1}{2}\epsilon_{ij}E_{i}E_{j}-\frac{1}{2}B_{klij}E_{k}E_{l}\varepsilon_{ij}-\mu_{ijkl}E_{i}\varepsilon_{{jk,l}}, \label{eqn:1} \\
	\sigma_{{ij}}=\frac{\partial G}{\partial \varepsilon_{{ij}}}=c_{{ijkl}}\varepsilon_{{kl}}-e_{{kij}}E_{k}, \label{eqn:01} \\
	\hat{\sigma}_{ijk}=\frac{\partial G}{\partial \varepsilon_{{jk,l}}}=-\mu_{ijkl}E_{i}=\mu_{lijk}E_{l}, \label{eqn:02}\\
	D_{{i}}=-\frac{\partial G}{\partial E_{i}}=\epsilon_{ij}E_{j}+e_{{ijk}}\varepsilon_{{jk}}+\mu_{ijkl}\varepsilon_{{jk,l}}. \label{eqn:03} 
\end{gather}

In \eqn\ref{eqn:1}, $\frac{1}{2}c_{{ijkl}}\varepsilon_{{ij}}\varepsilon_{{kl}}$ is the elastic deformation energy, $e_{{kij}}E_{k}\varepsilon_{{ij}}$ is piezoelectric energy, $\frac{1}{2}\epsilon_{ij}E_{i}E_{j}$ is dielectric energy (it is purely due to electric field and electric properties), $\frac{1}{2}B_{klij}E_{k}E_{l}\varepsilon_{ij}$ is electrostrictive energy, and $\mu_{ijkl}E_{i}\varepsilon_{{jk,l}}$ is the flexoelectric energy. In the present study, the electrostrictive energy term has been neglected.
Moreover, $\varepsilon_{{ij}}$ are the components of the mechanical strain tensor, $E_{k}$ are the components of the electric field vector, $c_{{ijkl}}$ are the linear elastic coefficients in the stiffness matrix, $e_{{ijk}}$ are the linear piezoelectric coefficients, and $\epsilon_{{jk}}$ are the dielectric permittivity coefficients, and $\mu_{ijkl} \varepsilon_{jk,l}$ is the fourth-order flexoelectric tensor.

By employing the two-dimensional biological cell model described above and assuming the presence of transversely isotropic piezoelectric material along the $x_3$-axis, the electromechanical phenomena can be investigated using either the $x_1$-$x_3$ or $x_2$-$x_3$ planes. Therefore, when considering the $x_1$-$x_3$ plane, using the Voigt notation, the constitutive relations (\eqns\ref{eqn:01}-\ref{eqn:03}) can be reduced into matrix form, as follows \cite{Krishnaswamy2019a,Krishnaswamy2019b,Krishnaswamy2019c,SUNDEEPSINGH2020}:
\begin{gather}
	\begin{bmatrix}
		\sigma_{11} \\ \sigma_{33}\\ \sigma_{13}
	\end{bmatrix}=
	\begin{bmatrix}
		c_{11} & c_{13} & 0 & \\
		c_{13} & c_{33} & 0 & \\
		0 & 0 & c_{44}  
	\end{bmatrix}
	\begin{bmatrix}
		\varepsilon_{11} \\ \varepsilon_{33}\\ 2\varepsilon_{13}
	\end{bmatrix}-
	\begin{bmatrix}
		0 & e_{31} \\
		0 & e_{33} \\
		e_{15} & 0  
	\end{bmatrix}
	\begin{bmatrix}
		E_{1} \\
		E_{3} 
	\end{bmatrix},
	\label{eqn:10}
\end{gather}
\begin{gather}
	\begin{bmatrix}
		D_{1} \\ D_{3}
	\end{bmatrix}=
	\begin{bmatrix}
		0 & 0 & e_{15} \\
		e_{31} & e_{33} & 0 
	\end{bmatrix}
	\begin{bmatrix}
		\varepsilon_{11} \\ \varepsilon_{33}\\ 2\varepsilon_{13}
	\end{bmatrix}+
	\begin{bmatrix}
		\epsilon_{11} & 0\\
		0 & \epsilon_{33} 
	\end{bmatrix}
	\begin{bmatrix}
		E_{1} \\
		E_{3} 
	\end{bmatrix}+
	\begin{bmatrix}
		\mu_{1111} & \mu_{1331}\\
		\mu_{3113} & \mu_{3333} 
	\end{bmatrix}
	\begin{bmatrix}
		\varepsilon_{11,1} & \varepsilon_{33,1}\\
		\varepsilon_{11,3} & \varepsilon_{33,3} 
	\end{bmatrix}.
	\label{eqn:11}
\end{gather}

The relationship between the strain field and displacement vector, electric potential, and electric field is mathematically expressed as 
\begin{gather}
	\varepsilon_{{ij}}=\frac{1}{2}\left(u_{{i,j}}+u_{{j,i}}\right),\\
	E_{i}=-V_{,i}. \label{eqn:05}
\end{gather}

Collagen and microtubule-associated tau proteins exhibit comparable crystalline homology \cite{Garcini1990}. As a result, the generation of piezoelectric potential is comparable in both proteins when subjected to similar load conditions \cite{Kushagra2015}. Therefore, in the present study, it has been assumed that the piezoelectric coefficients of the microtubules are equivalent to those of collagen. The hexagonal symmetry has been taken into consideration in order to determine each piezoelectric coefficient for collagen. The piezoelectric tensor can be represented using Voigt's notation as follows \cite{Denning2017,SUNDEEPSINGH2020}:
\begin{gather}
	d_{{ij}}=\begin{bmatrix}
		0 & 0 & 0 & d_{{14}} & d_{{15}} & 0 \\
		0 & 0 & 0 & d_{{15}} & -d_{{14}} & 0 \\
		d_{{31}} & d_{{31}} & d_{{33}} & 0 & 0 & 0
	\end{bmatrix}. 	\label{eqn:08}
\end{gather}
In the piezoelectric tensor ($d_{{ij}}$), the subscript $i$ signifies the direction of the electric field displacement, whereas the subscript $j$ represents the corresponding mechanical deformation.

Given that the current study is based on the stress form, the piezoelectric coefficients provided in \eqn\ref{eqn:08} in strain form have been transformed into the stress form using the following relationship:
\begin{gather}
	e_{{ijk}}=c_{{jklm}} d_{{ilm}}, \label{eqn:09}
\end{gather}
where $e_{{ijk}}$ are the piezoelectric stress coefficients, $c_{{jklm}}$ are the components of the elastic tensor, and $d_{{ilm}}$ are the piezoelectric strain coefficients.

The orientation of the microtubules and their effect on the piezoelectric tensor have also been taken into account in this work. The transformation of the piezoelectric coefficients has been formulated elsewhere \citep{SUNDEEPSINGH2020}.
\subsection{Boundary conditions} \label{BC}
\noindent The details of the biological cell geometry and the governing equations to understand the piezoelectric behaviour of the biological cell are provided previously. 
Here, we discuss the impact of mechanical displacements on the biological cell, specifically examining three different boundary conditions to understand the cell response comprehensively, as shown in \tab\ref{tab:2c}. The bottom surface of the two-dimensional biological cell model depicted in \fig\ref{fig:1b} has been subjected to fixed boundary conditions that are electrically grounded ($V=0$). These restrictions have been adopted from previous studies \citep{GARCIA2018,Garcia2018b,Li2018,Marcotti2019} and inspired by experimental investigations \citep{Roggen1966,Deurs1973,Dallai1992,Mollaret1997,Zhang2007,Lin2012,HoyerFender2013,Ishikawa2017,Nakamura2021,Bezler2022}. The direction in which the mechanical displacement is exerted on the piezoelectric material is as follows:
\begin{figure}[!t]
	\centering
	\subfloat[BC-1]{\includegraphics[width=0.49\linewidth]{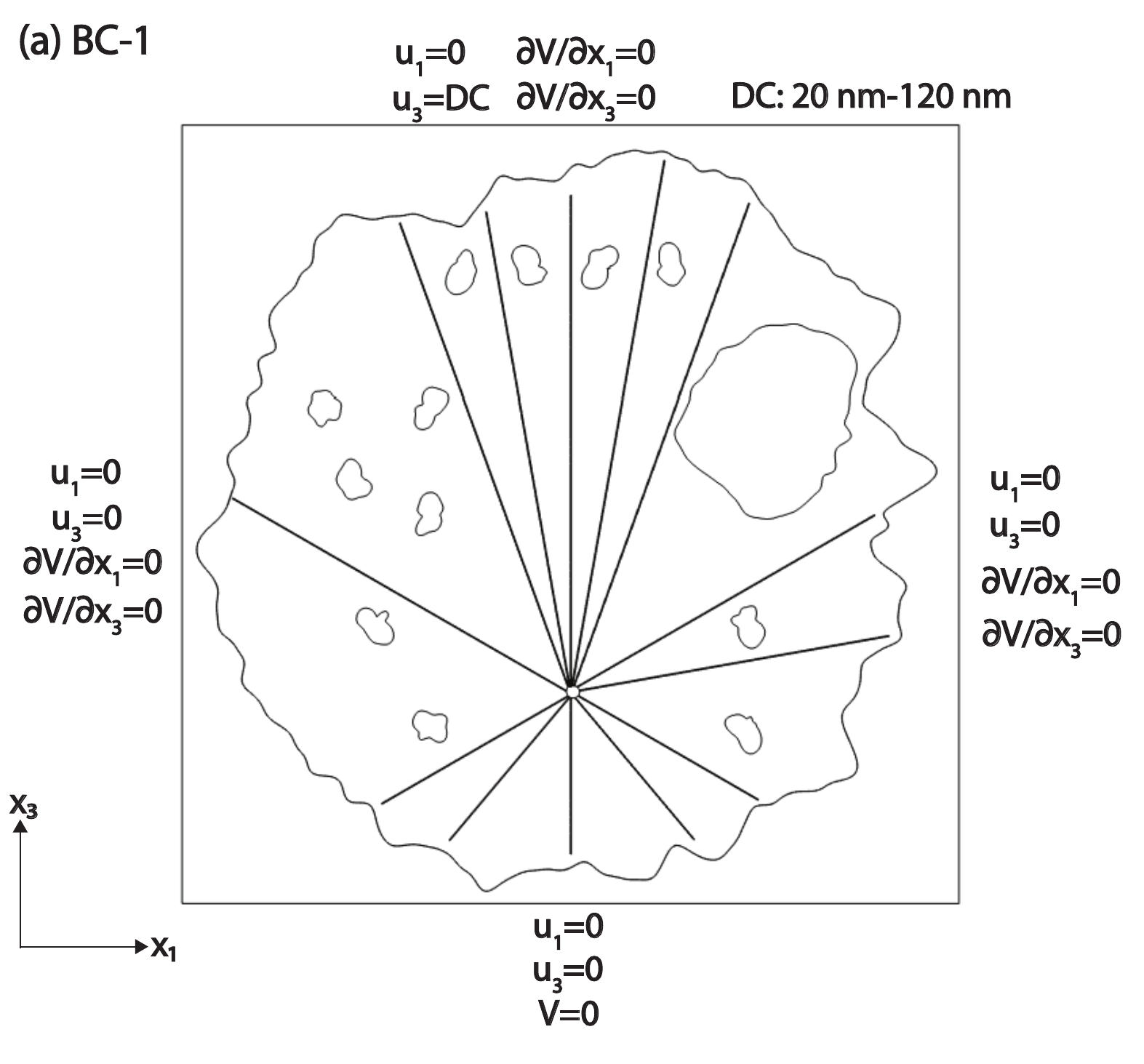}}
	\subfloat[BC-2]{\includegraphics[width=0.49\linewidth]{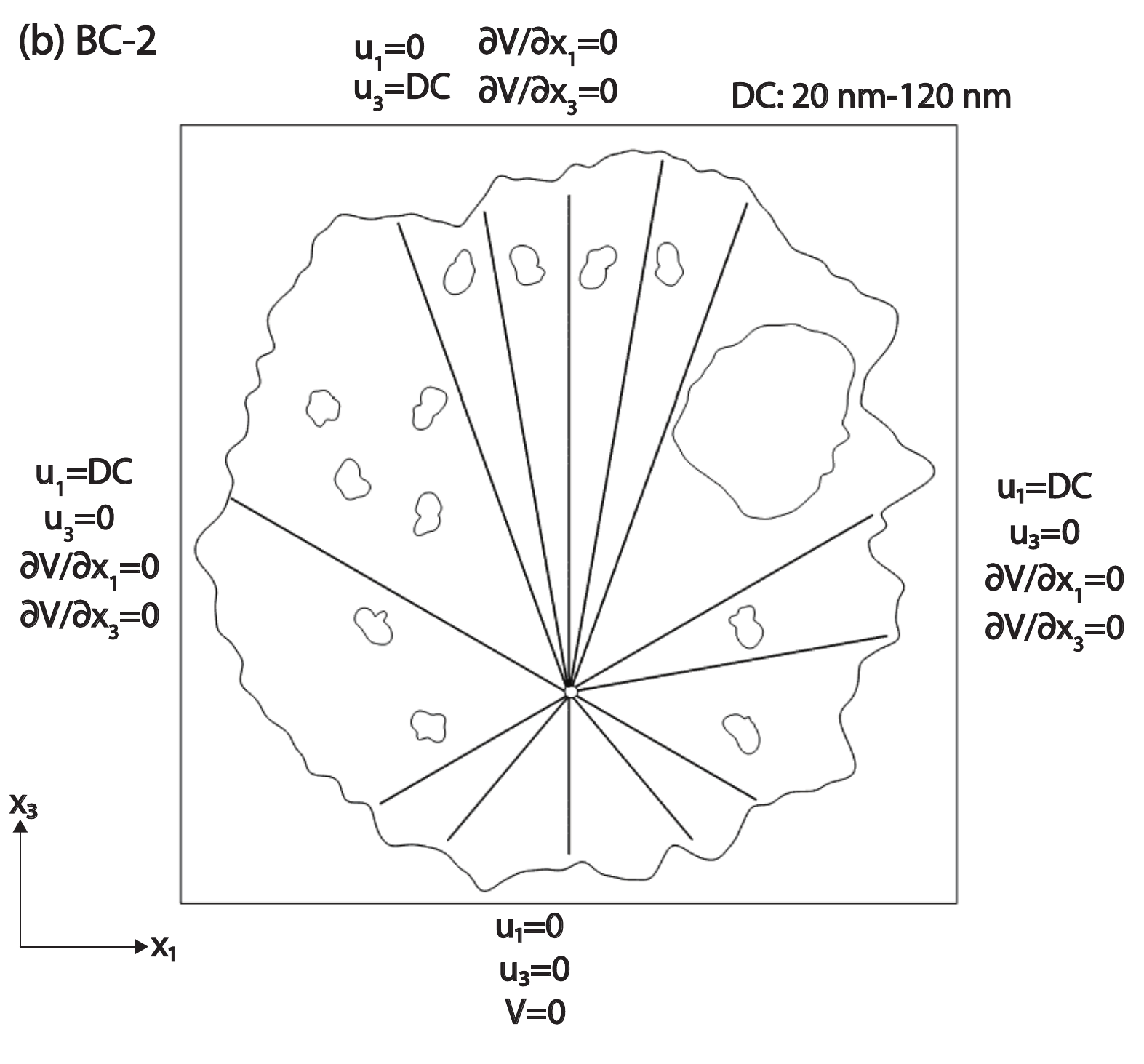}} \\
	\subfloat[BC-3]{\includegraphics[width=0.49\linewidth]{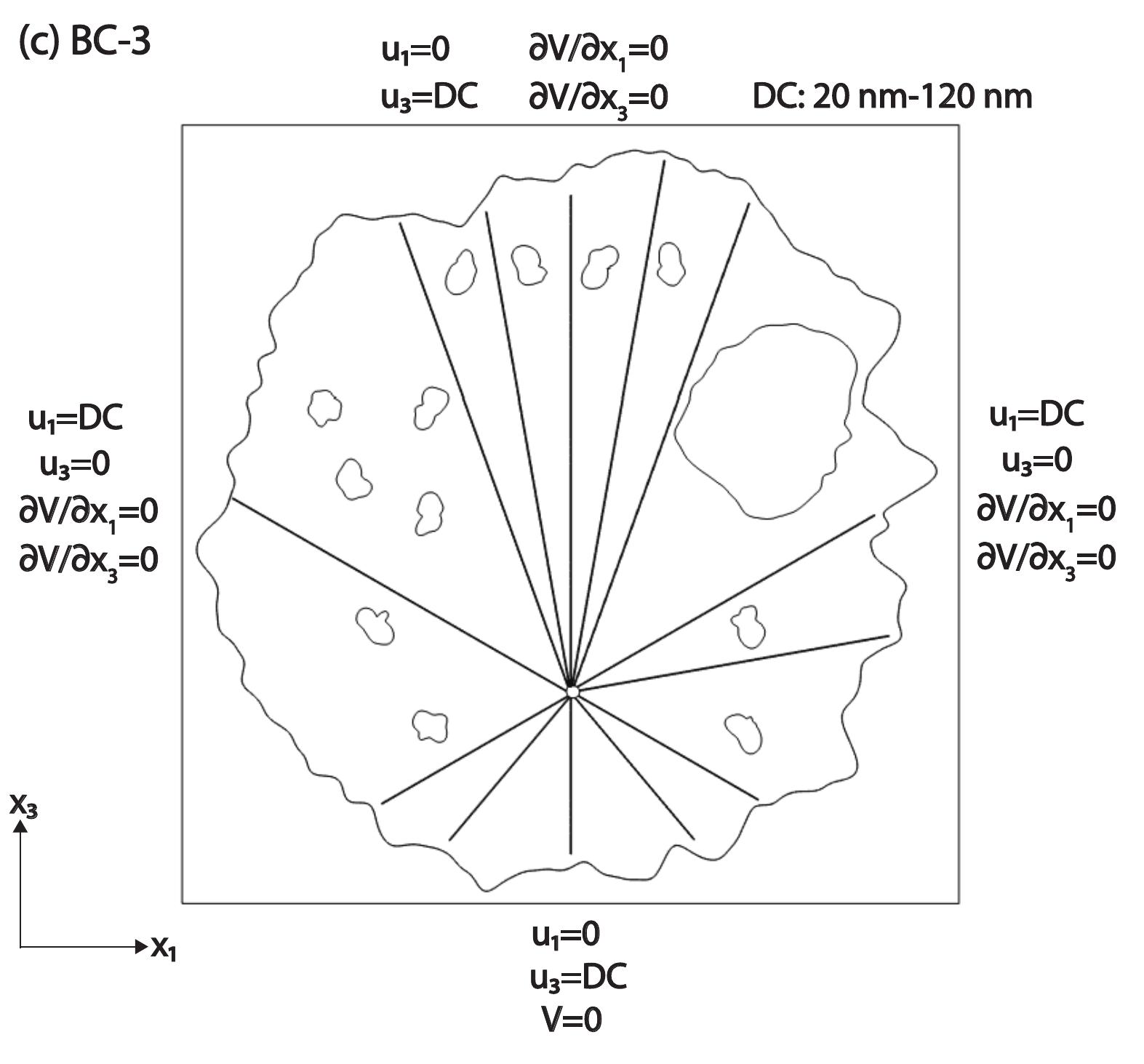}}
	\caption{A finite element model of a single cell is created with specific boundary conditions (BC). Compressive displacement (DC) in terms of $u_{1}$ and $u_{3}$ is applied to the cell, ranging from 20 to 120nm, in three different configurations: (a) on the top of the cell, (b) on the left and right sides of the cell, and (c) on the top, bottom, left, and right sides of the cell, while the bottom is electrically grounded in all the cases, i.e., the electric potential ($V$) is set to zero.}
	\label{fig:1b}
\end{figure}
\begin{table}[!t]
	\begin{center}
		\caption{Boundary conditions considered in the current study.}\label{tab:2c}
		\scalebox{0.9}
		{
			\begin{tabular}{|p{0.8in}|p{0.7in}|p{0.7in}|p{0.7in}|p{0.7in}|} \hline
			Boundary condition & Left & Right & Top & Bottom  \\ \hline
			BC-1 & $u_1=0$, $u_3=0$, $\frac{\partial V}{\partial x_1}=0 $  & $u_1=0$, $u_3=0$, $\frac{\partial V}{\partial x_1}=0 $  & $u_1=0$, $u_3=DC$, $\frac{\partial V}{\partial x_1}=0 $ & $u_1=0$, $u_3=0$, $V=0$ \\ \hline
			BC-2 & $u_1=DC$, $u_3=0$, $\frac{\partial V}{\partial x_1}=0 $  & $u_1=DC$, $u_3=0$, $\frac{\partial V}{\partial x_1}=0 $  & $u_1=0$, $u_3=DC$, $\frac{\partial V}{\partial x_1}=0 $ & $u_1=0$, $u_3=0$, $V=0$ \\ \hline
			BC-3 & $u_1=DC$, $u_3=0$, $\frac{\partial V}{\partial x_1}=0 $  & $u_1=DC$, $u_3=0$, $\frac{\partial V}{\partial x_1}=0 $  & $u_1=0$, $u_3=DC$, $\frac{\partial V}{\partial x_1}=0 $ & $u_1=0$, $u_3=DC$, $V=0$ \\ \hline
			\end{tabular}
		}
	\end{center}
\end{table}
\begin{itemize}
	\item Boundary condition-1 (BC-1): A mechanical displacement is exerted at the top of the boundary by altering the compressive displacement (DC) within the range of 20 nm to 120 nm, as depicted in \fig\ref{fig:1b}(a).
	\item Boundary condition 2 (BC-2) involves the application of a mechanical displacement at the left, right, and top boundaries by altering the compressive displacement within the range of 20 nm to 120 nm, as depicted in \fig\ref{fig:1b}(b).
	\item Boundary condition-3 (BC-3) deals with the imposition of a mechanical displacement at the left, right, top, and bottom boundaries of the system. This force is exerted by altering the compressive displacement values within the range of 20 nm to 120 nm, as depicted in \fig\ref{fig:1b}(c).
\end{itemize}

These boundary conditions are very important to understand the cell mechanics in many applications, including bone repair \citep{Wang2007,Tichy2010,Nakamura2021,Zhang2021,Chunyu2022}.

To investigate the mechanical response and piezoelectric behavior of the biological cell more precisely, effective coefficients have been determined by creating many load instances with varying boundary conditions. Each load instance should be structured in such a way that only one value in the strain/electric field vector is non-zero, with all other values being zero. To determine the effective coefficient, use an average non-zero value in the strain/electric field vector and the average values in the stress/electrical displacement vector \citep{Berger2005,Dascalescu2014,Saputra2017,Hamdia2024}.
The effective elastic coefficients and effective piezoelectric coefficients are computed in the principal plane directions ($x_1$-$x_3$) under the specified boundary conditions, i.e., the mechanical ($c_{11,eff}$, $c_{13,eff}$, and $c_{33,eff}$) and piezoelectric response ($e_{31,eff}$, and $e_{33,eff}$) of the biological cell in the transverse and longitudinal directions \citep{Krishnaswamy2019a,Krishnaswamy2019c}. 
The volume average of quantity $A$ is denoted by $\langle A \rangle$ in the calculation below. It is determined as
\begin{gather}
	\langle A \rangle=\frac{1}{(a_{w} a_{w})} \int_{\Omega} A d \Omega,
	\label{eqn:1_1}
\end{gather}
where $\Omega$ represents the volume of the domain in which the integration is performed. $A$ denotes the area of the biological cell, while $a_{w}$ represents the width of the cell.

The effective elastic and piezoelectric coefficients are calculated by utilizing the volume averages for the boundary conditions specified in Section \ref{BC}, which are as follows \citep{Krishnaswamy2019a}:
\begin{gather}
	c_{11,eff}=\frac{{\langle \sigma_{11} \rangle}}{\varepsilon_{\text 11}}, \quad c_{13,eff}=\frac{{\langle \sigma_{11} \rangle}}{\varepsilon_{\text 33}}, \quad c_{33,eff}=\frac{{\langle \sigma_{33} \rangle}}{\varepsilon_{\text 33}}, \nonumber \\
	e_{31,eff}=\frac{\langle D_{3} \rangle }{\varepsilon_{\text 11}}, \quad e_{33,eff}=\frac{\langle D_{3} \rangle }{\varepsilon_{\text 33}},
	\label{eqn:1_2}
\end{gather}
where $\langle D_{3} \rangle$ is the volume average of the $D_{3}$ component of the electric flux density vector.

The finite element analysis has been performed to numerically implement the coupled electromechanical model described above using COMSOL multiphysics while applying boundary conditions to the biological cell.
\section{Results and discussion} \label{Section:RandD}
\noindent Cell mechanics refers to the division and merging of cell organelles, which are crucial for maintaining their structure, distribution, and size. In order to investigate the cell behaviour when it is subjected to mechanical displacement, three boundary conditions (BC-1 to BC-3) are employed for the compressive displacement ranging from 20 nm to 120 nm. The current study presents a comparative analysis of the piezoelectric effect in the cell, and analyzing the outcomes both with and without consideration of flexoelectric effects for the strain distribution, electric potential distribution, and determination of the effective elastic and piezoelectric coefficients. 
\subsection{Impact of piezoelectric and flexoelectric effects on strain distribution in biological cells}
\begin{figure}[hpbt]
	\centering
	\subfloat[CS-1, BC-1]{\includegraphics[width=0.43\linewidth]{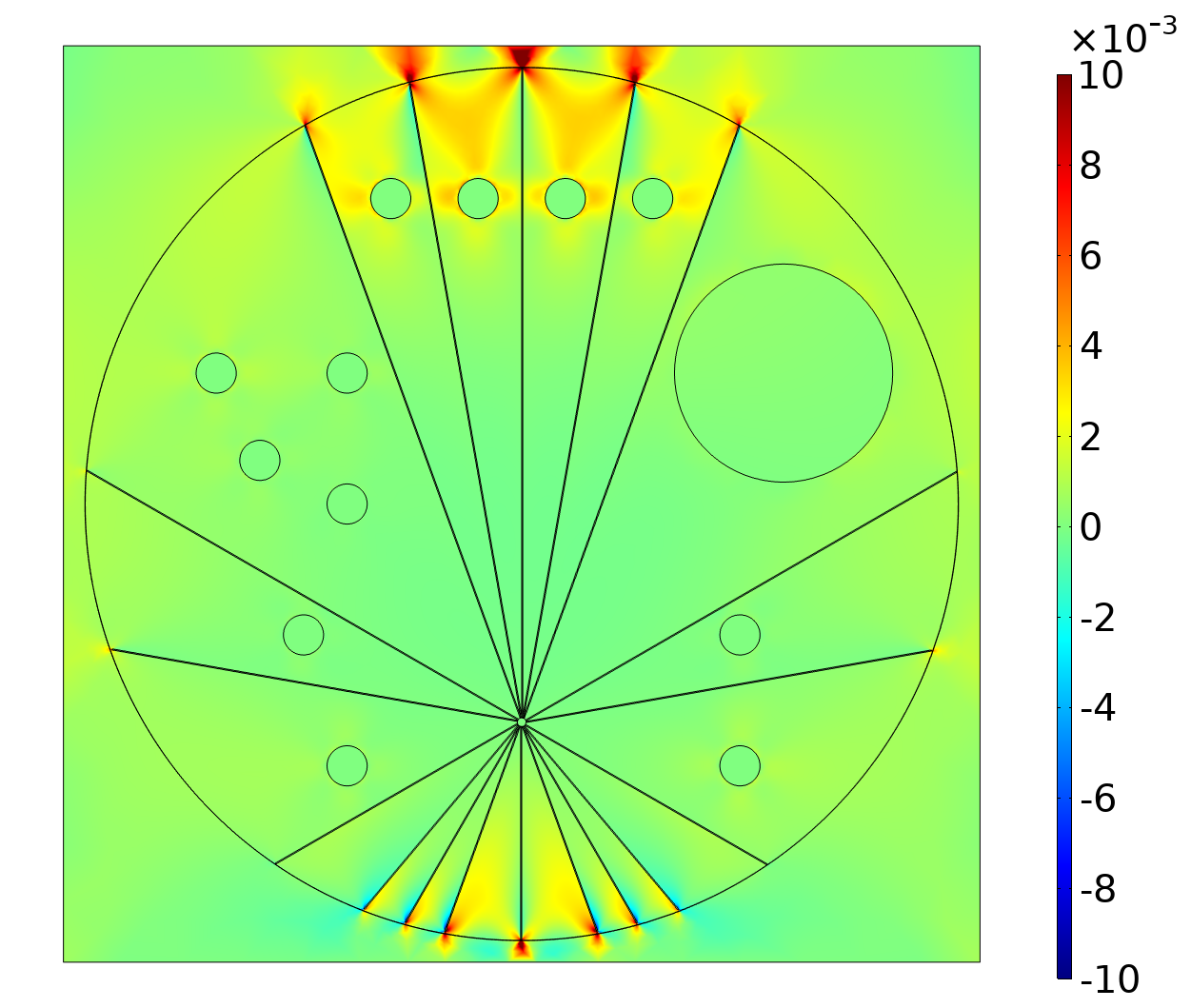}}
	\subfloat[CS-2, BC-1]{\includegraphics[width=0.43\linewidth]{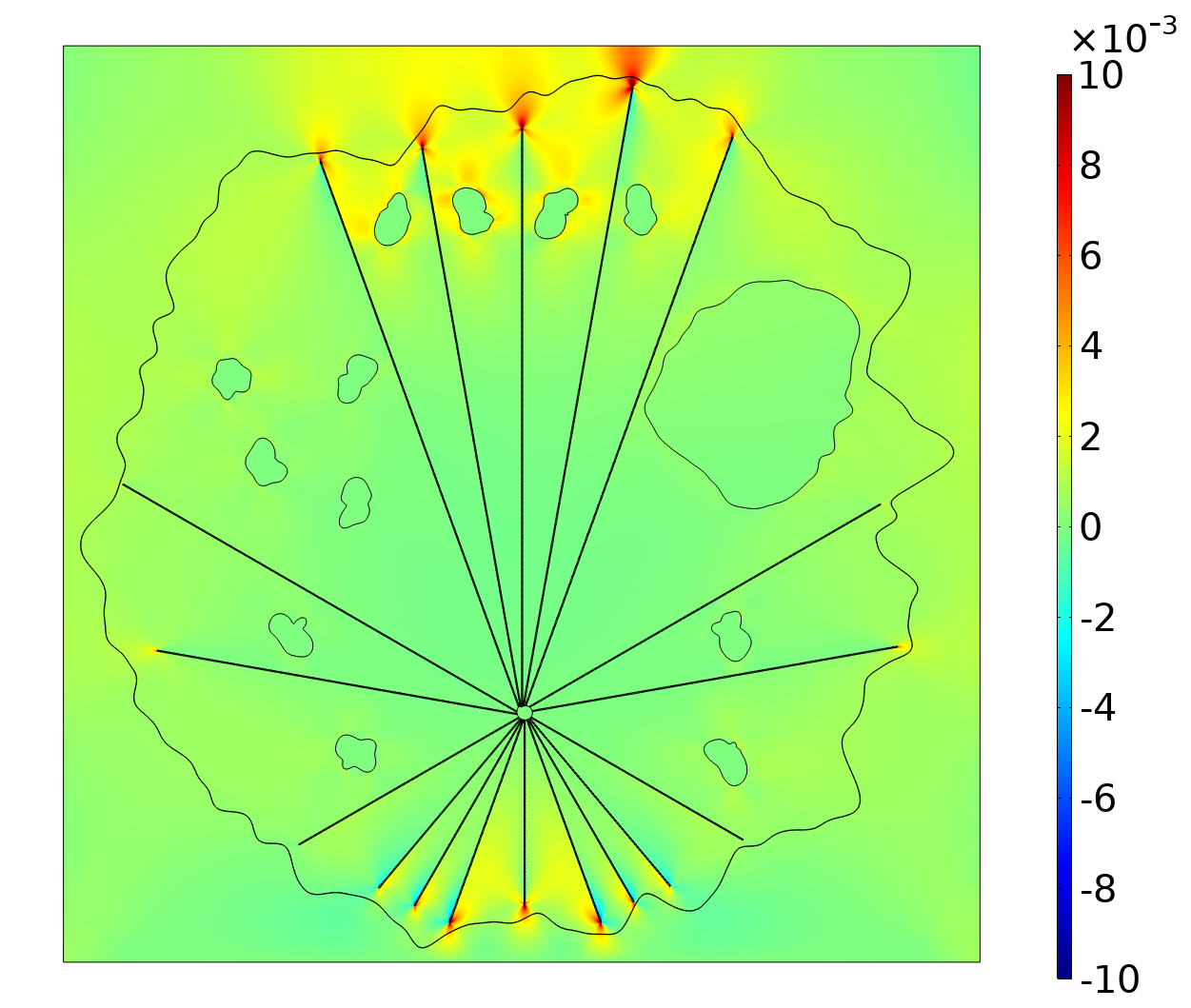}}\\
	\subfloat[CS-1, BC-2]{\includegraphics[width=0.43\linewidth]{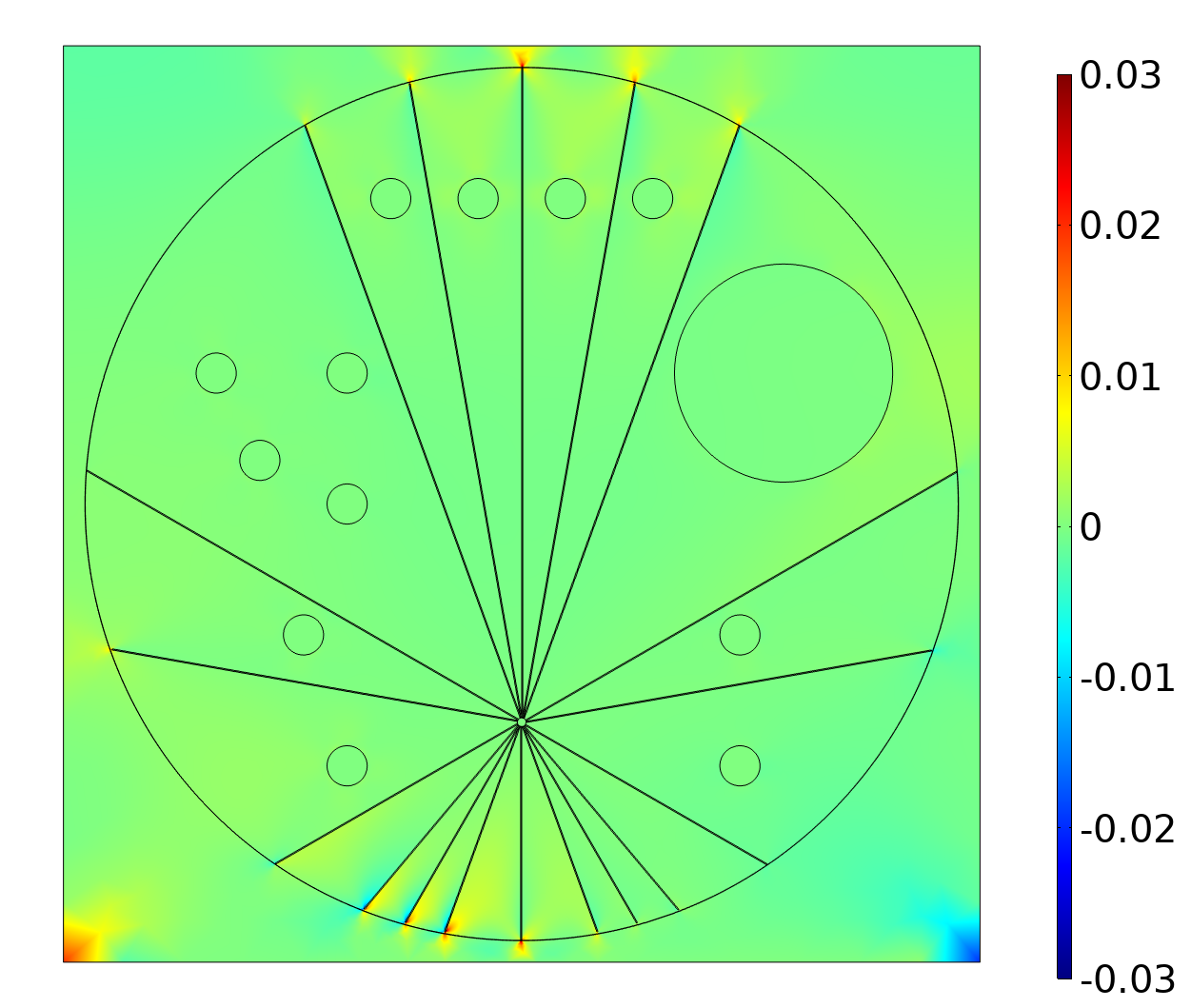}}
	\subfloat[CS-2, BC-2]{\includegraphics[width=0.43\linewidth]{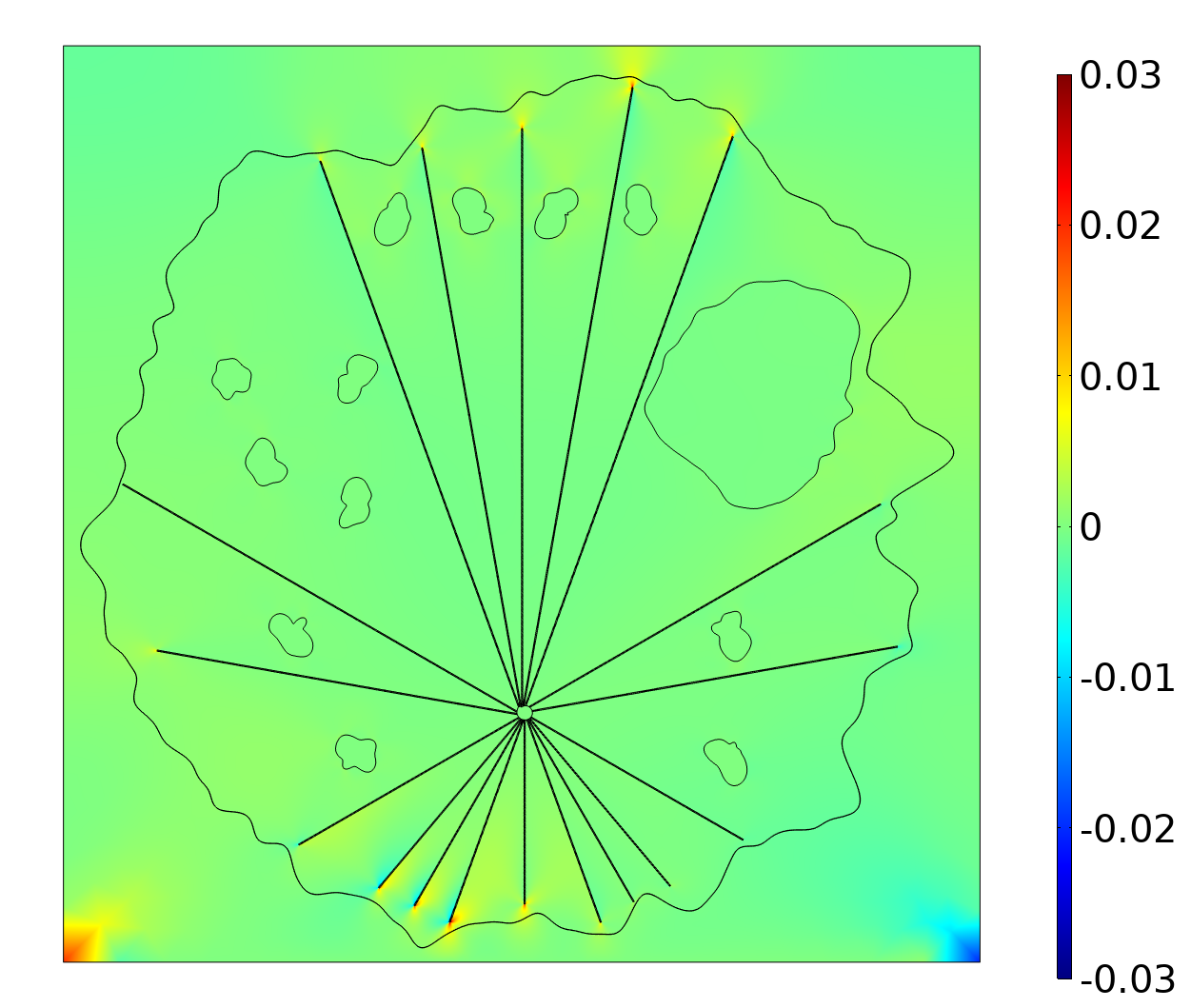}}\\
	\subfloat[CS-1, BC-3]{\includegraphics[width=0.43\linewidth]{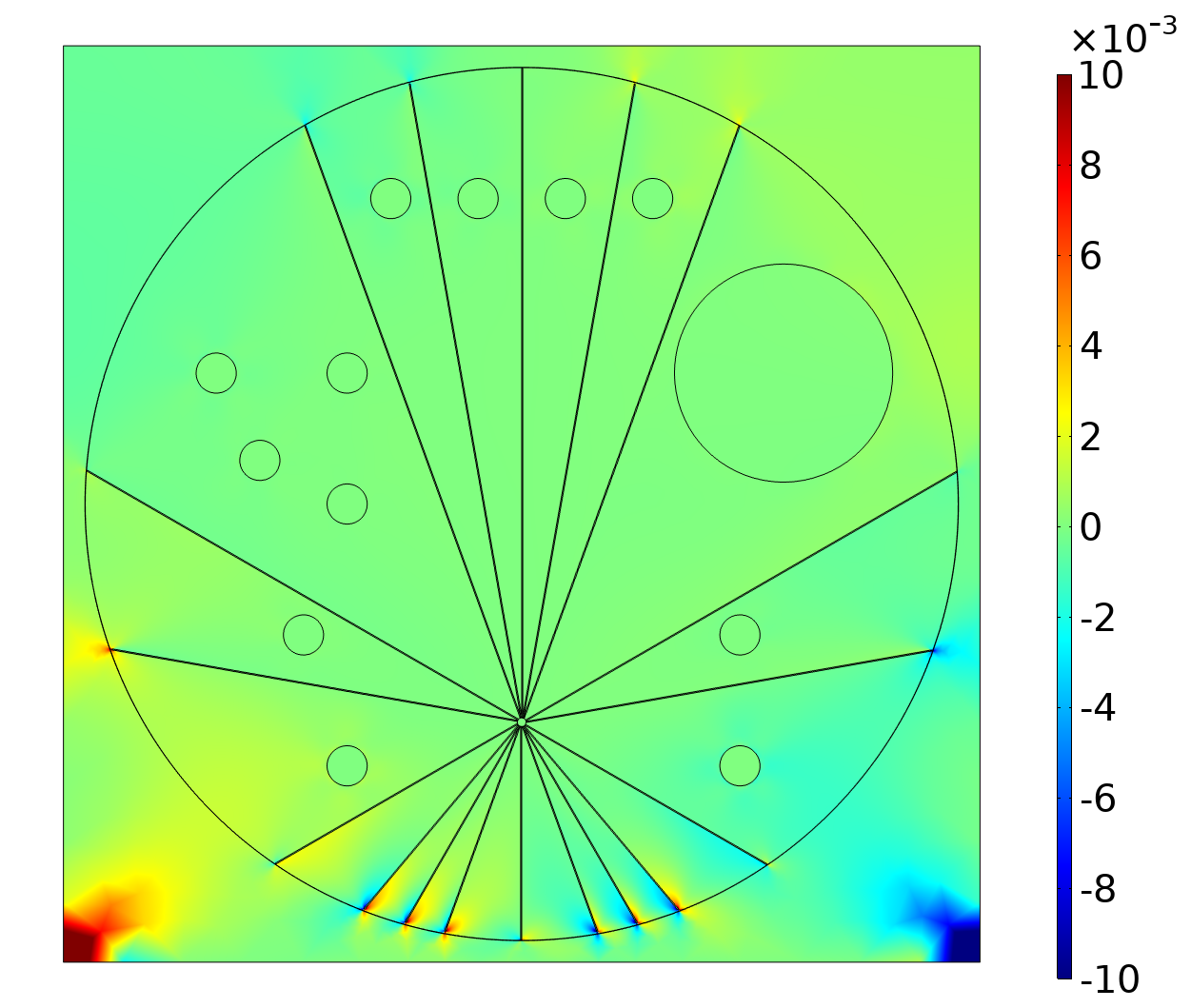}}
	\subfloat[CS-2, BC-3]{\includegraphics[width=0.43\linewidth]{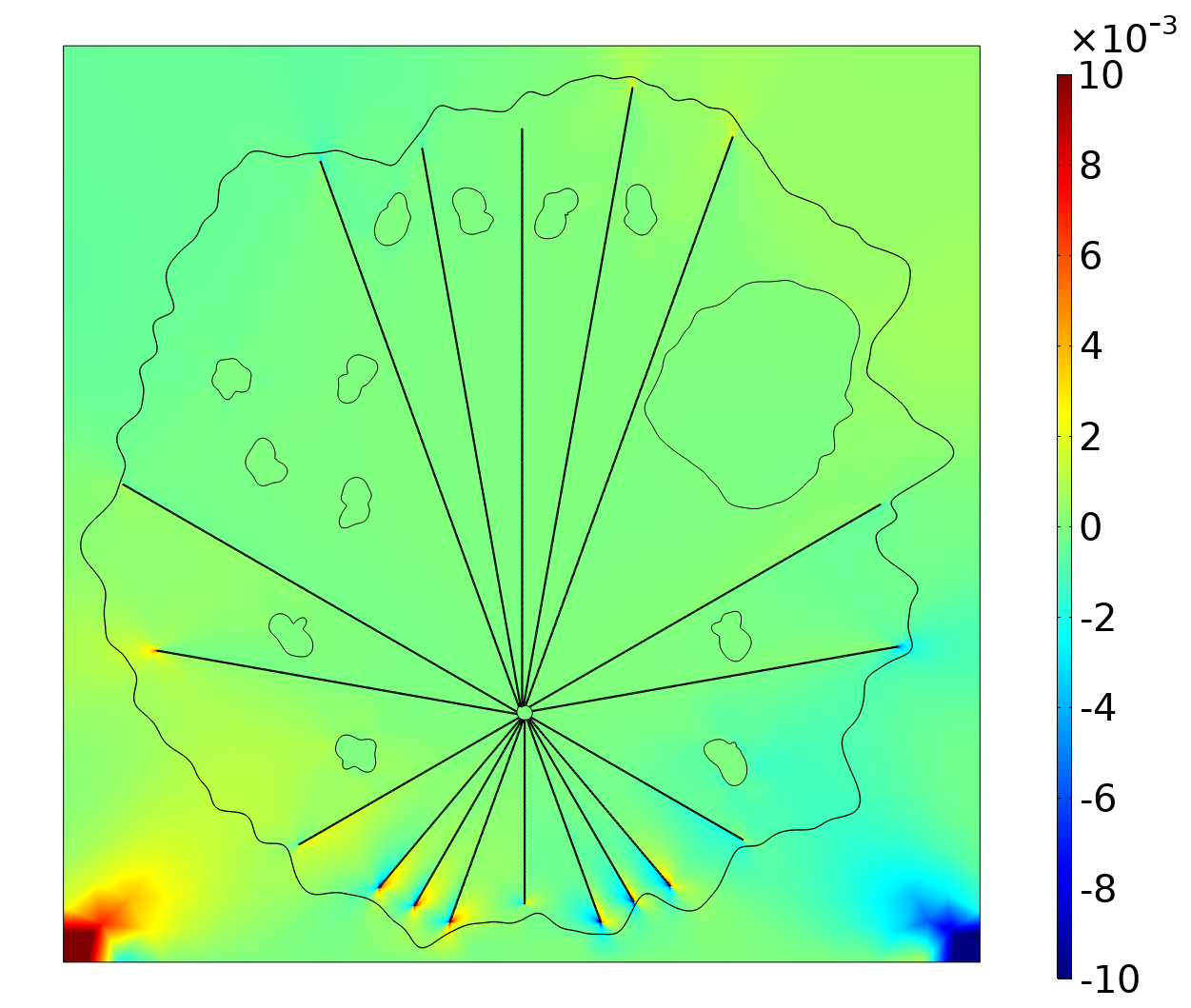}}
	\caption{The distribution of the strain component in the $x_1$-direction ($\varepsilon_{11}$) for two cell structures, namely cell structure-1 (a, c, e) and cell structure-2 (b, d, f), under three different boundary conditions considering the piezoelectric effect alone and the combined effect of the piezoelectricity and flexoelectricity for the compressive displacement of 20 nm.}
	\label{fig:3_32}
\end{figure}

\noindent The strain component ($\varepsilon_{11}$) distribution in the $x_1$-direction inside the biological cell as a result of the combined effects of piezoelectricity and flexoelectricity is shown in \fig\ref{fig:3_32}. The distribution has been derived for two distinct cell structures subjected to a compressive displacement of 20 nm, while accounting for three applied boundary conditions (BC-1 to BC-3). For BC-1, the magnitude of $\varepsilon_{11}$ is concentrated in close proximity to the cell membrane, as well as the top and bottom walls of the cell, as depicted in \fig\ref{fig:3_32}(a)-(b). This distribution of strain is a result of the application of compressive mechanical displacement in the transverse direction. For BC-2 and BC-3, the magnitude of $\varepsilon_{11}$ is concentrated toward the corners of the bottom walls of the cell. 
\begin{figure}[hpbt]
	\centering
	\subfloat[CS-1, BC-1]{\includegraphics[width=0.43\linewidth]{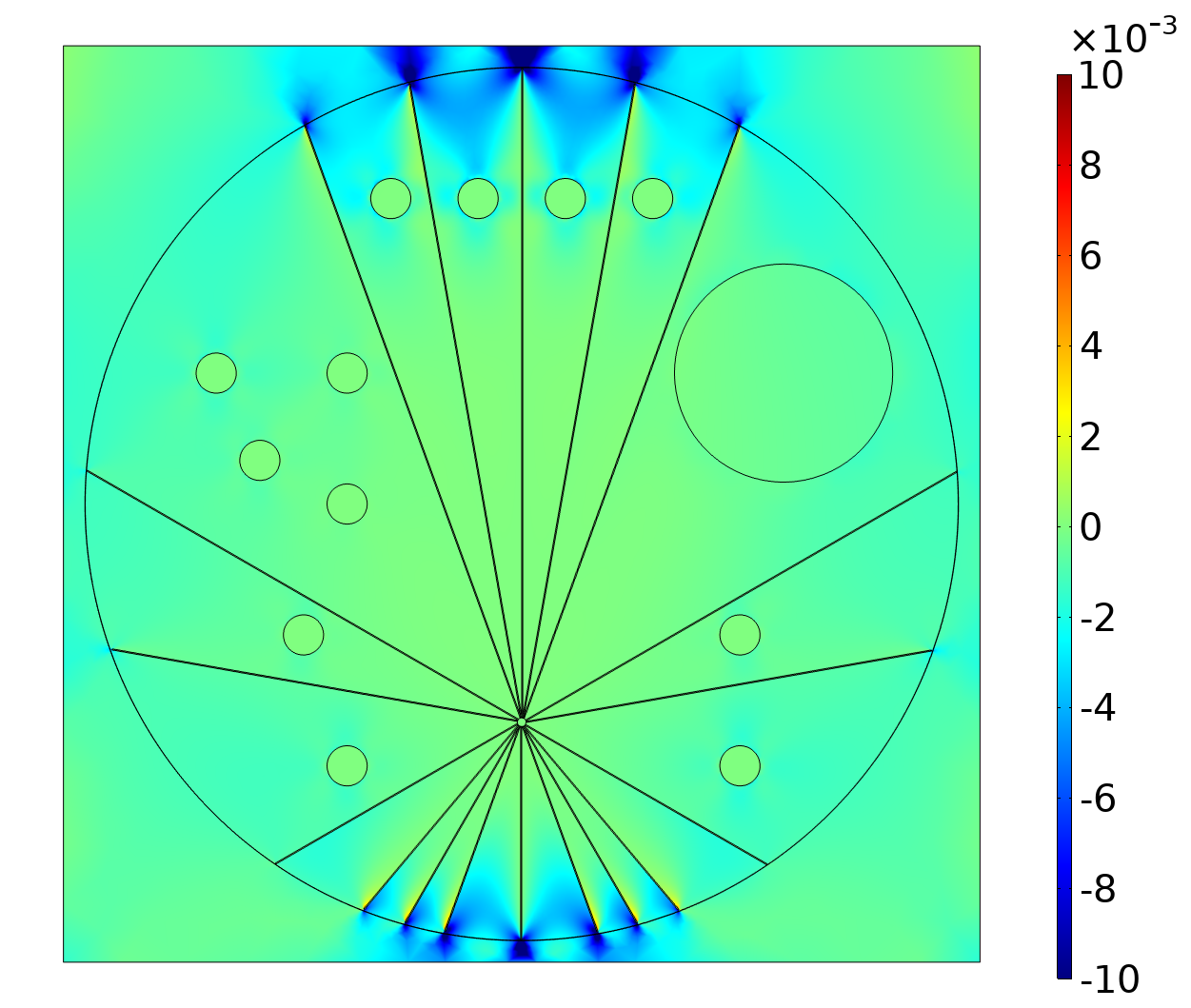}}
	\subfloat[CS-2, BC-1]{\includegraphics[width=0.43\linewidth]{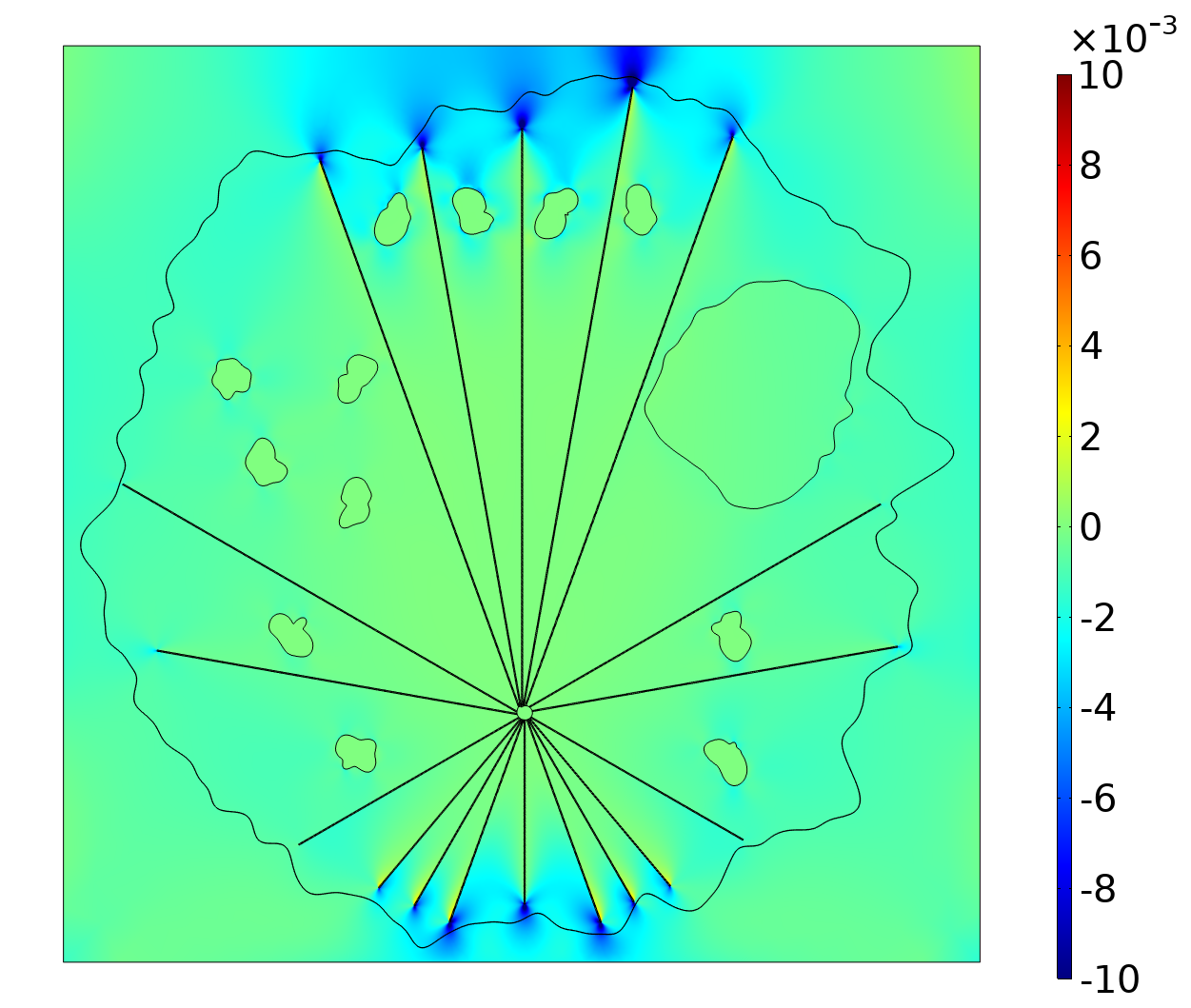}}\\
	\subfloat[CS-1, BC-2]{\includegraphics[width=0.43\linewidth]{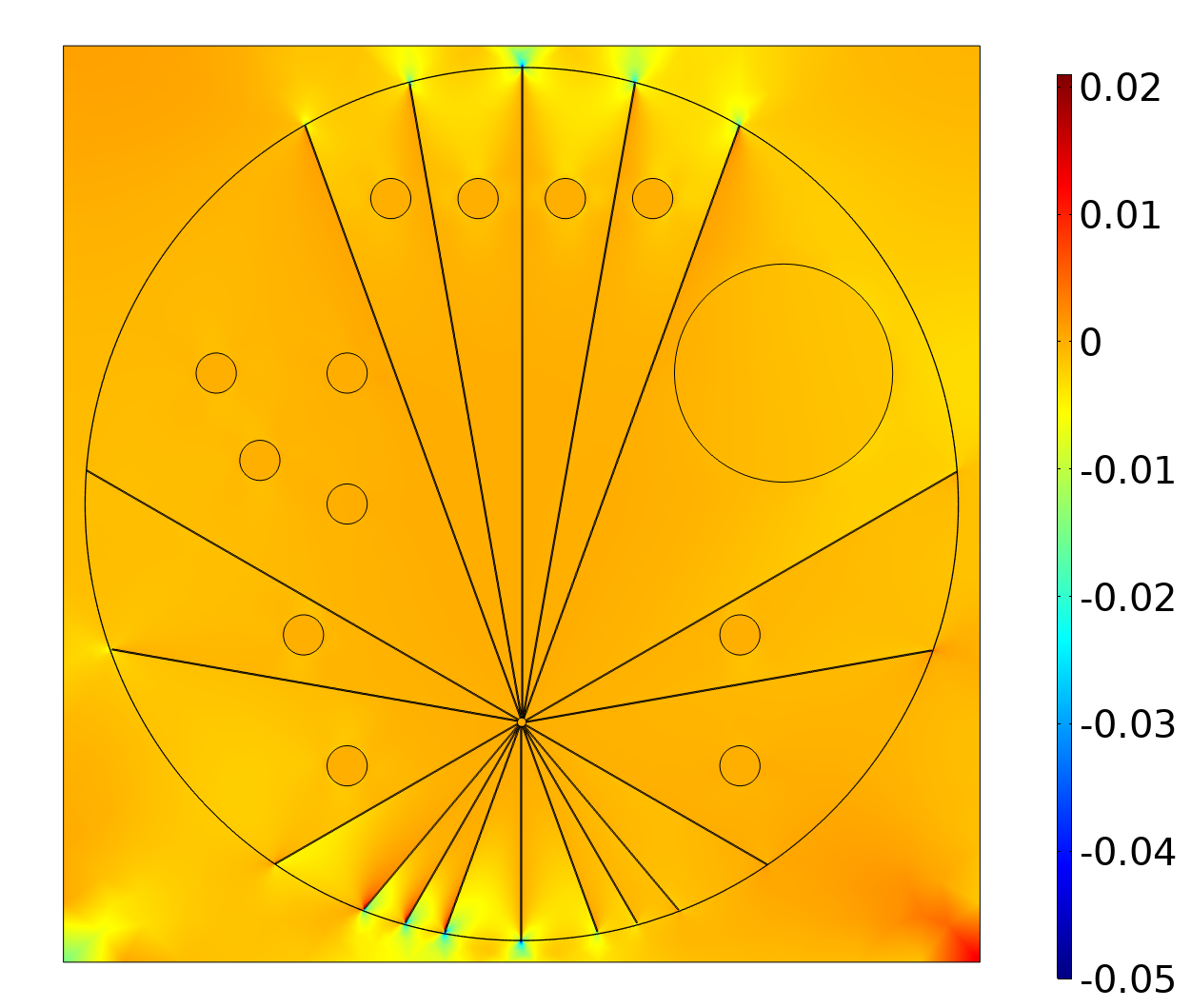}}
	\subfloat[CS-2, BC-2]{\includegraphics[width=0.43\linewidth]{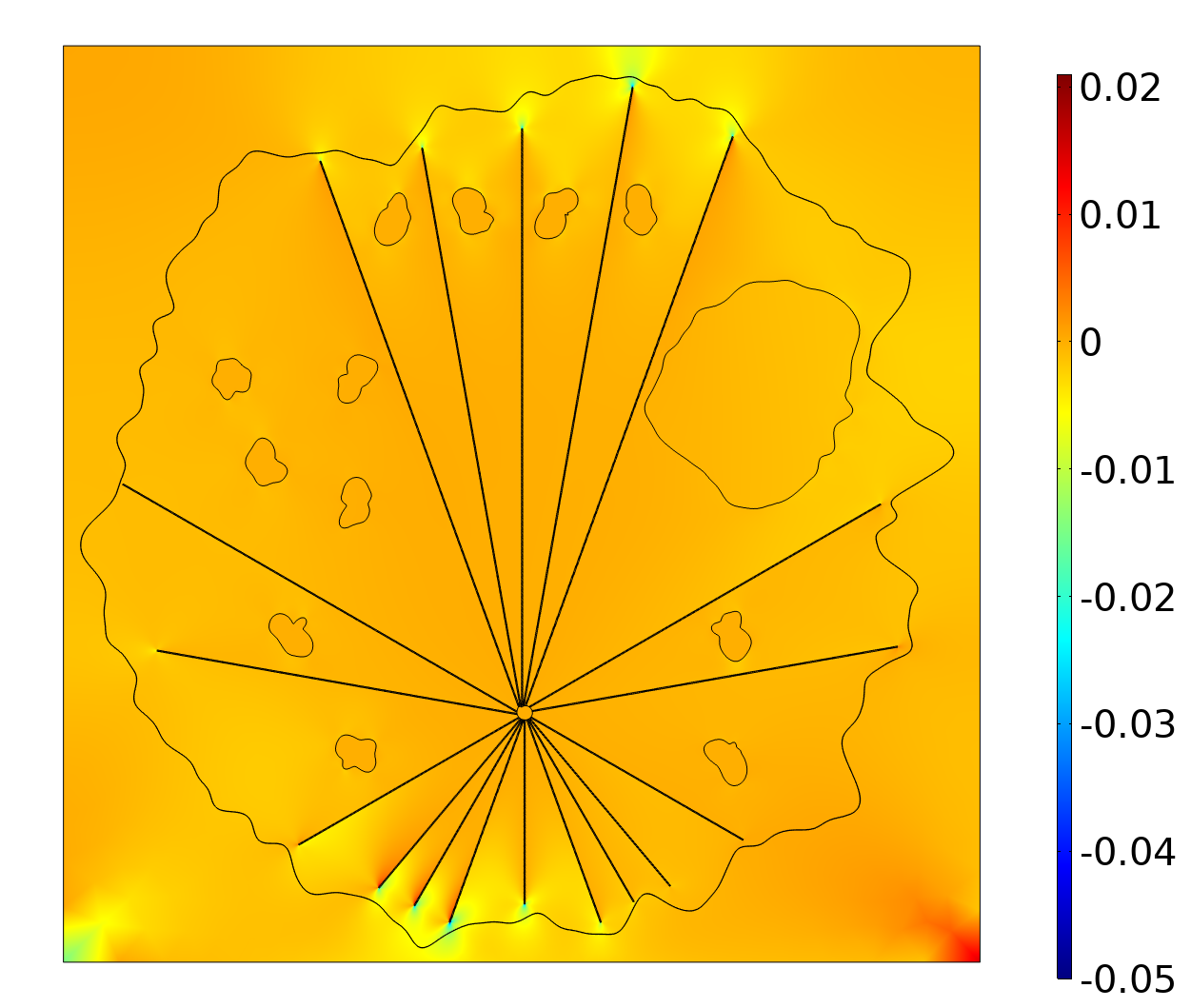}}\\
	\subfloat[CS-1, BC-3]{\includegraphics[width=0.43\linewidth]{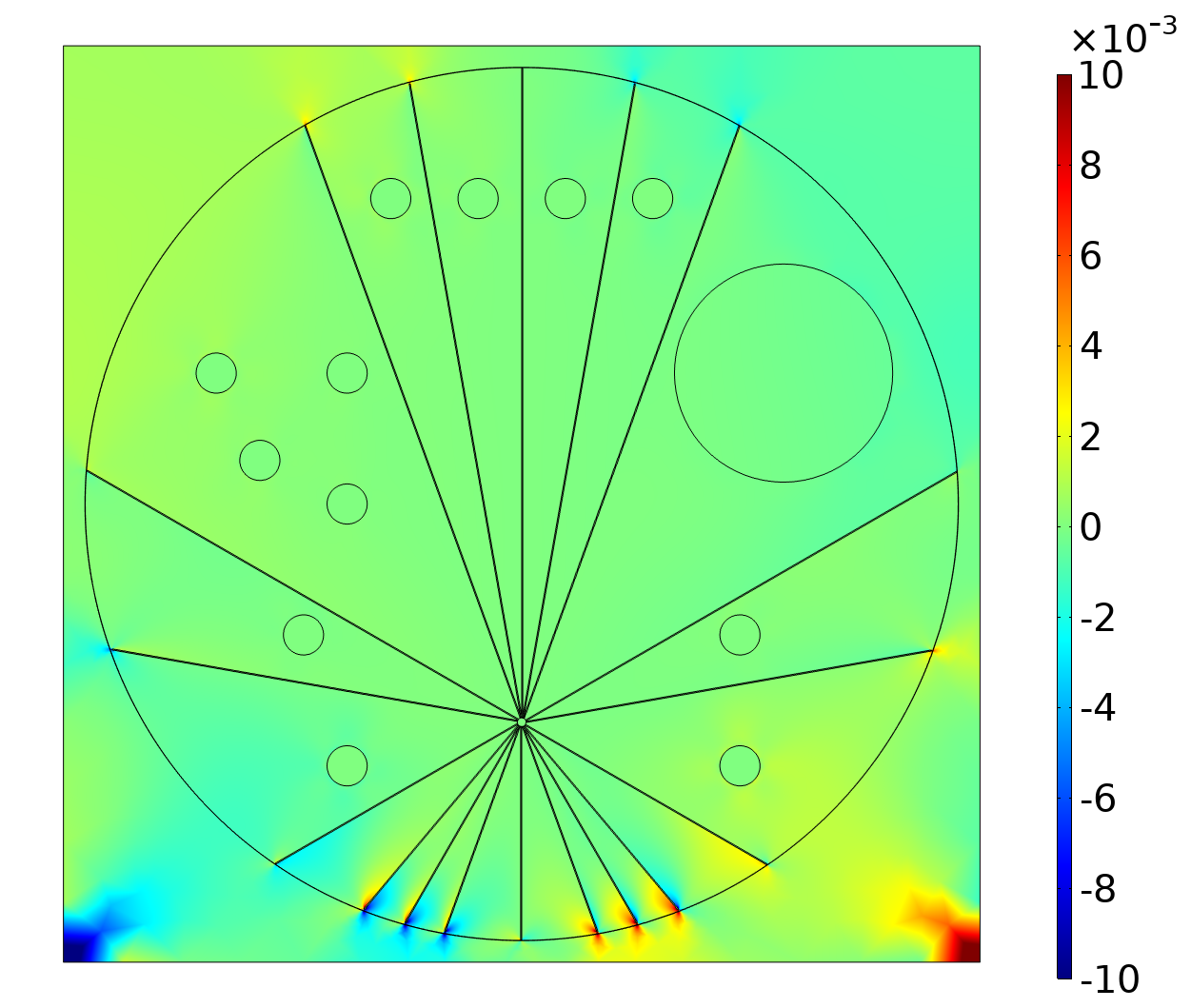}}
	\subfloat[CS-2, BC-3]{\includegraphics[width=0.43\linewidth]{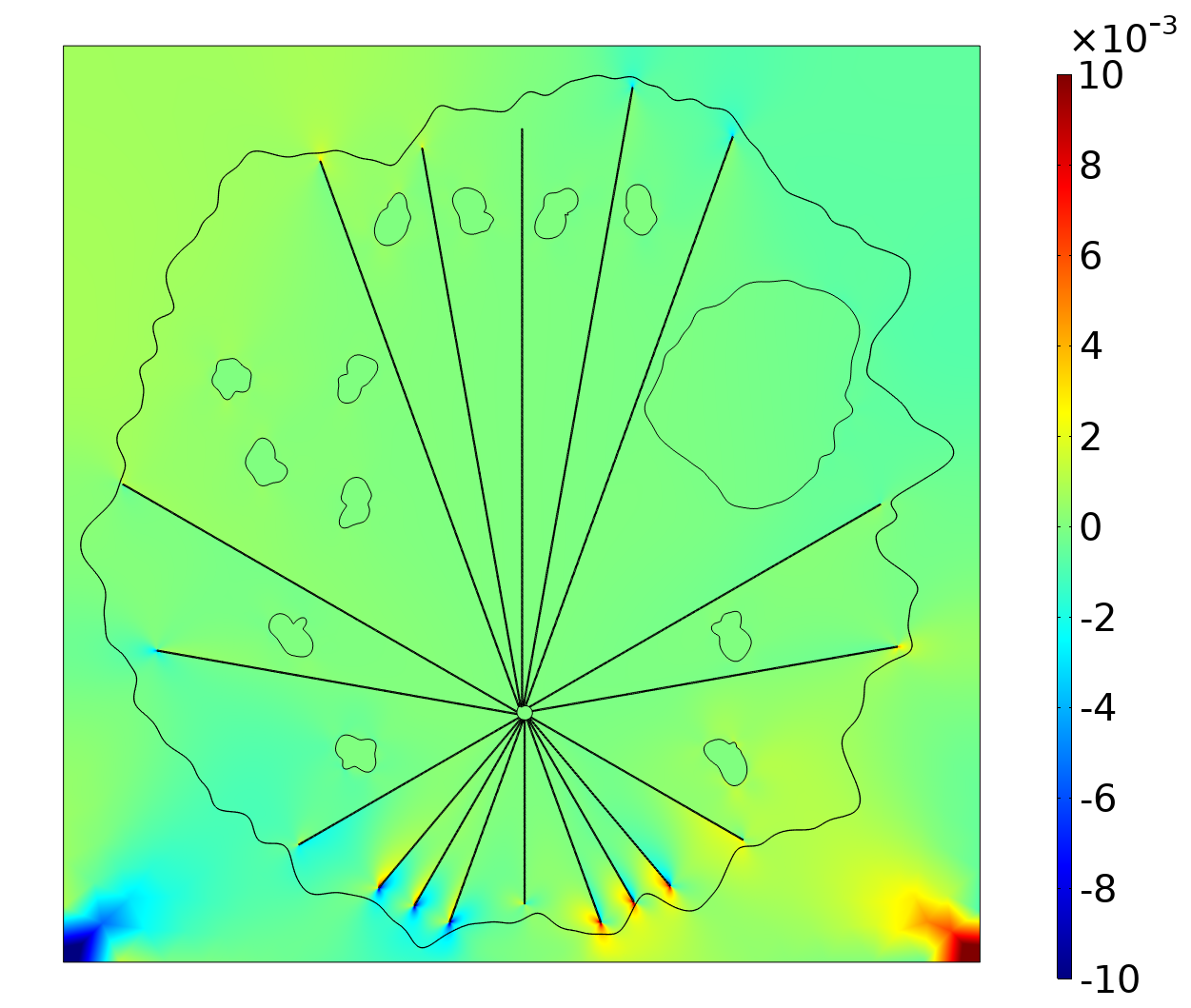}}
	\caption{The distribution of the strain component in the $x_3$-direction ($\varepsilon_{33}$) for the piezoelectric effect alone and the combined effect of piezoelectricity and flexoelectricity for cell structure-1 (a, c, e) and cell structure-2 (b, d, f) under three different boundary conditions for the compressive displacement of 20 nm.}
	\label{fig:3_34}
\end{figure}
	
On the other hand, there is a noticeable difference in the distribution of the strain component ($\varepsilon_{33}$) in the $x_3$-direction compared to the $x_1$-direction within the biological cell. This difference is attributed to the combined influence of piezoelectricity and flexoelectricity, as depicted in \fig\ref{fig:3_34}. 
	
The impact of flexoelectric stress components on the strain distribution is insignificant. Therefore, the strain distribution for the effect of piezoelectricity only and for the combined effect of piezoelectricity and flexoelectricity show a remarkable similarity and have nearly comparable magnitudes, as depicted in \figs\ref{fig:3_32} to \ref{fig:3_34}. Nevertheless, the strain distribution in the $x_3$-direction is slightly greater than in the $x_1$-direction, as the flexoelectric stress component in the $x_3$-direction is 1.80 times larger than its component in the $x_1$-direction, regardless of the boundary conditions (BC-1 to BC-3). The flexoelectric stress component in cell structure-2 has exhibited a notable alteration in comparison to CS-1 as a result of the modifications in the shape, size, and orientation of the microorganelles, particularly the microtubules.
\subsection{Electrical potential distribution}
\noindent The electric potential distribution of a biological cell under compressive displacement has been determined by solving the governing equations described in Section \ref{Sec:PoM}. Accordingly, the electric potential distribution for each boundary condition applied to the cell considering only the piezoelectric effect and the combined piezoelectric-flexoelectric effects is provided below.
\subsubsection{Consideration of piezoelectric effects alone}
\begin{figure}[hpbt]
	\centering
	\subfloat[CS-1, BC-1]{\includegraphics[width=0.43\linewidth]{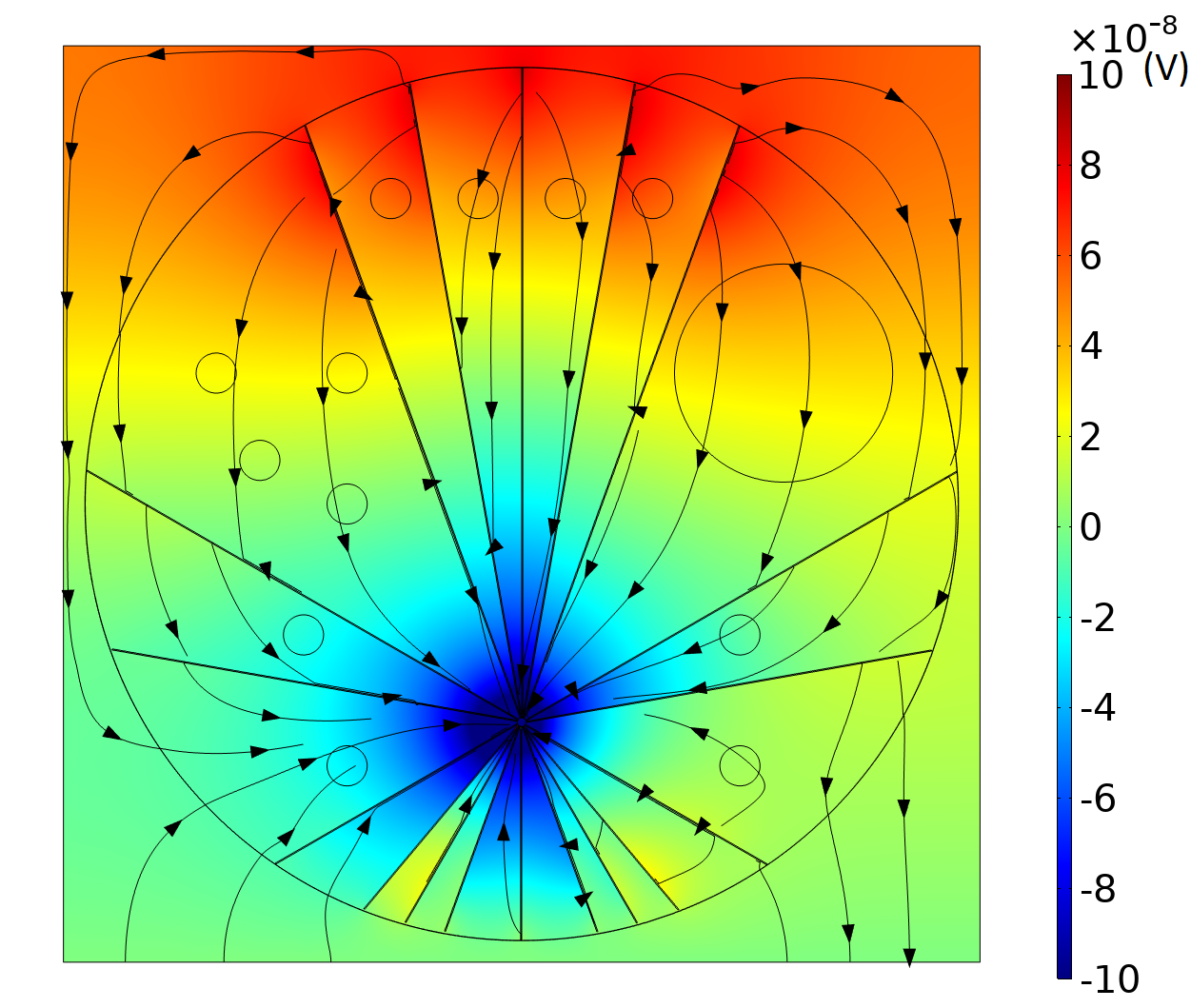}}
	\subfloat[CS-2, BC-1]{\includegraphics[width=0.43\linewidth]{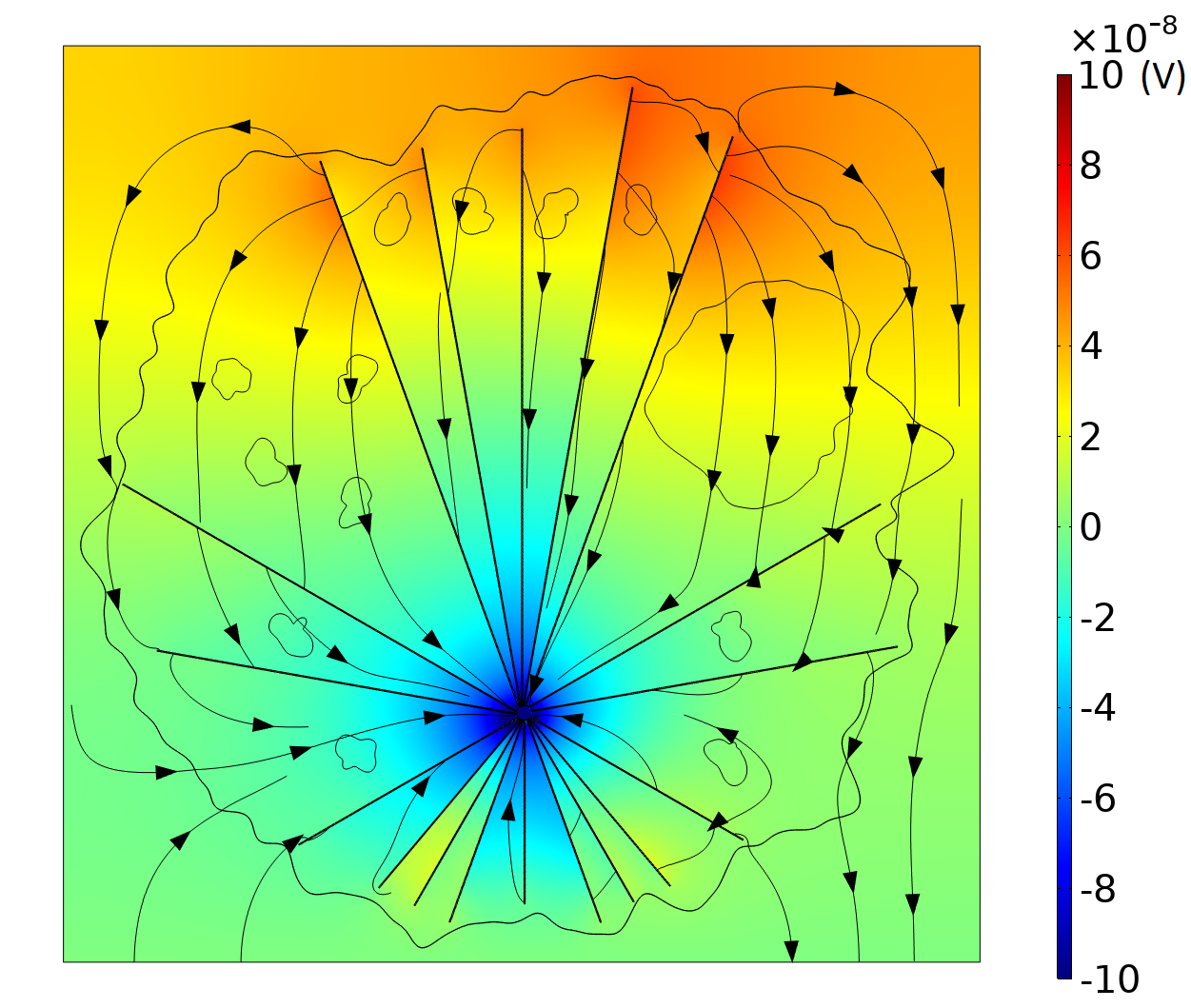}}\\
	\subfloat[CS-1, BC-2]{\includegraphics[width=0.43\linewidth]{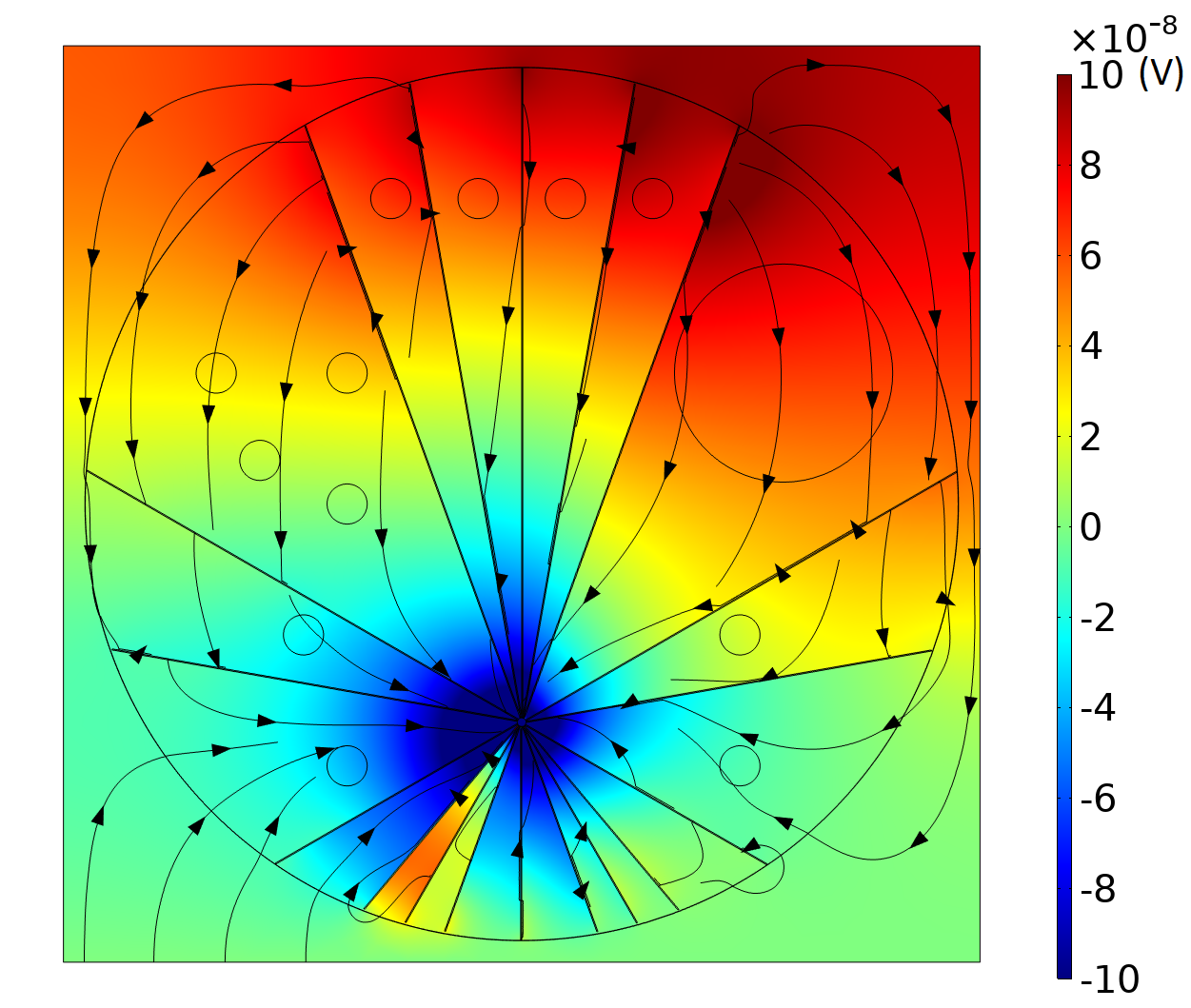}}	
	\subfloat[CS-2, BC-2]{\includegraphics[width=0.43\linewidth]{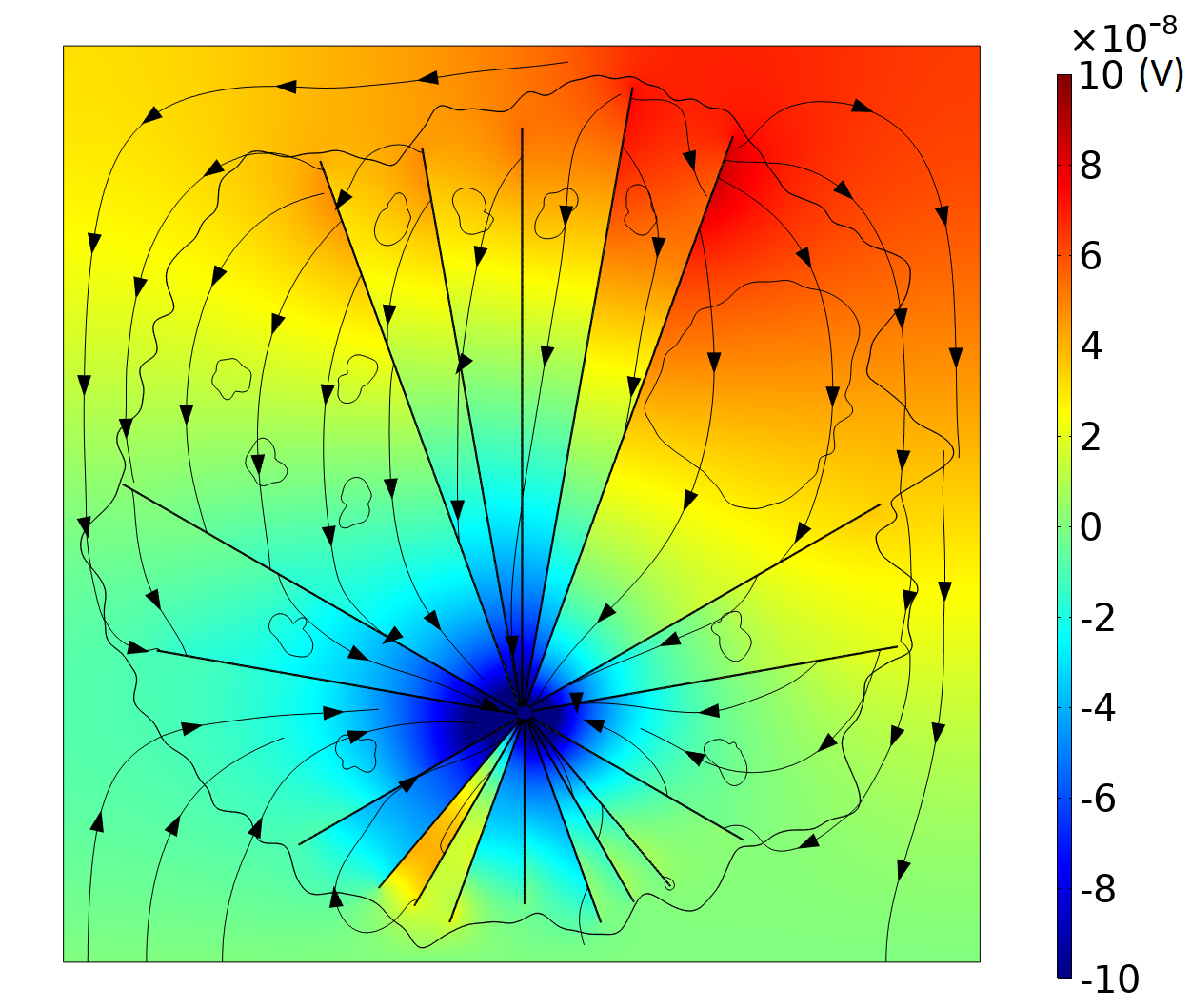}} \\
	\subfloat[CS-1, BC-3]{\includegraphics[width=0.43\linewidth]{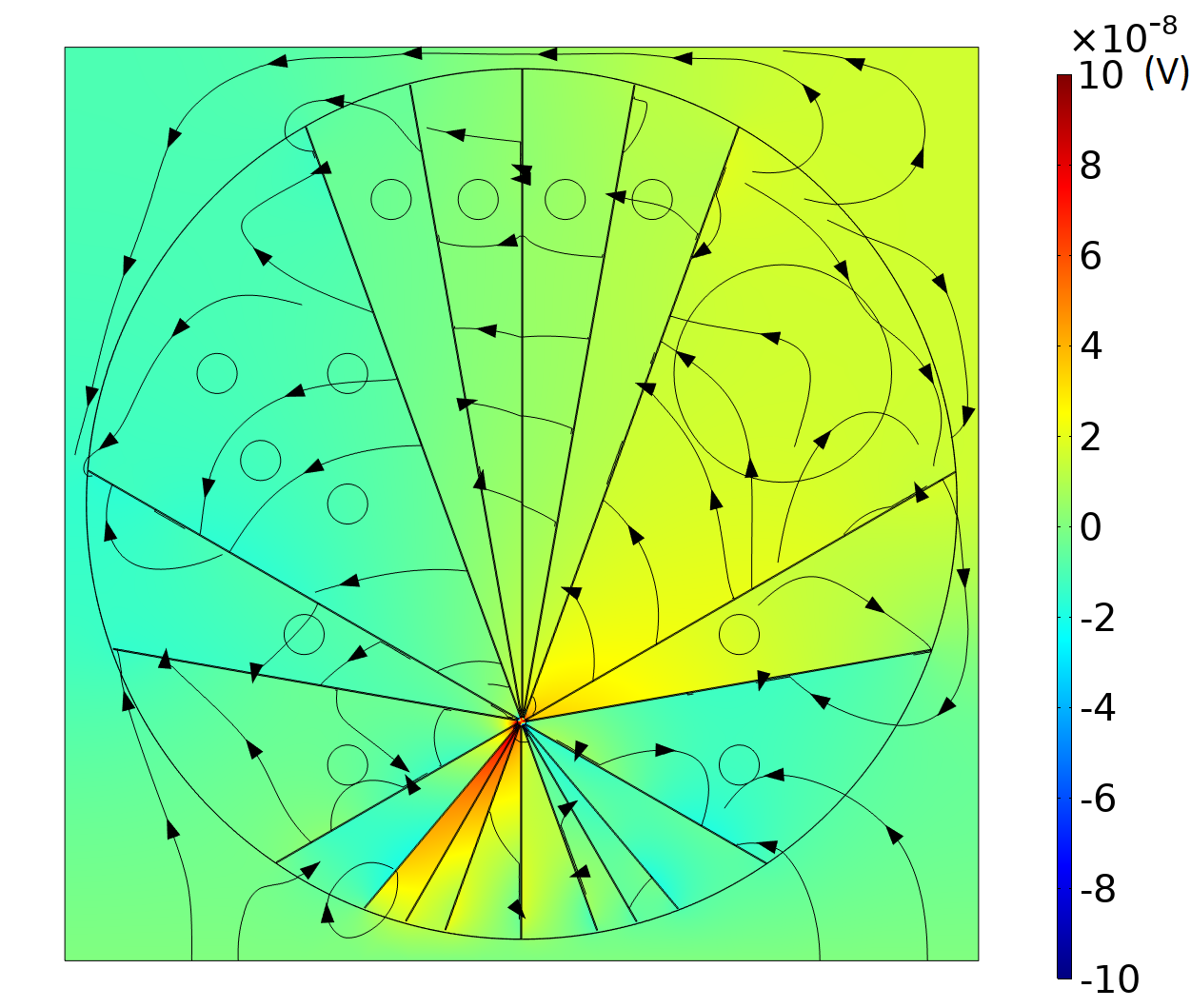}}
	\subfloat[CS-2, BC-3]{\includegraphics[width=0.43\linewidth]{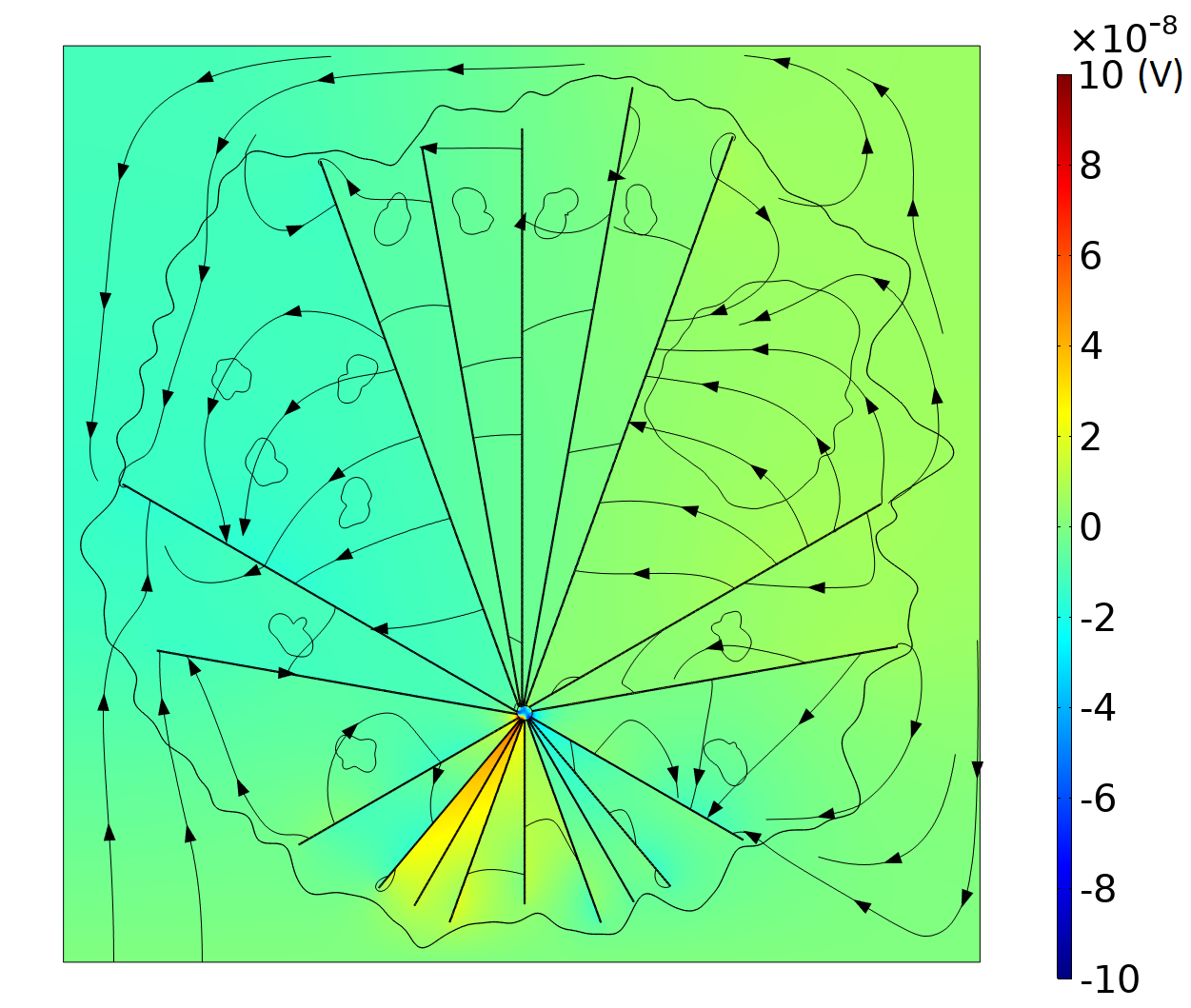}}
	\caption{The electric potential distribution in the biological cell is being analyzed under the influence of a 20nm compressive displacement, taking into consideration only the piezoelectric effect. The analysis has been conducted for two cells: cell structure-1 (a, c, e) and cell structure-2 (b, d, f).}
	\label{fig:3_21}
\end{figure}

\noindent The electric potential (V) distribution within the biological cell, resulting from a 20 nm mechanical (compressive) displacement, accounting for the piezoelectric effect alone, is illustrated in \fig\ref{fig:3_21}. The distribution is studied under three applied boundary conditions, marked as BC-1 to BC-3.
The arrows indicate the direction of the electric field lines within the cell. 
In all the cases, BC-1 and BC-3, it is evident from \fig\ref{fig:3_21}(a)-(d) that the electric potential generated is higher for CS-1 in comparison to CS-2. 
The strain distribution is no more streamlined along the boundary of the microtubules due to the localization of the strain near to the curved edges. As a result, the intensity of the strain distribution in the vicinity of the centrosome is reduced. Hence, the overall electric potential being generated in the cell is reduced. Consequently, a weak piezoelectric effect leads to the generation of a lower electric potential. The electric potential being generated has been reduced by a factor of 10 when BC-3 is applied to the biological cell compared to BC-1 and BC-2 (see \fig\ref{fig:3_21}(e)-(f)).
\subsubsection{Consideration of flexoelectric effects in addition to piezoelectric effects }
\begin{figure}[hpbt]
	\centering
	\subfloat[CS-1, BC-1]{\includegraphics[width=0.43\linewidth]{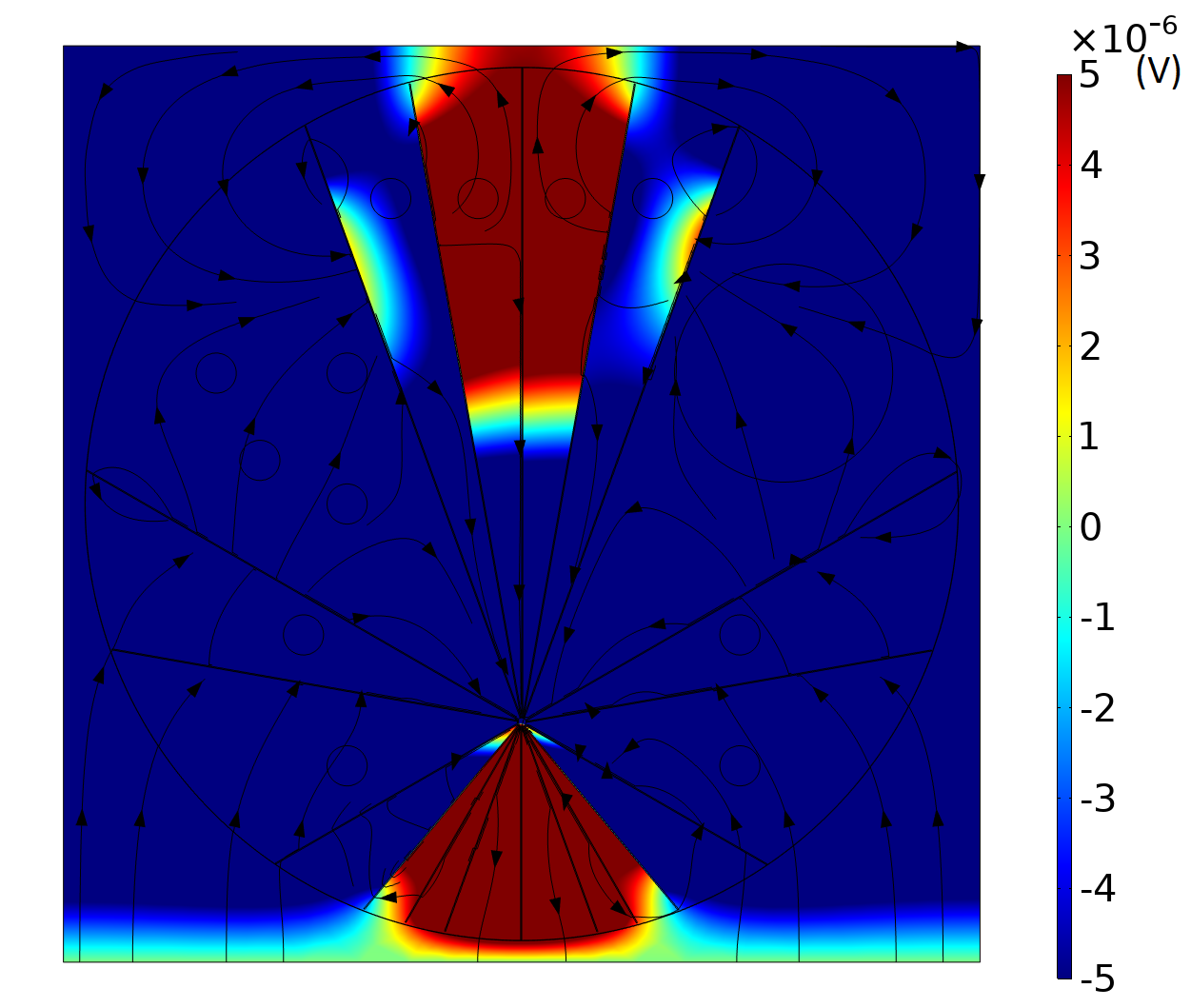}}	
	\subfloat[CS-2, BC-1]{\includegraphics[width=0.43\linewidth]{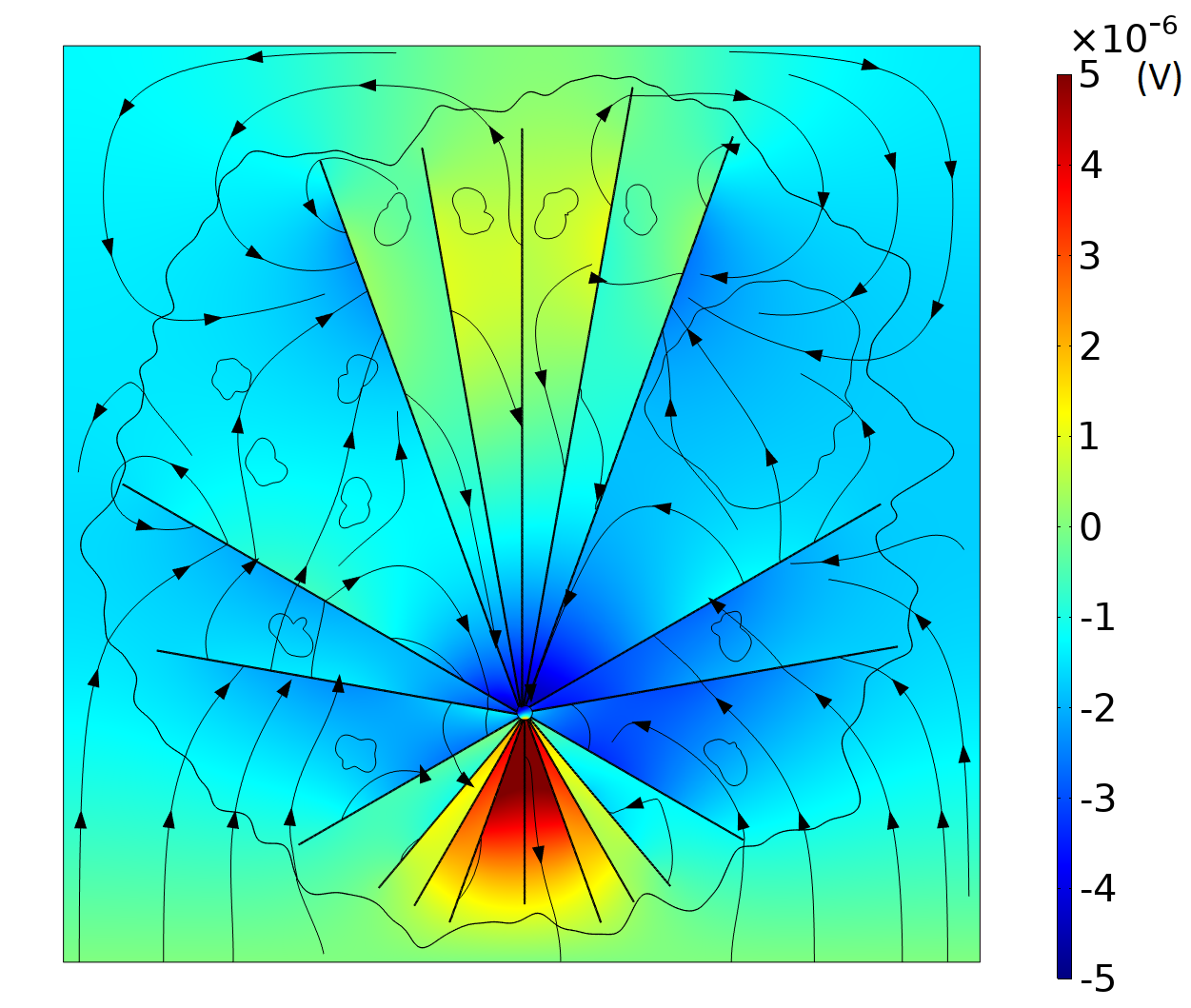}}\\
	\subfloat[CS-1, BC-2]{\includegraphics[width=0.43\linewidth]{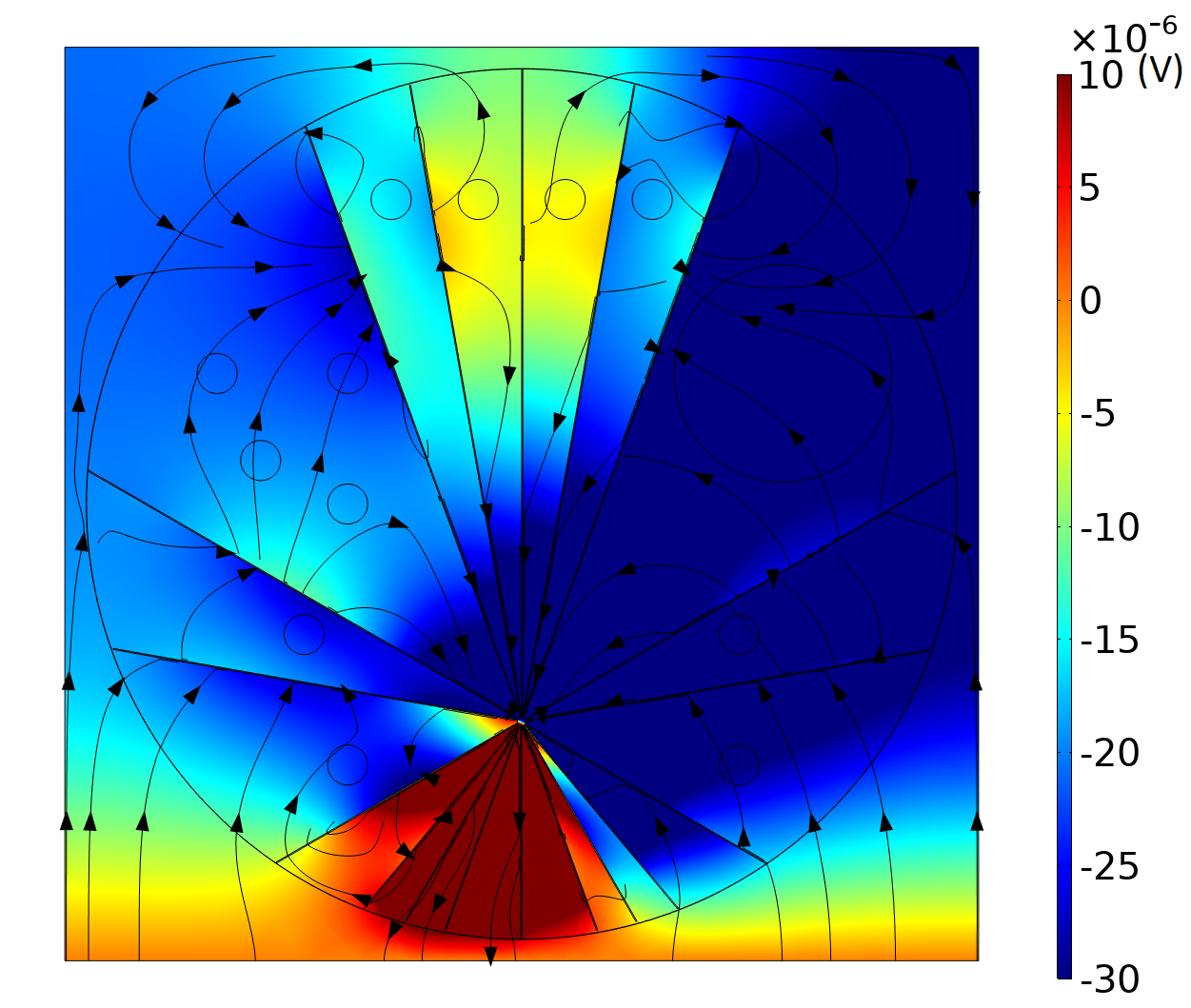}}	
	\subfloat[CS-2, BC-2]{\includegraphics[width=0.43\linewidth]{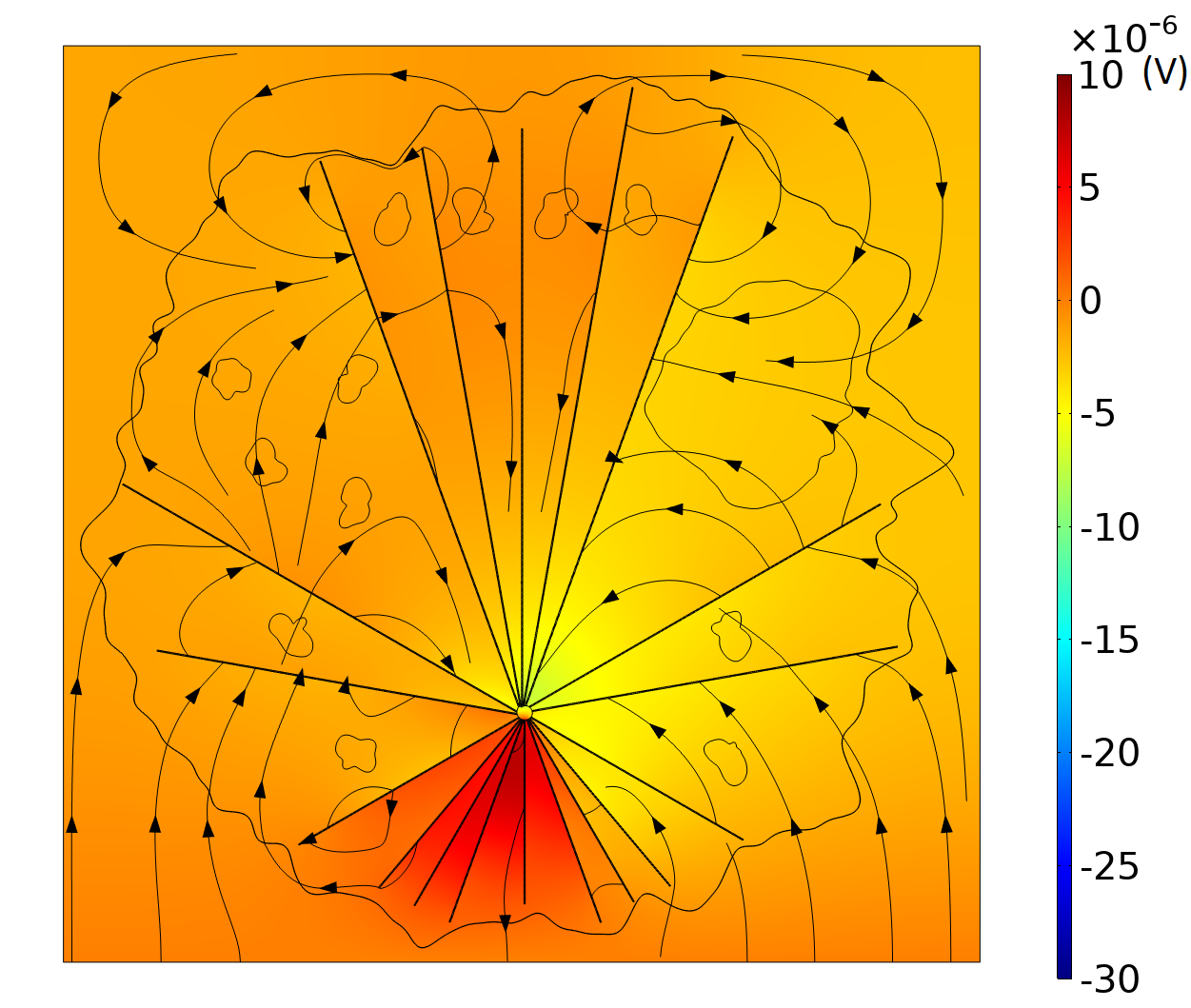}}\\
	\subfloat[CS-1, BC-3]{\includegraphics[width=0.43\linewidth]{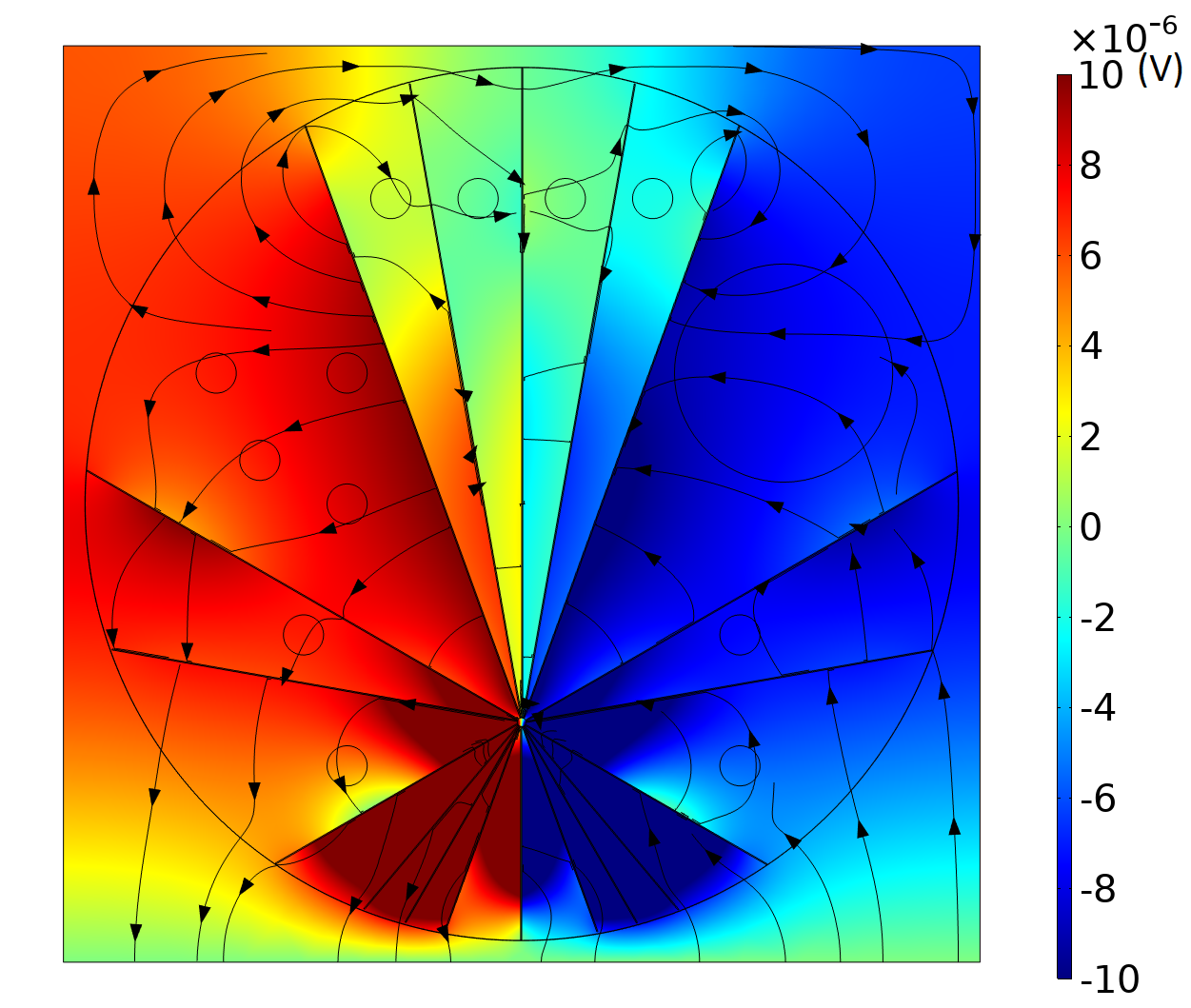}} 
	\subfloat[CS-2, BC-3]{\includegraphics[width=0.43\linewidth]{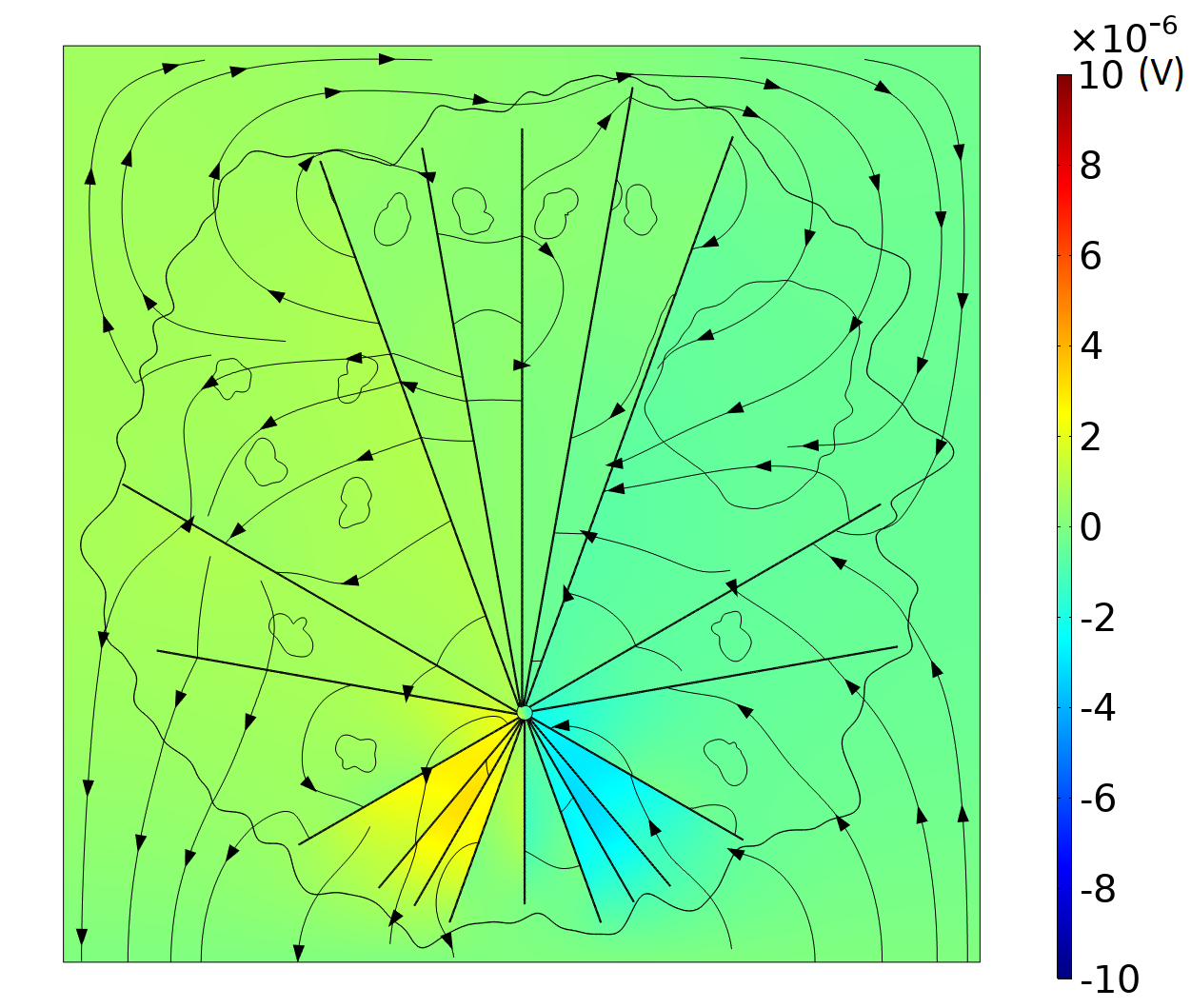}}
	\caption{The electric potential distribution in the biological cell is analyzed for a compressive displacement of 20 nm, taking into account both the piezoelectric and flexoelectric effects for two cells: cell structure-1 (a, c, e) and cell structure-2 (b, d, f).}
	\label{fig:3_22}
\end{figure}

\noindent There is a significant enhancement of the electric potential distribution within  the cell when considering the application of non-local flexoelectric effects in addition to the linear piezoelectric effects. \figs\ref{fig:3_22}(a), \ref{fig:3_22}(c) and \ref{fig:3_22}(e) depict the electric potential distribution in CS-1, whereas \figs\ref{fig:3_22}(b), \ref{fig:3_22}(d) and \ref{fig:3_22}(f) for the CS-2. The contour plots displayed in \figs\ref{fig:3_22}(a)-(d) demonstrate that, under all three boundary conditions (BC-1 to BC-3), the electric potential generated in CS-1 is greater than that of CS-2. The strain gradients are stronger for the regular shapes of organelles in the cell (CS-1).
The electric potential generated in BC-2 is observed to increase due to the absence of strain gradients in the longitudinal directions. The transverse direction ($x_3$) exhibits a strain gradient that is responsible for the increased electric potential compared to BC-1 and BC-3. The combined effect of piezoelectricity and flexoelectricity induces a higher electric potential when considering only the piezoelectric effect. 
\subsubsection{Consideration of the number of microtubules and the shape of the organelles}
\noindent It is observed that the maximum electrical potential obtained for each boundary condition and cell structure is also different, as depicted in \figs\ref{fig:3_21} and \ref{fig:3_22}. This is attributed to the shape and arrangement of the microtubules within the biological cell that have a significant impact on the distribution of the electric potential \citep{Denning2017}.
The absence of longitudinal and transverse strains in BC-3 can be attributed to the uniform displacement boundary conditions in the principal planes, i.e., $x_1$ and $x_3$. Nevertheless, only a small amount of transverse strain is generated near the corners of the bottom side due to the applied mechanical displacement. This can be attributed to the negligible piezoelectric effect, resulting in a low generation of electric potential in this case. 

The growth, maintenance, and regulation of cells and tissues depend on the strength of electric fields. Moreover, studying electric potential distribution within the cell is essential for tissue regeneration, embryonic development, and wound healing \citep{Titushkin2009,Schofield2020}. For instance, electric fields are used in tissue engineering for characterizing artificial tissues, forming tissue-like materials, and micro-manipulating cells. Electrical stimulation has emerged as a novel and cutting-edge method in the field of regenerative medicine \citep{Markx2008,Chen2019}.
\subsubsection{Maximum induced electric potential considering the piezoelectric effect and the combined piezoelectric-flexoelectric effects}
\begin{figure}[hpbt]
	\centering
	\subfloat[BC-1, Piezo]{\includegraphics[width=0.42\linewidth]{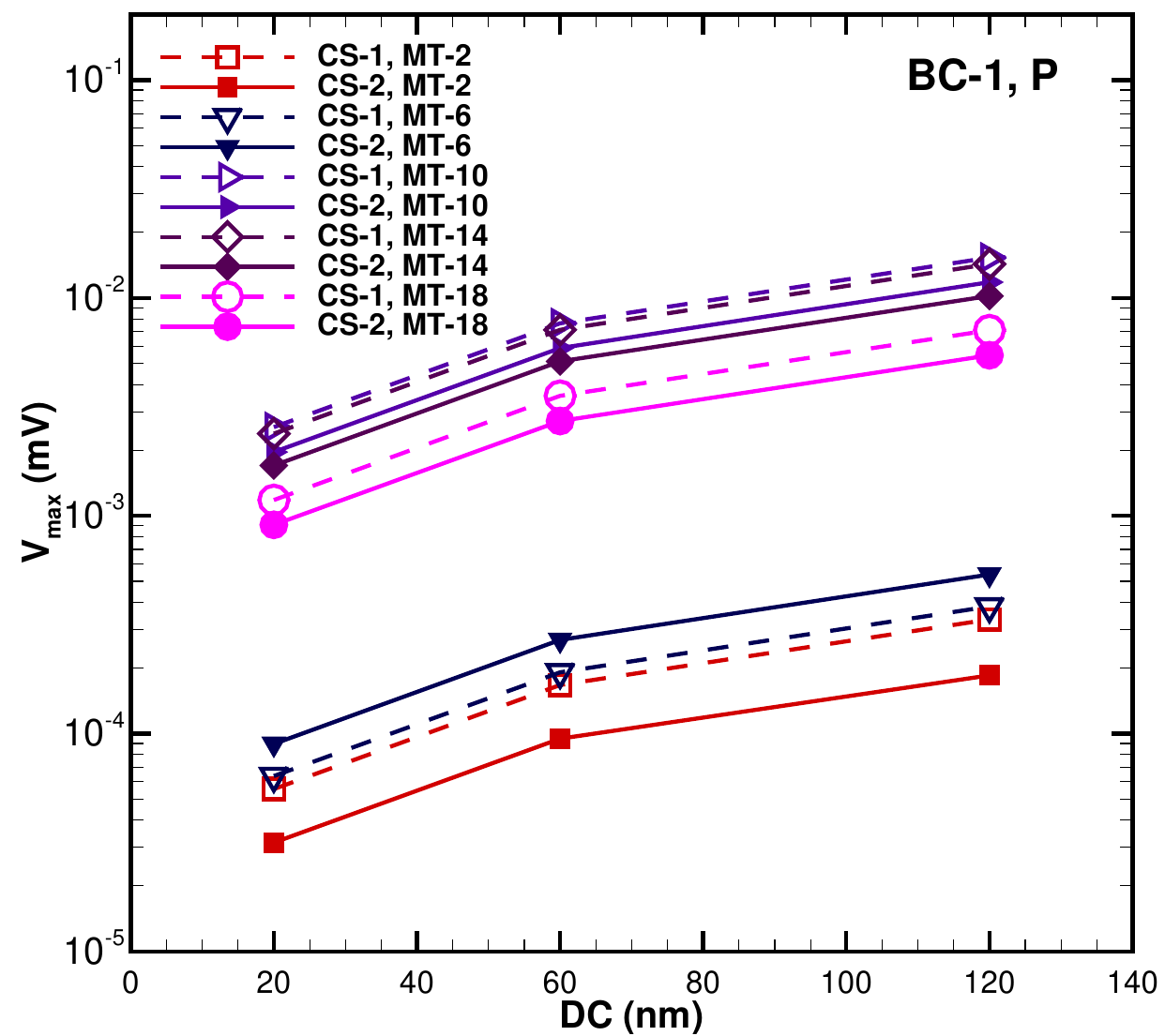}}
	\subfloat[BC-1,P+F]{\includegraphics[width=0.42\linewidth]{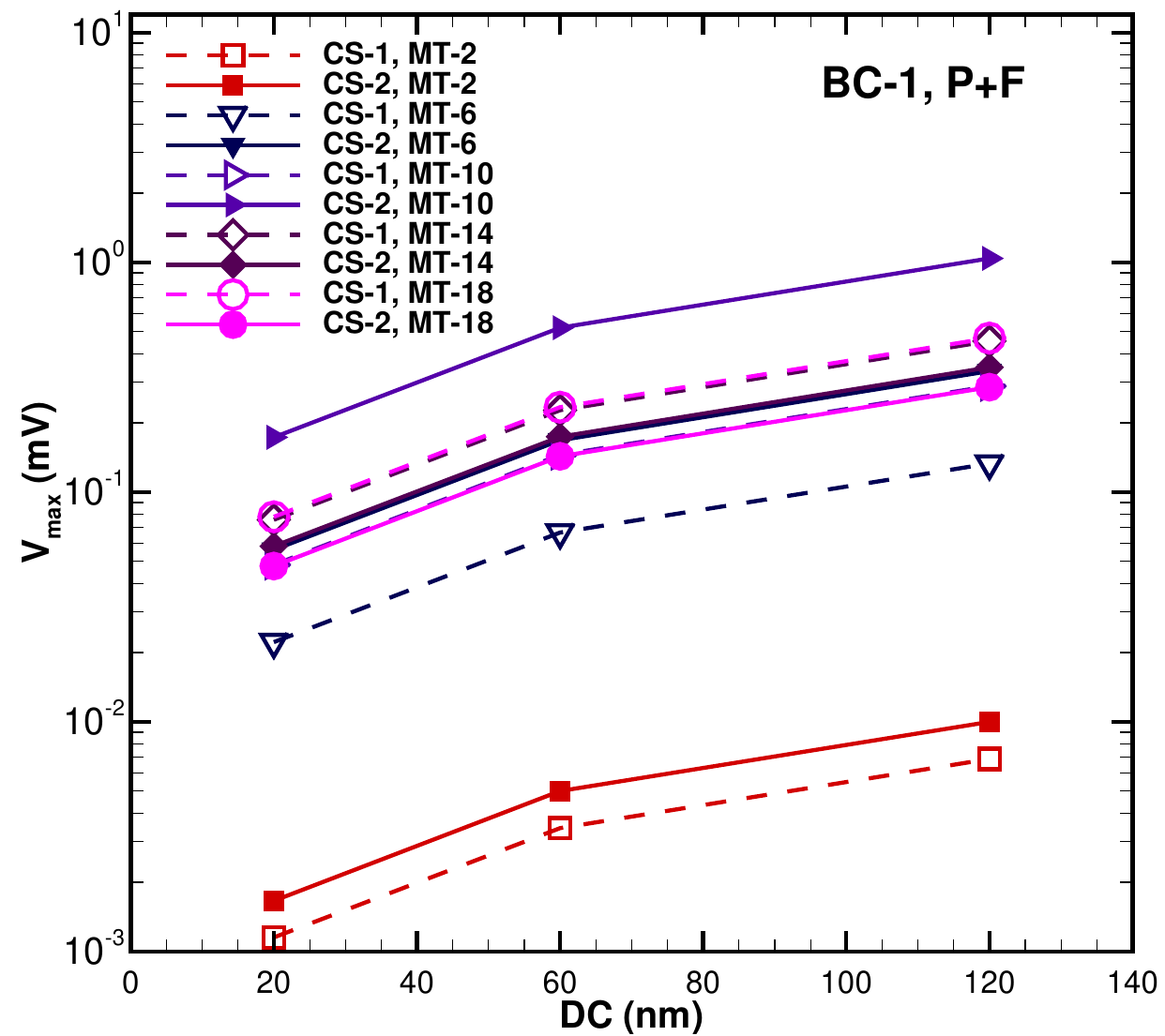}}\\
	\subfloat[BC-2, Piezo]{\includegraphics[width=0.42\linewidth]{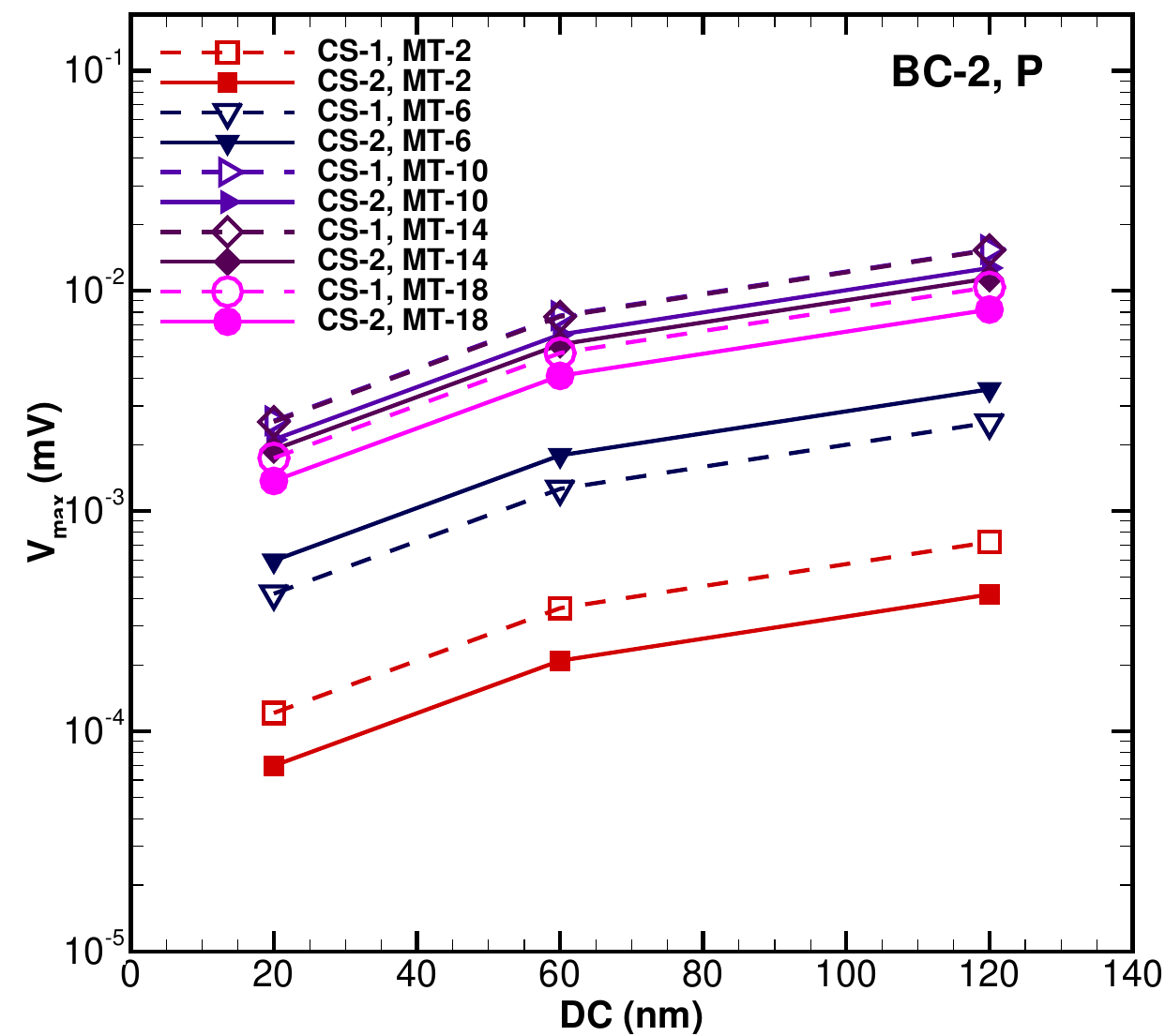}}
	\subfloat[BC-2,P+F]{\includegraphics[width=0.42\linewidth]{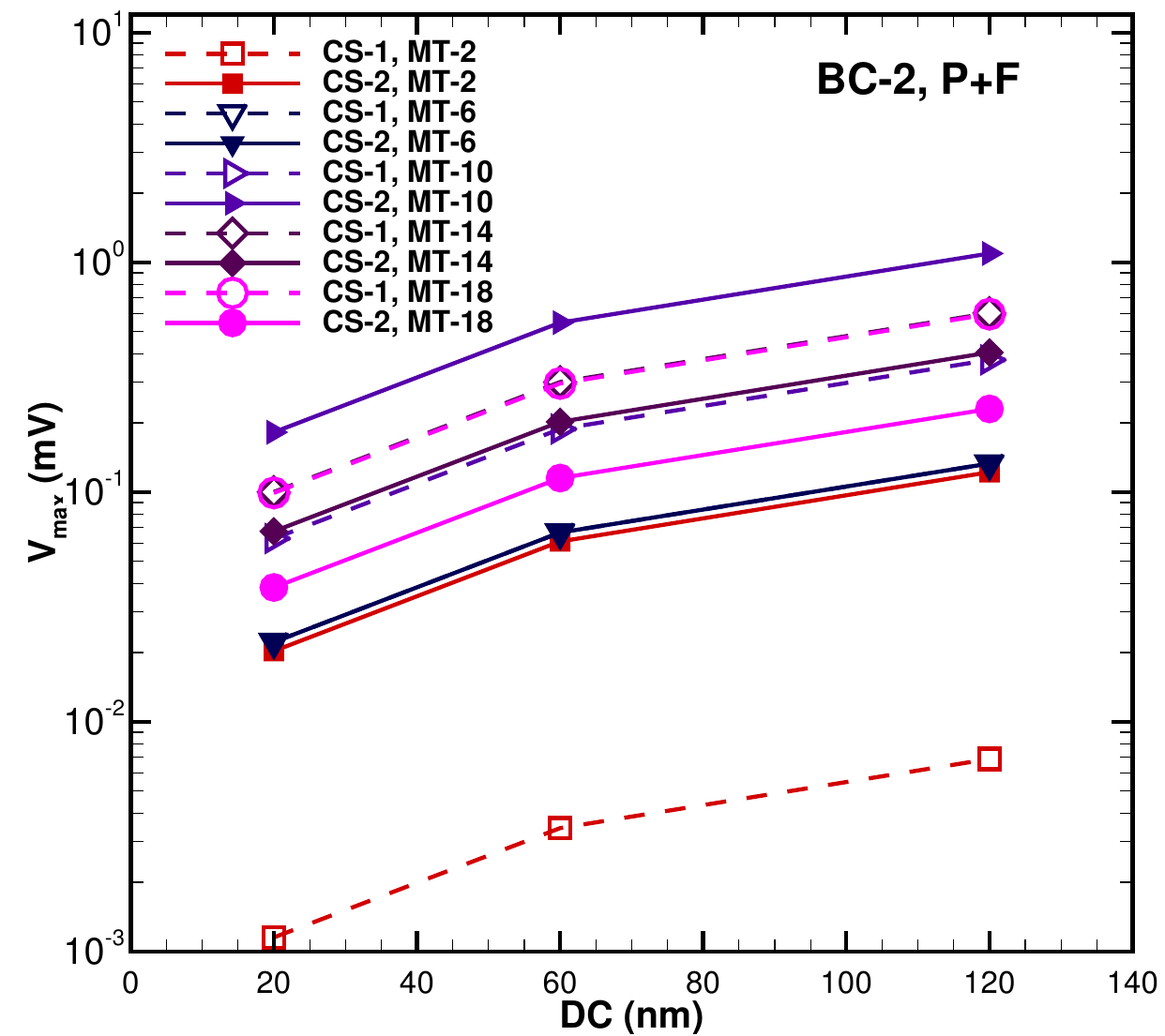}}\\
	\subfloat[BC-3, Piezo]{\includegraphics[width=0.42\linewidth]{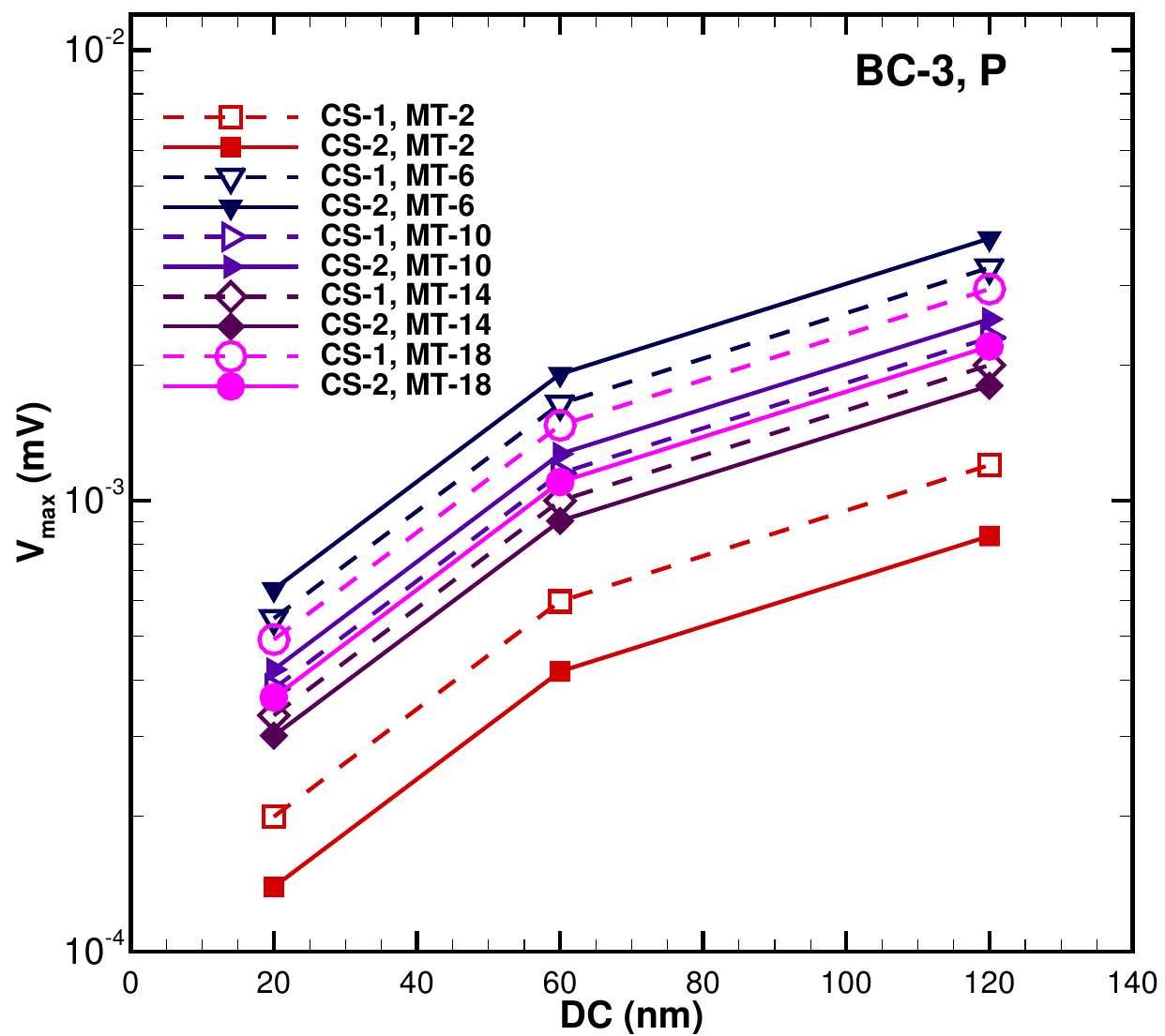}}
	\subfloat[BC-3,P+F]{\includegraphics[width=0.42\linewidth]{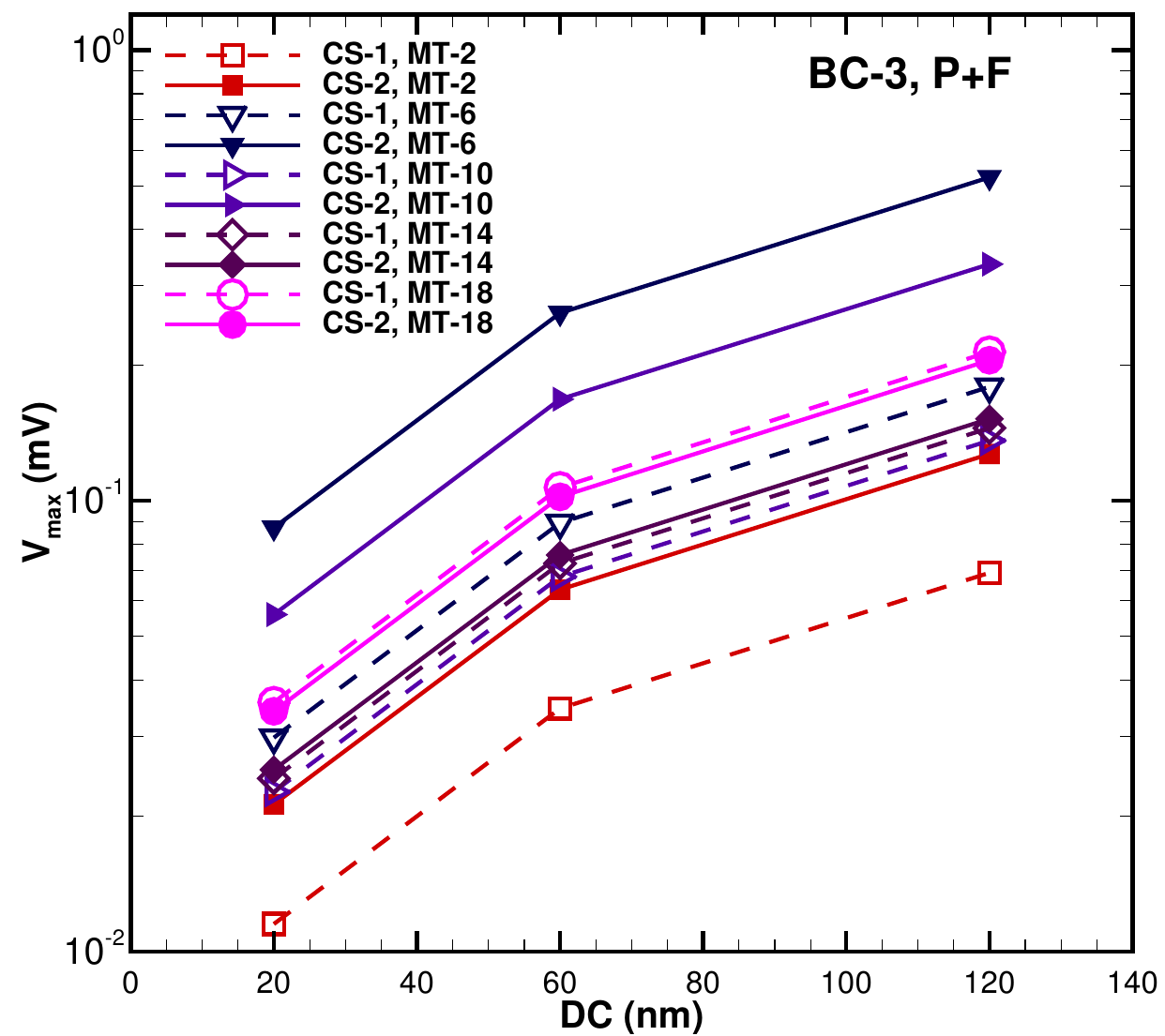}}\\
	\caption{Maximum electric potential ($V_{\text{max}}$) as a function of the compressive displacement and number of microtubules considering only for the piezoelectric effect (a,c,e) and for the combined effect of piezoelectricity-flexoelectricity (b, d, f) under the specific boundary conditions.}
	\label{fig:3_23}
\end{figure}

\noindent A parametric analysis has been carried out to quantify the maximum electric potential ($V_{\text{max}}$) obtained when a compressive displacement of 20 nm is applied for BC-1 to BC-3. The number of microtubules in the present two-dimensional biological cell considered are 18, as shown in \fig\ref{fig:3_23}. When considering purely the piezoelectric effect, CS-1 generates a $V_{\text{max}}$ larger than CS-2 for BC-1, as illustrated in \fig\ref{fig:3_23}(a). On the other hand, $V_{\text{max}}$ shows a significant rise in the case of CS-1 when both the piezoelectric and flexoelectric effects are taken into account together for BC-1, as shown in \fig\ref{fig:3_23}(b) and a similar pattern is observed for BC-2 and BC-3.
An increase in the number of microtubules in the cellular structure results in an increase of $V_{\text{max}}$. This increase is evident from the compressive displacement range of 20 nm to 120 nm observed in BC-1 to BC-3, as shown in \fig\ref{fig:3_23}(a)-(f). This occurs due to the orientation, shape, and number of microtubules. Recall that, in the present study, the flexoelectric coefficients are considered only for the microtubules. 

\begin{figure}[hpbt]
	\centering
	\subfloat[BC-1, Piezo]{\includegraphics[width=0.42\linewidth]{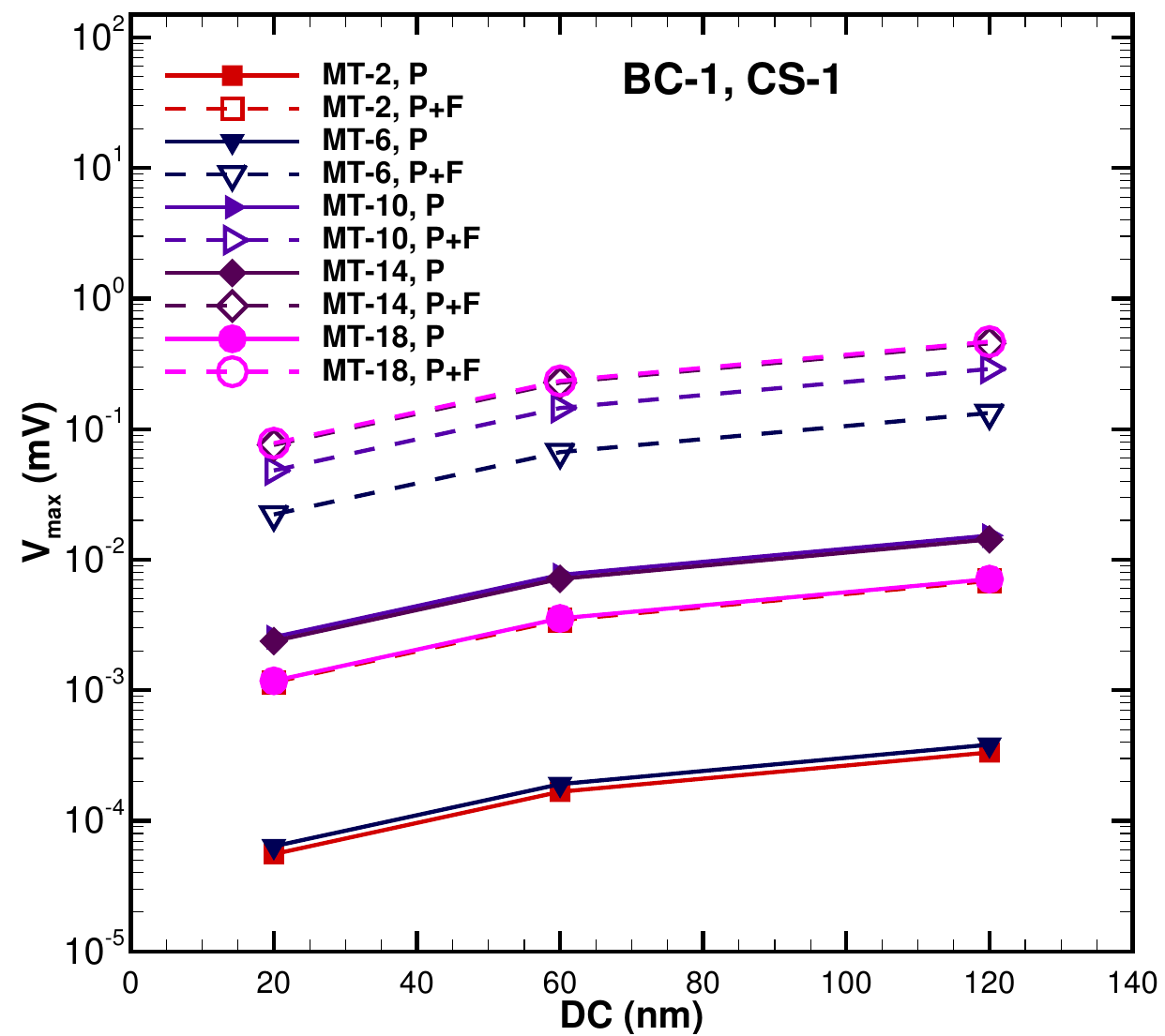}}
	\subfloat[BC-1,P+F]{\includegraphics[width=0.42\linewidth]{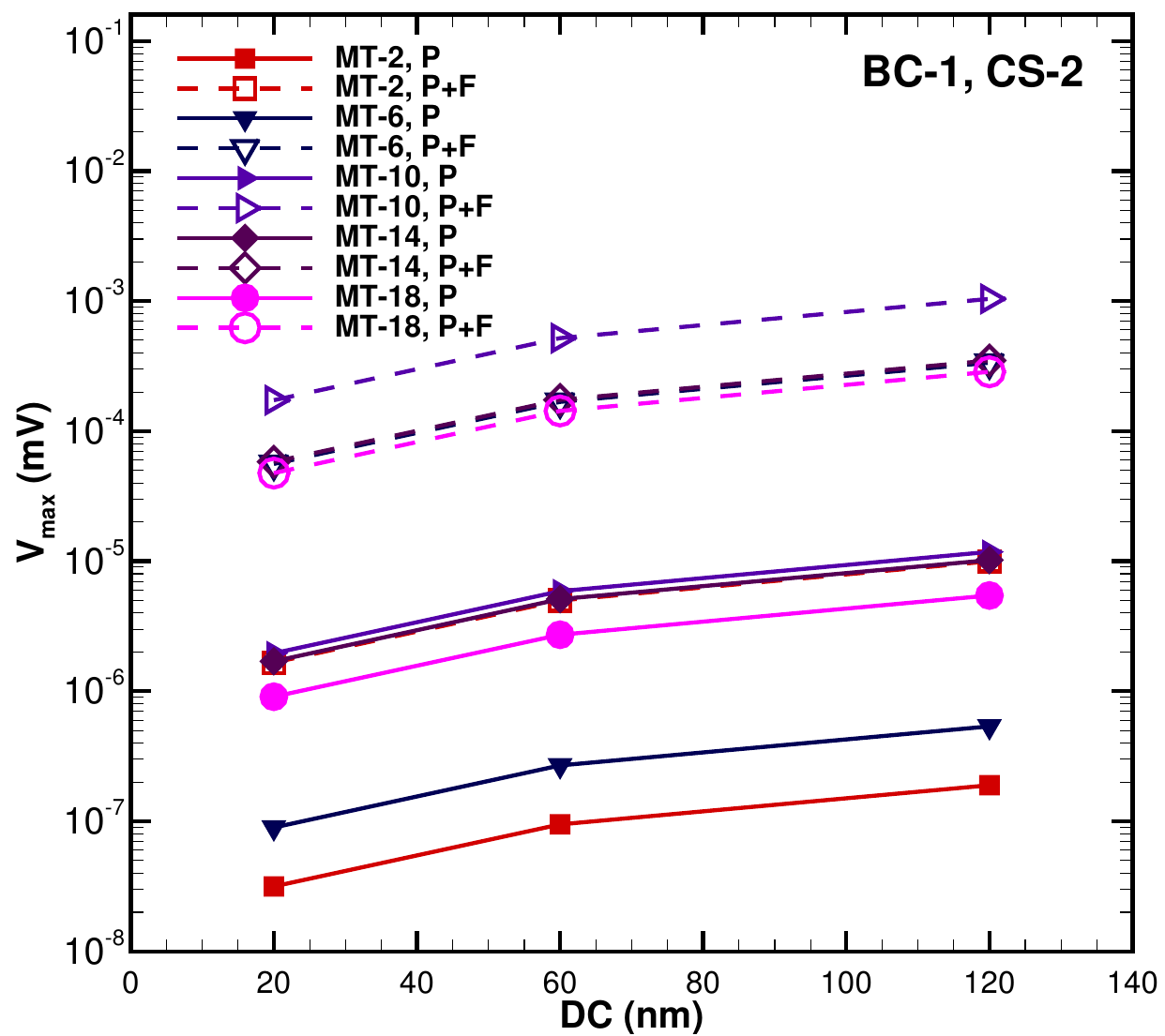}}\\
	\subfloat[BC-2, Piezo]{\includegraphics[width=0.42\linewidth]{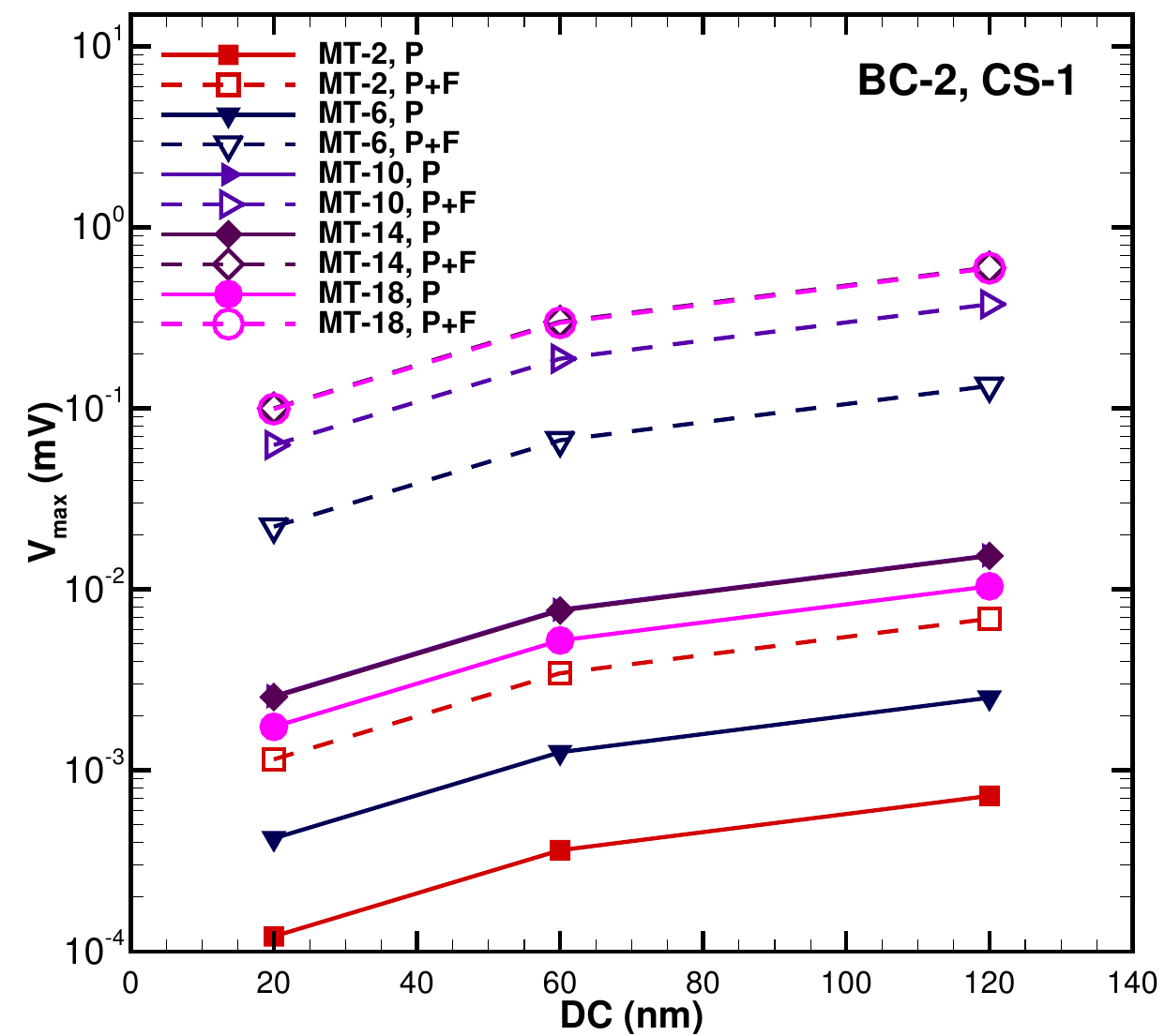}}
	\subfloat[BC-2,P+F]{\includegraphics[width=0.42\linewidth]{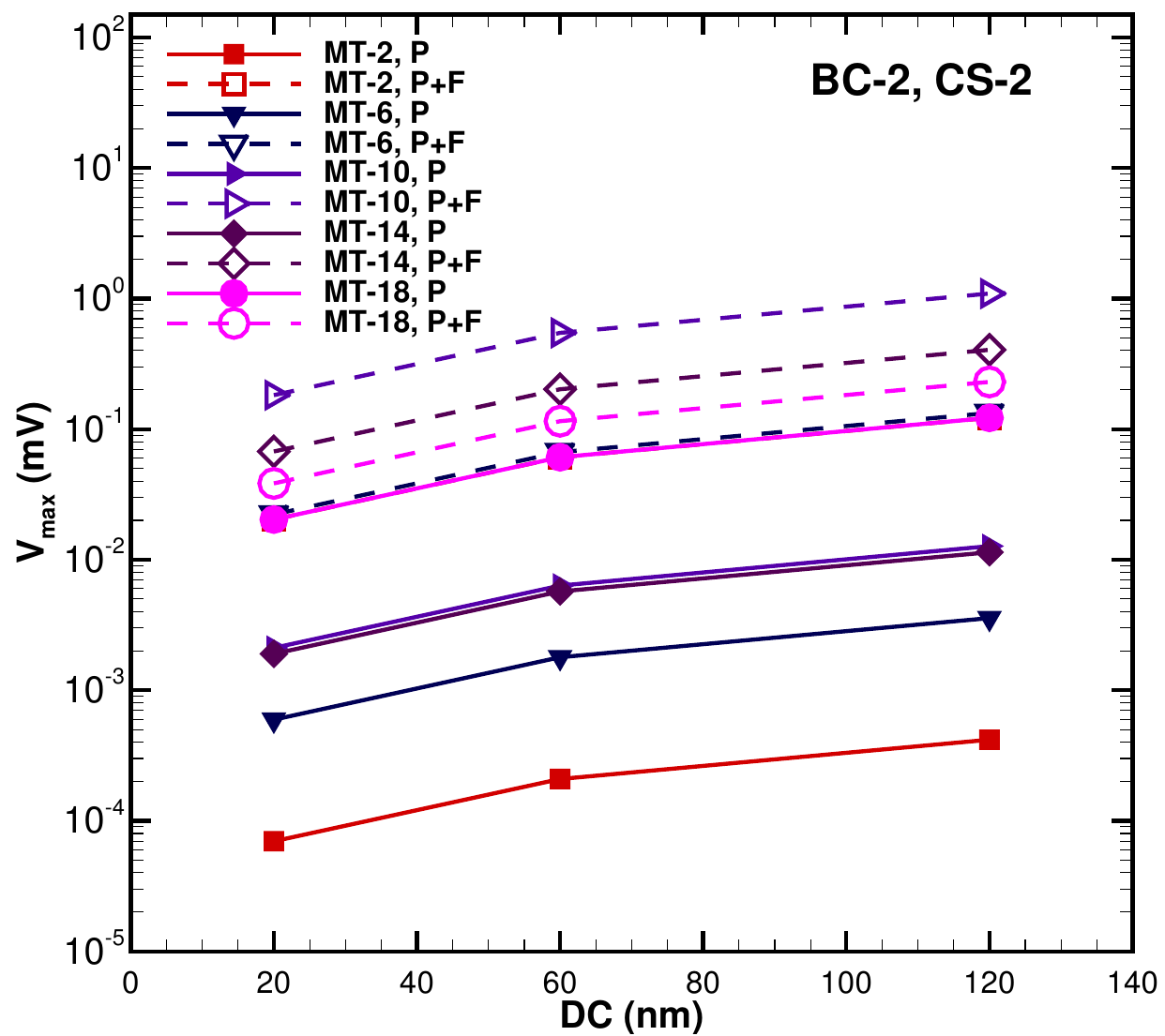}}\\
	\subfloat[BC-3, Piezo]{\includegraphics[width=0.42\linewidth]{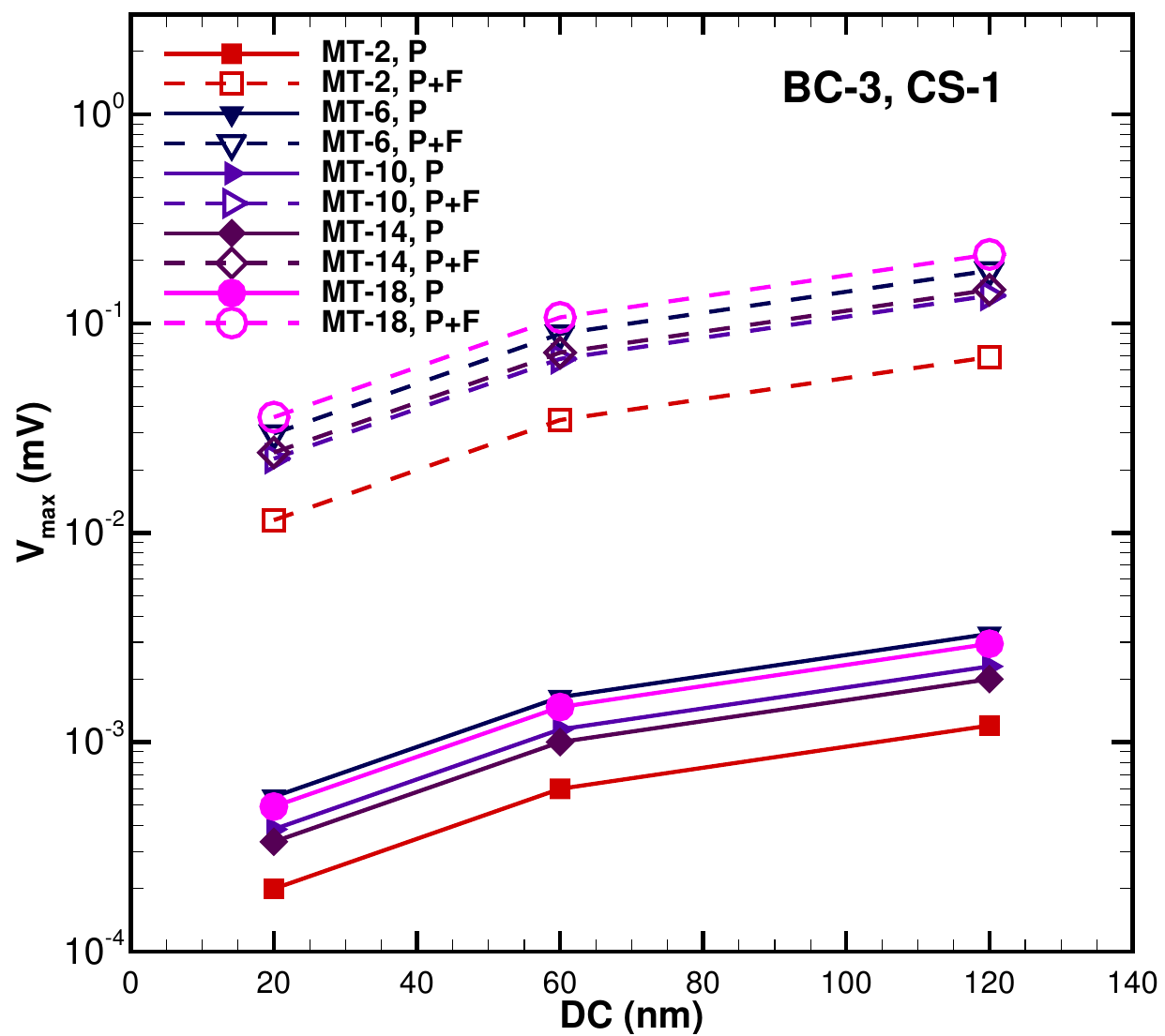}}
	\subfloat[BC-3,P+F]{\includegraphics[width=0.42\linewidth]{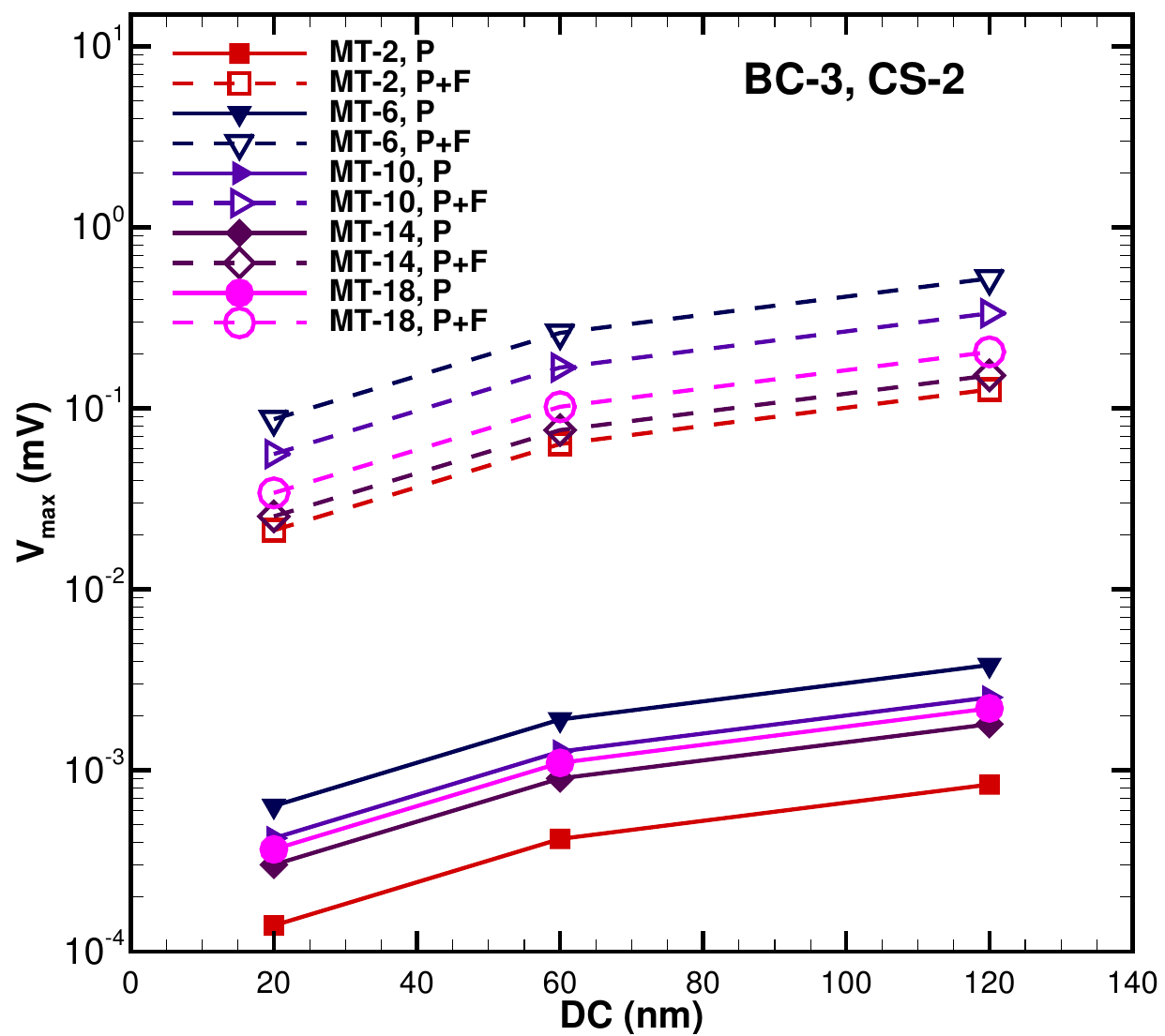}}\\
	\caption{Maximum electric potential distribution ($V_{\text{max}}$) and the compressive displacement and number of microtubules in the case of the piezoelectric (a,c,e) and flexoelectric (b, d, f) effects, considering the three boundary conditions.}
	\label{fig:3_24}
\end{figure}

\fig\ref{fig:3_24} illustrates the variation in $V_{\text{max}}$ distribution resulting from the piezoelectric-flexoelectric effects within the biological cell, while maintaining boundary conditions and cell structures. This variation is observed as a function of the compressive displacement and the number of microtubules. In CS-1 for BC-1, the induced $V_{\text{max}}$ is significantly higher for the piezoelectric-flexoelectric effects compared to the piezoelectric effect alone.
It can be observed from \fig\ref{fig:3_24}(a)-(f) that an increase in the number of microtubules leads to an increase in the generated $V_{\text{max}}$ for all the boundary conditions. This is attributed due to the orientation of the microtubules in the cells. Experimentally, it has been found that the orientation and alignment of microtubules significantly impact the shape and functionality of keratinocyte cells \citep{Kobori2024}. 

\figs\ref{fig:3_23} and \ref{fig:3_24} also clearly demonstrate that the magnitude of $V_{\text{max}}$ follows a trend of increasing from BC-1 to BC-2 and thereafter decreasing at BC-3. For instance, when the piezoelectricity alone is considered,  $V_{\text{max}}$ for CS-1 is found to be $1.18\times 10^{-6}$ V (BC-1), $1.74\times 10^{-6}$ V (BC-2), and $4.92 \times 10^{-7}$ V (BC-3).
Experimental findings confirm that the maximum induced electric potential in the bone by the piezoelectric effect is on the order of microvolts (\micro V) \citep{Nakamura2021}.
Further, a similar pattern has been noticed in the case of CS-2. This occurs because, in BC-1, the electric potential is solely generated by the transverse strain from the top side, without any contribution from the longitudinal strain. In BC-2, the applied compressive displacement on both the left and right sides is balanced due to the absence of a strain gradient. The gradient of BC-3 is symmetrical from all directions, resulting in a minimal net value of the electric potential.
 
\begin{figure}[!t]
	\centering	
	\subfloat[BC-1]{\includegraphics[width=0.49\linewidth]{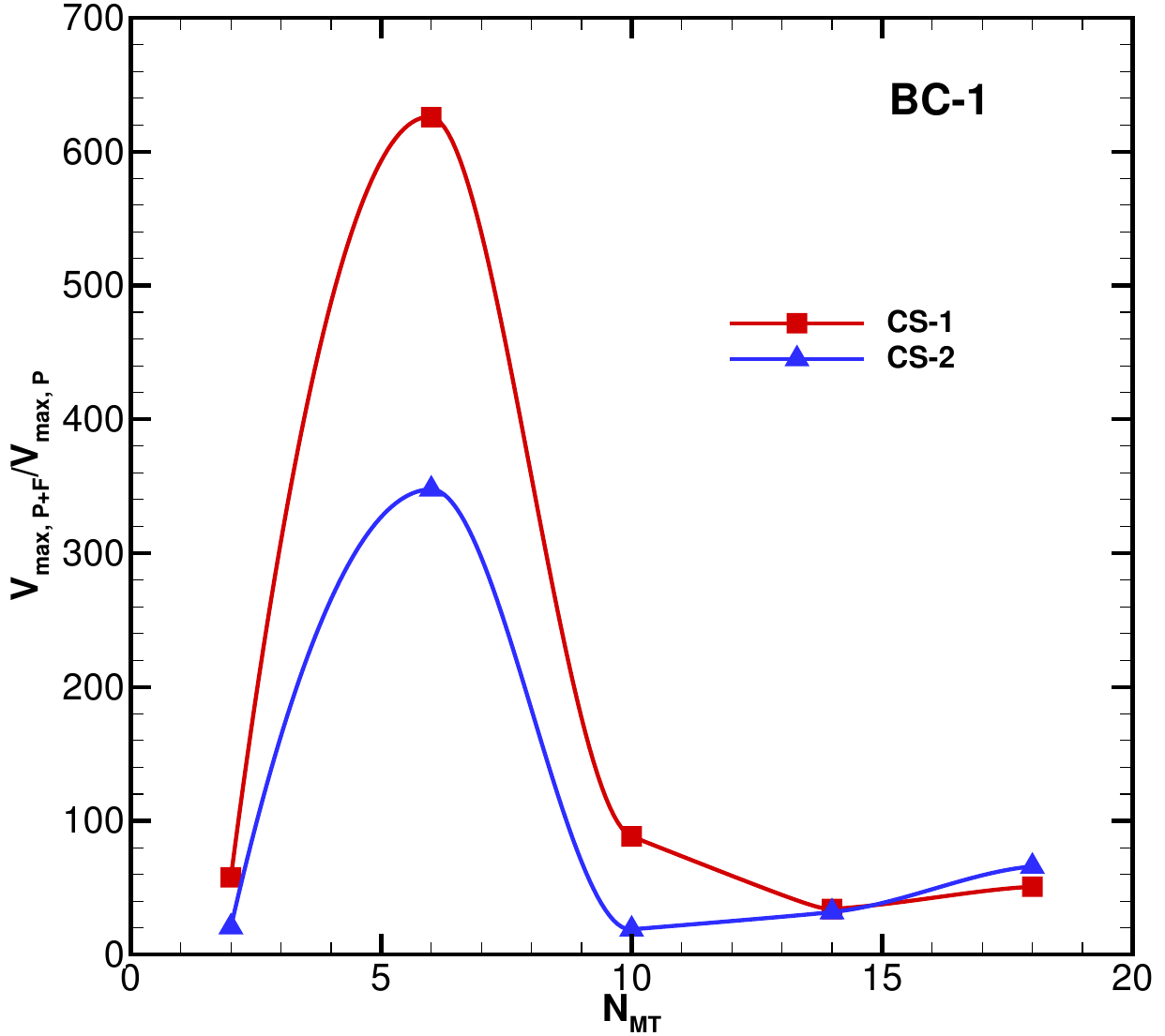}}
	\subfloat[BC-2]{\includegraphics[width=0.49\linewidth]{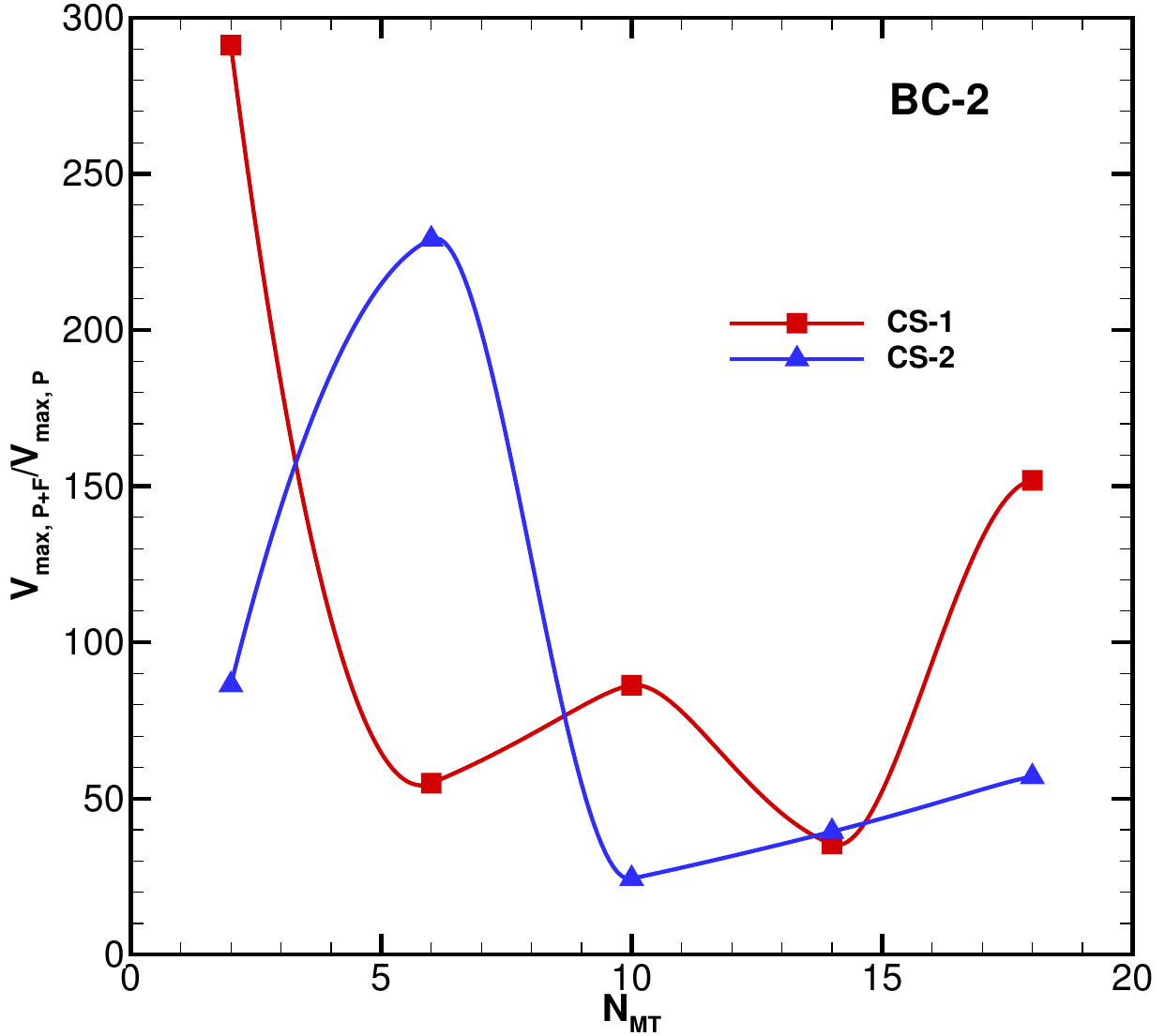}}\\
	\subfloat[BC-3]{\includegraphics[width=0.49\linewidth]{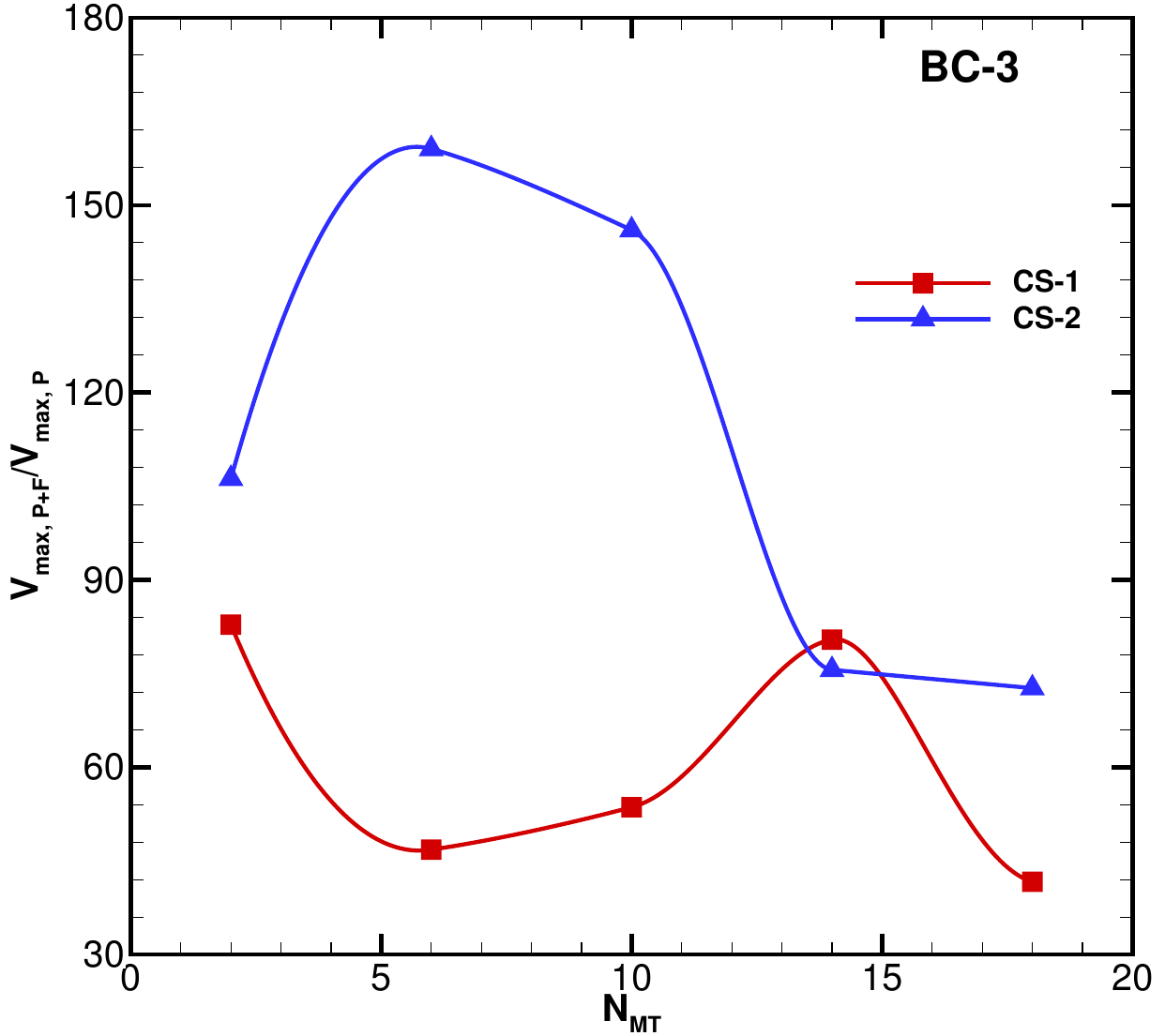}}
	\caption{Ratio of the maximum electric potential for the piezoelectric and flexoelectric effects to the piezoelectric effects, denoted as $V_{\text{R,max}}=V_{\text{max,P+F}}/V_{\text{max,P}}$, as a function of the number of microtubules under three boundary conditions.}
	\label{fig:3_25}
\end{figure}

Flexoelectricity exerts a substantial influence on the biological cell. It is additionally characterized by the ratio of the maximum electric potential resulting from the combined effects of piezoelectricity and flexoelectricity to the electric potential arising from piezoelectricity alone ($V_{\text{R,max}}=V_{\text{max,P+F}}/V_{\text{max,P}}$). This ratio has been examined in connection with the total number of microtubules under the three specified boundary conditions, as illustrated in \fig\ref{fig:3_25}. It is evident that the highest value of $V_{\text{R,max}}$ occurs when there are 6 microtubules in the case of CS-2, under the three boundary conditions BC-1 to BC-3. It has been found that when BC-1 is applied, $V_{\text{R,max}}$ for CS-1 takes precedence over CS-2. Whereas $V_{\text{R,max}}$ has shown an uneven trend for CS-1 on imposing BC-2 and BC-3. The primary factor influencing the behaviour is the angular orientation and shape of the microtubules \citep{Boehm2005,Nakamura2021}.

This knowledge is very useful to our better understanding of the induced electric potential resulting from piezoelectricity in the healing mechanism for bone fractures and other medical therapies. For example, the primary building blocks of bone, collagen and hydroxyapatite, are oriented strongly in the direction of the bone axis. The direction in which ultrasonic radiation or a load is applied greatly influences the induced electric potential in the bone resulting from piezoelectricity. The piezoelectric properties of bone have a role in the healing process of the Low-intensity pulsed ultrasound (LIPUS) therapy \citep{Fukada1957,Antebi2012,Nakamura2021}.
The two-dimensional biological cell is significantly influenced by the quantity and loading conditions of microtubules. The current study predicts a significant increase in electromechanical effects, potentially useful in bio-compatible nano-biosensors, drug delivery, noninvasive diagnostic methods, and treatment strategies \citep{DENG2014,Peppas2016,Ashammakhi2018,Kamel2022}.

The stimulation of cells depends on the generation of electric potential resulting from the mechanical load being applied. For instance, the glial cells undergo changes in structure when stimulated electrically and align themselves perpendicular or parallel to the electric field \citep{Tsui2022}. Hence, cell stimulation depends on the intensity of the electric potential, which is dependent on the boundary condition.
The biological cell having a structure with regular-shaped organelles exhibited over cellular stimulation than the arbitrary shapes of the organelles. The previous study utilized a cell structure \citep{SUNDEEPSINGH2020} model that exhibited regular shapes of the organelles. Therefore, in the present study, we carry out a more realistic analysis where the shapes of the organelles have been chosen to be arbitrary, which are non-spherical in nature. Moreover, the cell stimulation is controlled by varying the mechanical displacement as described previously (Section \ref{BC}).
\subsection{Determination of the effective elastic coefficients}
\noindent Effective elastic coefficients are computed using \eqns\ref{eqn:1_1} and \ref{eqn:1_2} to elucidate the mechanical response of the cell, which show that the principal plane stresses are the only factors influencing these coefficients. The compressive displacement has an impact on the effective elastic coefficients, $c_{11,eff}$ and $c_{33,eff}$ in the principal planes ($x_1$ and $x_3$). We also have another elastic coefficient, $c_{13,eff}$, defined as the ratio of the average principal stress in the $x_3$-direction to the principal strain in the $x_1$-direction.
The following sub-section discusses the numerical studies conducted to investigate the variations in the effective elastic coefficients influenced by biological cells' local geometry and physical properties, considering only the piezoelectric and combined piezoelectric-flexoelectric effects.
\subsubsection{Consideration of piezoelectric effects alone and combination of piezoelectric-flexoelectric effects}
\noindent 

\begin{figure}[!t]
	\centering
	\subfloat[BC-1]{\includegraphics[width=0.49\linewidth]{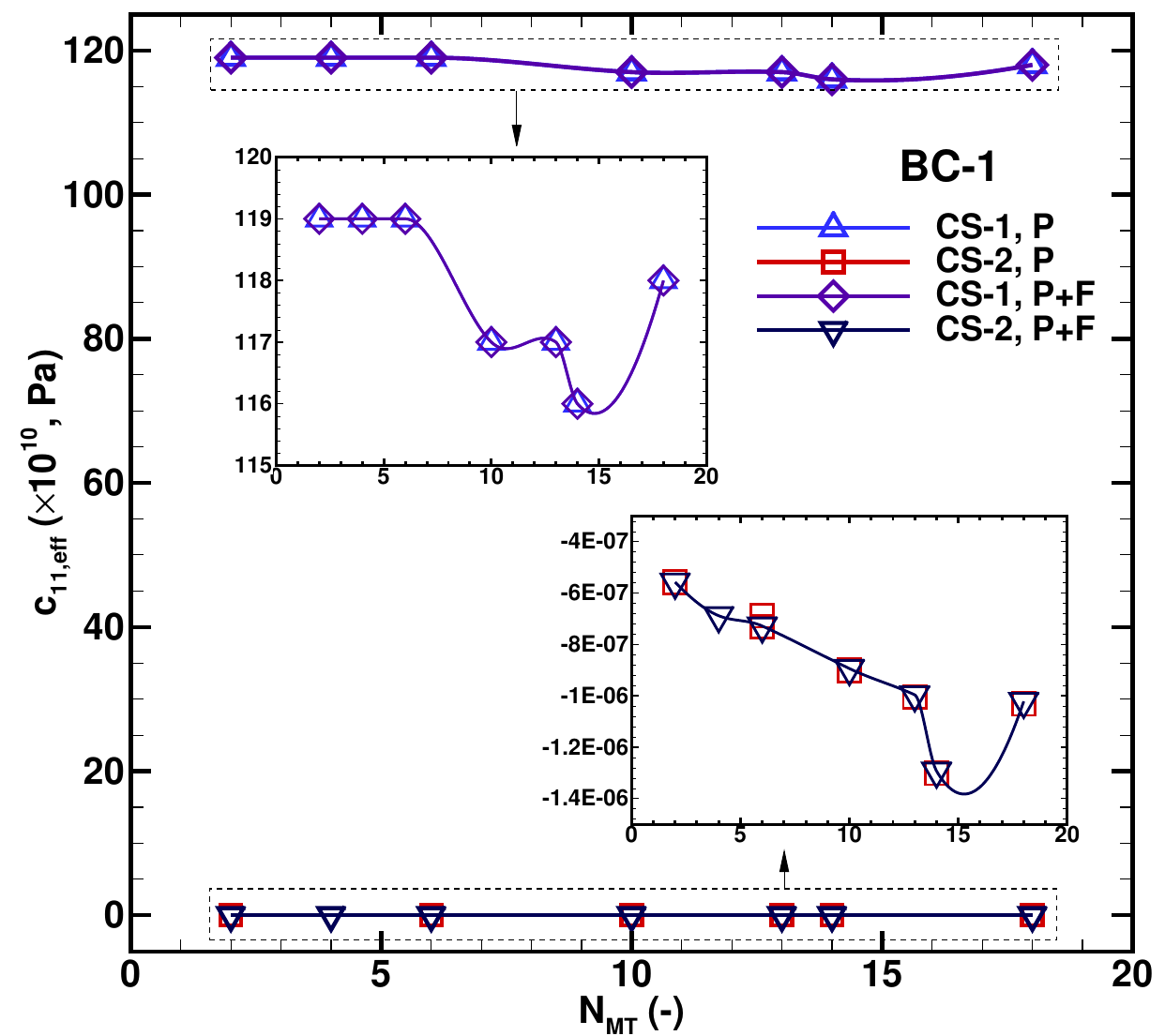}}
	\subfloat[BC-2]{\includegraphics[width=0.49\linewidth]{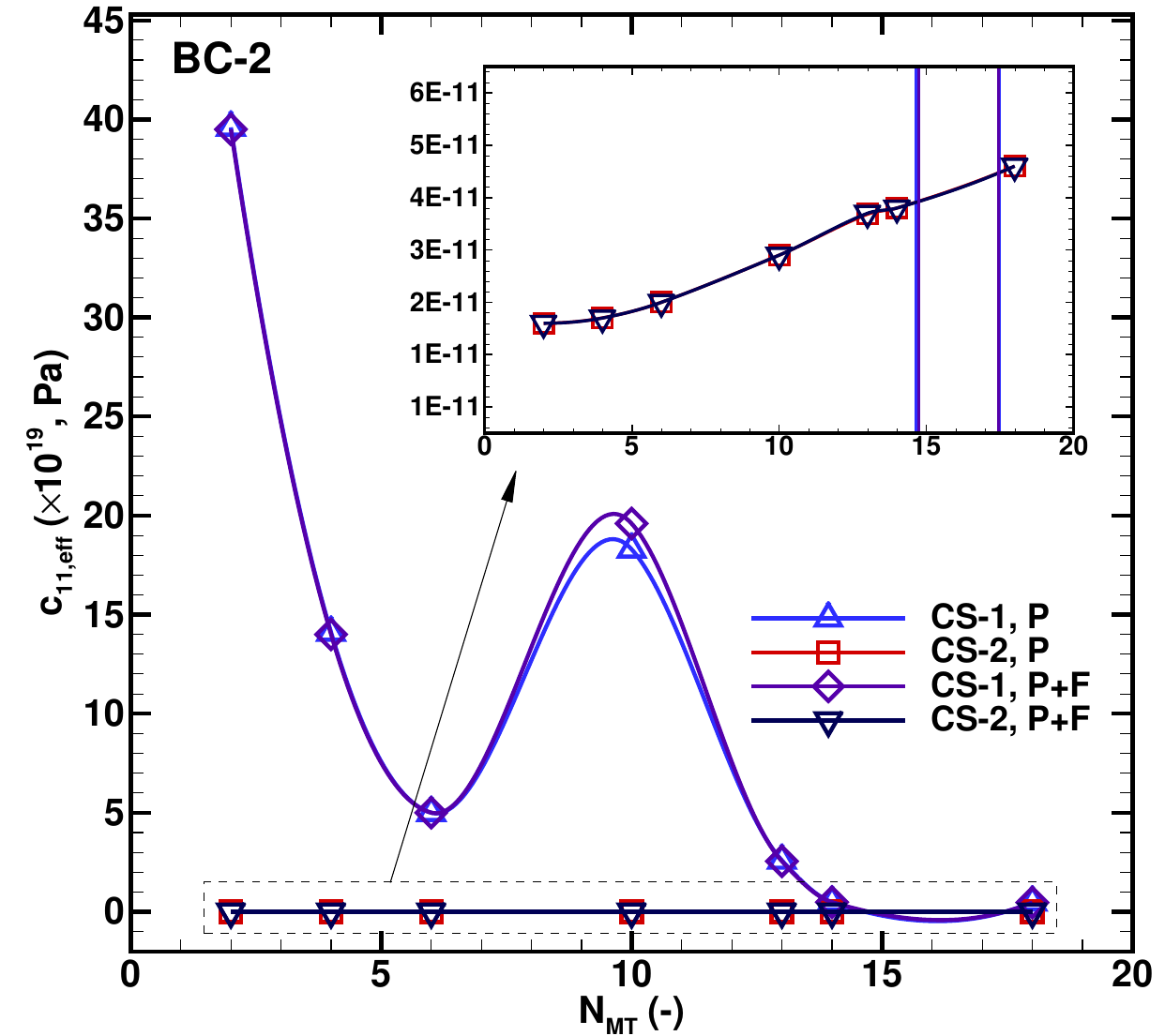}}\\
	\subfloat[BC-3]{\includegraphics[width=0.49\linewidth]{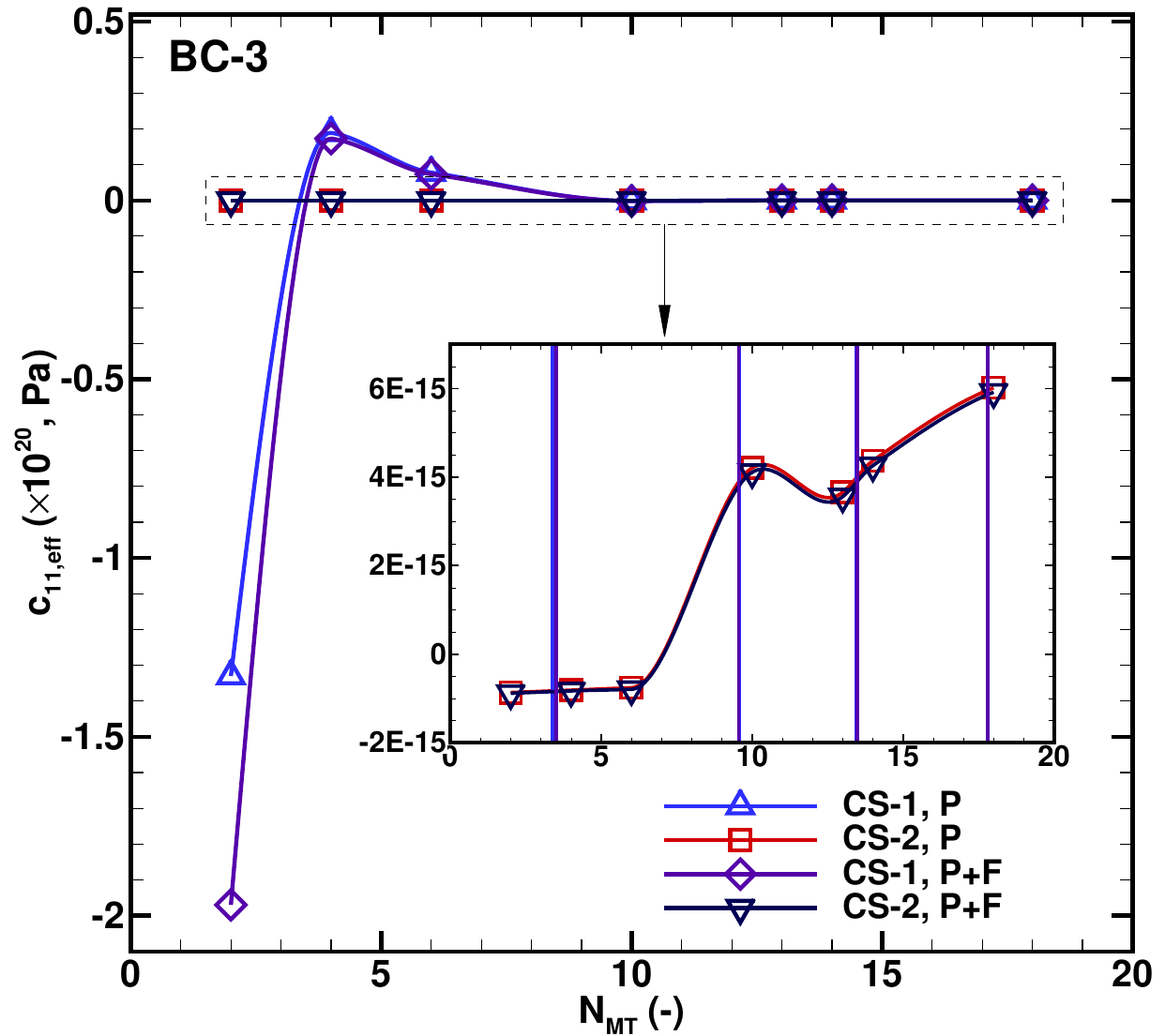}}
	\caption{The effective elastic coefficient ($c_{11,eff} $) of cell structure-1 (CS-1) and cell structure-2 (CS-2) as a function of the number of microtubules for the piezoelectric (P) and flexoelectric (P+F) effects are plotted against the compressive displacement at 20 nm for (a) BC-1, (b) BC-2, and (c) BC-3.}
	\label{fig:3_41}
\end{figure}

BC-1 clearly demonstrates that the mechanical response of the cell in the $x_1$-direction, $c_{11,eff}$, is 100 times greater for CS-1 than for CS-2, considering solely the piezoelectric effect, as shown in \fig\ref{fig:3_41}(a). Because the compressive displacement is given to the upper side of the cell. This shows that the predicted value of $c_{11,eff}$ depends not only on the elastic coefficients but also on the geometry of the cell \citep{Yaojin2012}. Additionally, it should be noted that the value of $c_{11,eff}$ remains constant for both the piezoelectric and flexoelectric effects. 

For BC-2, there is no fluctuation in the value of $c_{11,eff}$ for CS-2 when only the piezoelectric effect has been taken into account. However, the value of $c_{11,eff}$ is lower for CS-1 compared to CS-2 and displays variation for CS-1, as depicted in \fig\ref{fig:3_41}(b). The compressive displacement is exerted on three sides of the cell, while the bottom side is electrically grounded. Therefore, there is no variation in stress along the $x_3$-direction, and it is only present in the $x_1$-direction. $c_{11,eff}$ exhibits a declining trend in the context of the combined piezoelectric and flexoelectric phenomena as the number of microtubules is augmented from 2 to 6. However, it experiences a rapid surge at 10 microtubules, followed by a subsequent decrease (see \fig\ref{fig:3_41}(b)).

Upon analyzing \fig\ref{fig:3_41}(c), it becomes apparent that when the compressive displacement is applied to all sides of the cell (BC-3), the effective coefficient of $c_{11,eff}$ displays lower values for CS-1 initially, but eventually reaches the same value as CS-2 when just the piezoelectric effect is taken into account. The effective elastic constant $c_{11,eff}$ is larger for CS-1 compared to CS-2, particularly for up to 4 microtubules, considering the combination of the piezoelectric and flexoelectric effects, and $c_{11,eff}$ shows a declining pattern in CS-1. This drop is noticed when the number of microtubules is increased from 4 to 6. Moreover, the effective coefficient of $c_{11,eff}$ converges to a comparable value as that of CS-2 when the number of microtubules surpasses 10.
This is because the stress that is being generated is seen only in the $x_1$-direction due to the impact of the mechanical displacement.

\begin{figure}[!t]
	\centering
	\subfloat[BC-1]{\includegraphics[width=0.49\linewidth]{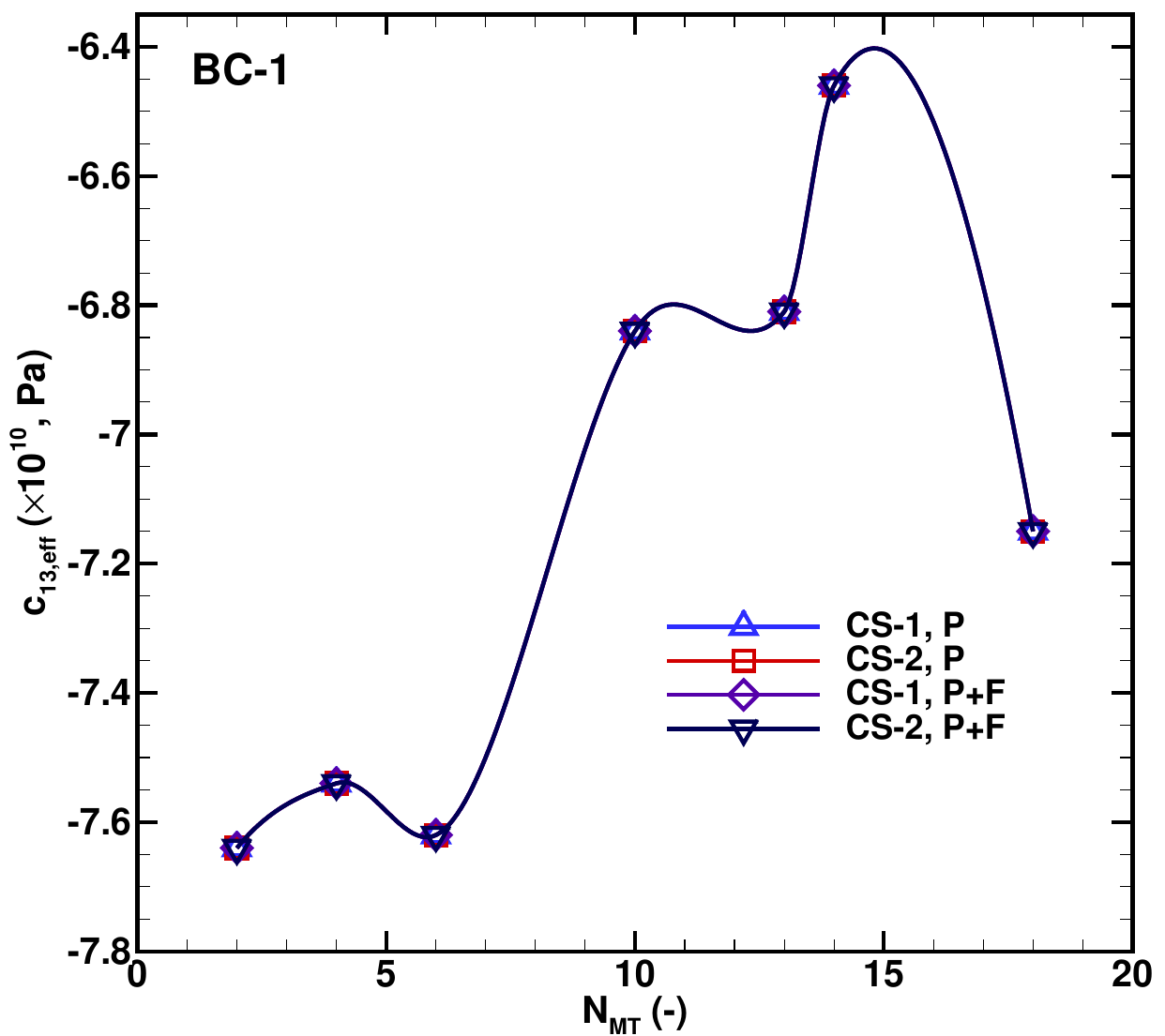}}
	\subfloat[BC-2]{\includegraphics[width=0.49\linewidth]{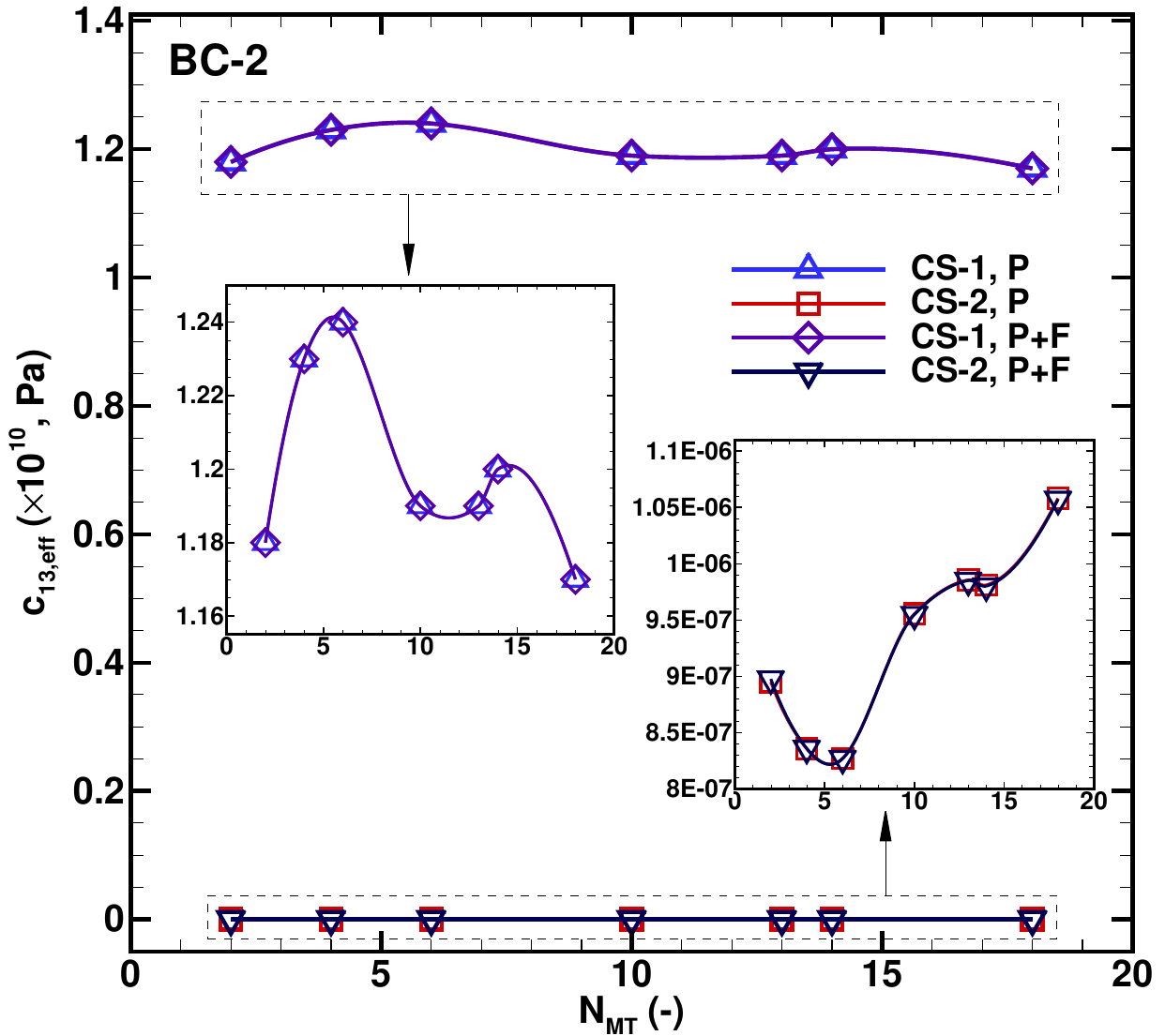}}\\
	\subfloat[BC-3]{\includegraphics[width=0.49\linewidth]{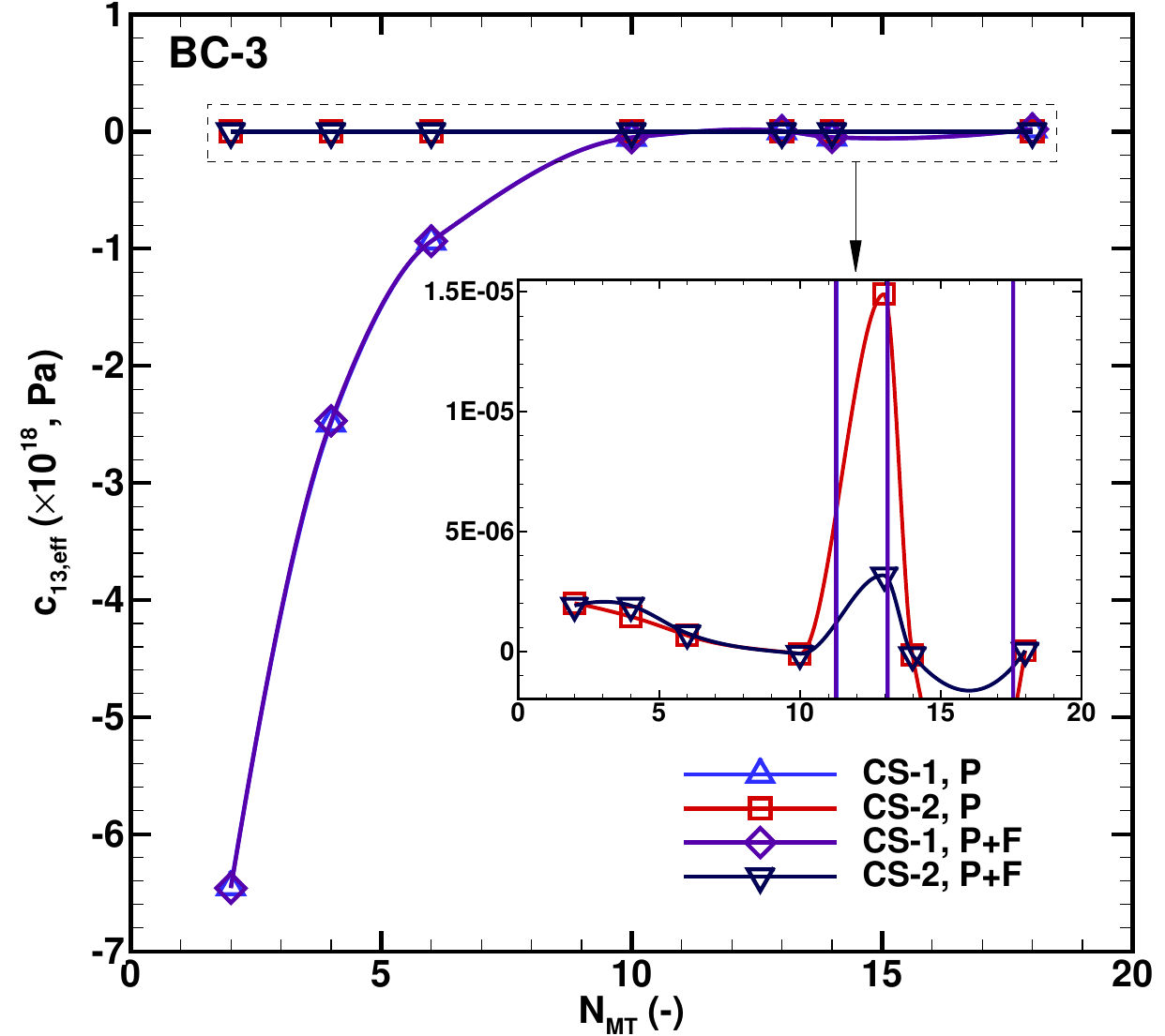}}
	\caption{The effective elastic coefficient ($c_{13,eff} $) of cell structure-1 (CS-1) and cell structure-2 (CS-2) as a function of the number of microtubules considering only the piezoelectric effect (P) and for combined piezoelectric-flexoelectric (P+F) effects plotted against the compressive displacement at 20 nm for (a) BC-1, (b) BC-2, and (c) BC-3.}
	\label{fig:3_42}
\end{figure}

\fig\ref{fig:3_42} depicts the changes in $c_{13,eff}$ across the biological cell, as computed by taking into account the piezoelectric effects. For BC-1, it is apparent from \fig\ref{fig:3_42}(a) that $c_{13,eff}$ exhibits a consistent pattern for both CS-1 and CS-2 when only the piezoelectric effect is considered. However, it is evident from \fig\ref{fig:3_42}(b) that $c_{13,eff}$ value remains unchanged for both the piezoelectric and flexoelectric effects. $c_{13,eff}$ is shown to decrease initially as the number of microtubules is increased from 2 to 6. 
This is because the average stress values in the x3-direction and strain in the x1-direction are the same for both cell structures.
In the case of BC-2, $c_{13,eff}$ is higher for CS-2 in comparison to CS-1 because the principal stresses and strains for these cell structures are different. Moreover, the value of $c_{13,eff}$ remains unchanged for both the piezoelectric and flexoelectric effects, as depicted in \fig\ref{fig:3_42}(b). It can be observed that the value of $c_{13,eff}$ shows no variation while the number of microtubules is increased from 2 to 6. However, there is a dramatic decrease in $c_{13,eff}$ thereafter, followed by a period of no apparent variation.
In the case of BC-3, $c_{13,eff}$ value is lower for CS-2 in comparison to CS-1, specifically for up to 10 microtubules, when considering the piezoelectric and flexoelectric effects. Subsequently, the value of $c_{13,eff}$ tends to converge with that of CS-1 when considering 10 microtubules and beyond, as depicted in \fig\ref{fig:3_42}(c).

\begin{figure}[!t]
	\centering
	\subfloat[BC-1]{\includegraphics[width=0.49\linewidth]{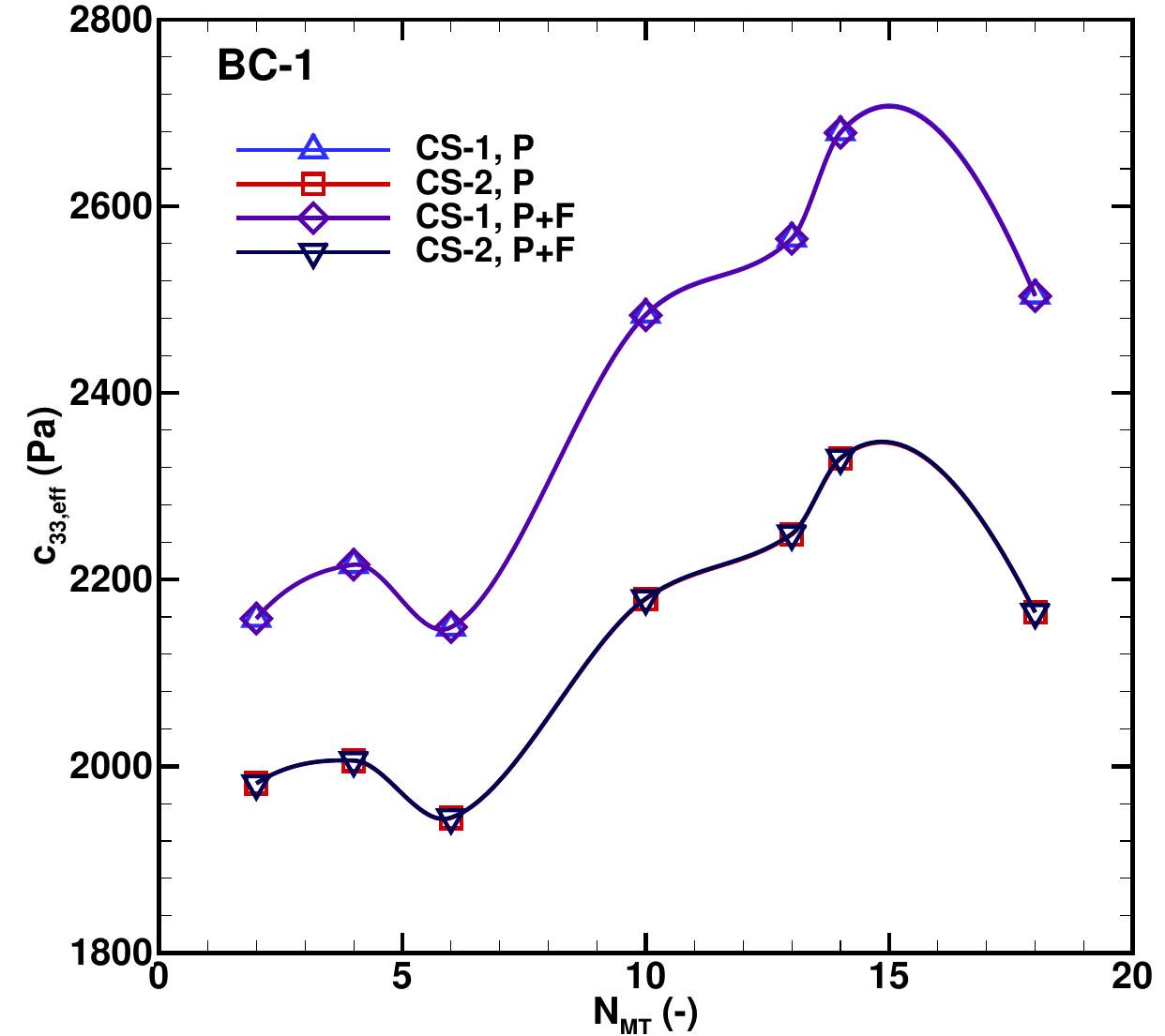}}
	\subfloat[BC-2]{\includegraphics[width=0.49\linewidth]{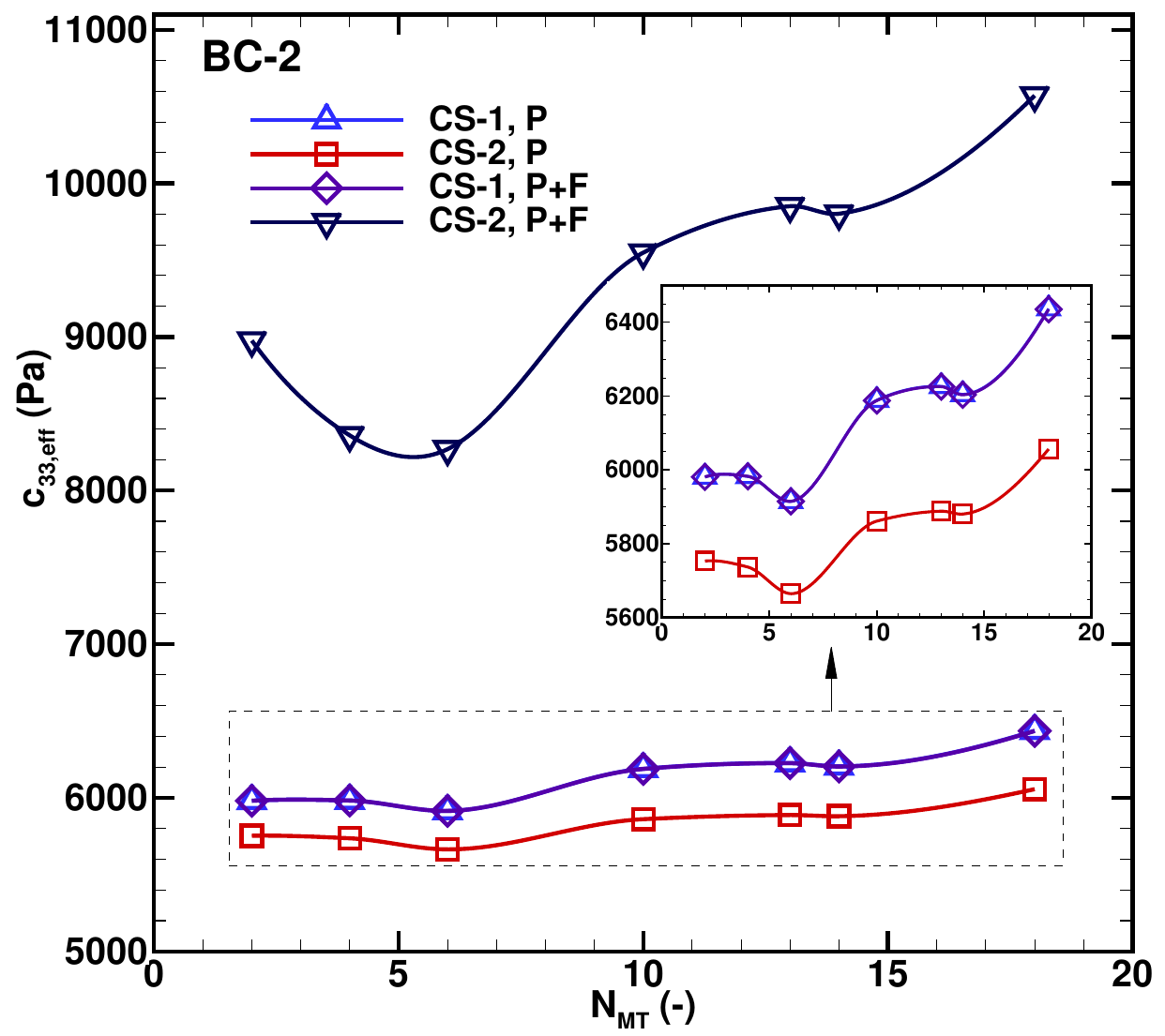}}\\
	\subfloat[BC-3]{\includegraphics[width=0.49\linewidth]{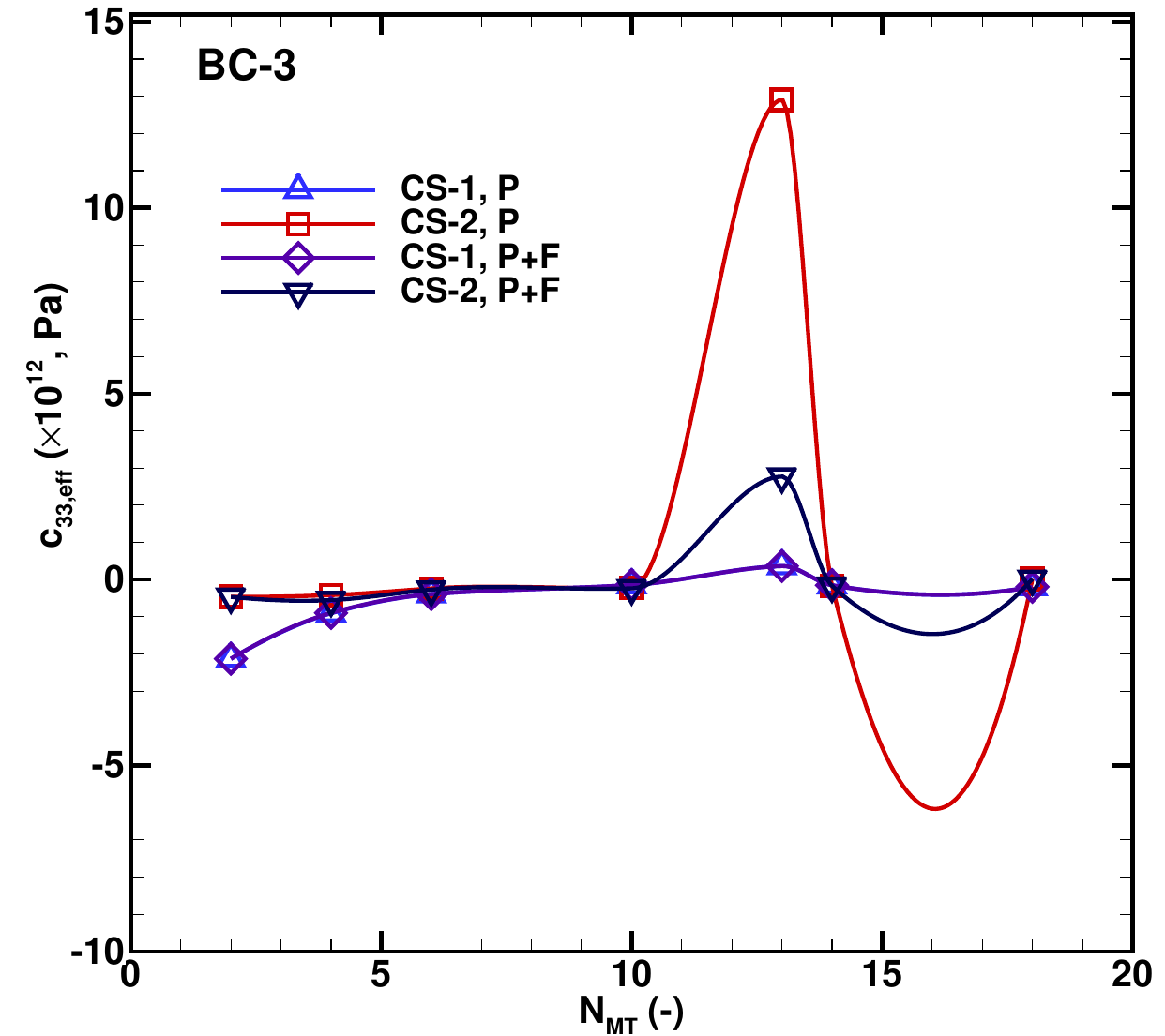}}
	\caption{The effective elastic coefficient ($c_{33,eff} $) of cell structure-1 (CS-1) and cell structure-2 (CS-2) as a function of the number of microtubules considering only the piezoelectric effect (P) and for combined piezoelectric-flexoelectric (P+F) effects are plotted against the compressive displacement at 20 nm for (a) BC-1, (b) BC-2, and (c) BC-3.}
	\label{fig:3_43}
\end{figure}

\fig\ref{fig:3_43}(a) shows that the effective value of $c_{33}$ (the mechanical response of the biological cell in $x_3$-direction) is greater for CS-1 than for CS-2 in BC-1. Both cell structures exhibit an increasing trend in $c_{33,eff}$ with an increase in the number of microtubules. $c_{33,eff}$ value remains unchanged for both the piezoelectric and flexoelectric phenomena. The value of $c_{33,eff}$ exhibits an upward trend when the number of microtubules increases from 2 to 4. However, it experiences an abrupt decline when the number of microtubules reaches 6. The values of $c_{33,eff}$ are shown to increase for both the piezoelectric and flexoelectric effects, as well as for CS-1 and CS-2, extending beyond microtubules 6. The stress and strain are present only in the $x_1$ and $x_3$-directions.
As seen in \fig\ref{fig:3_43}(b), both cell structures exhibit a consistent pattern with an increase in microtubule number for BC-2 when considering the piezoelectric effect alone, and $c_{33,eff}$ is slightly larger for CS-1 than CS-2. It is apparent that $c_{33,eff}$ value remains unchanged for both the piezoelectric and flexoelectric phenomena. $c_{33,eff}$ is declining as the number of microtubules increases from 2 to 6. However, it experiences a dramatic decline at 10 microtubules. The values of $c_{33,eff}$ are observed to increase for both the piezoelectric and flexoelectric effects considering the cell structures, CS-1 and CS-2, extending the microtubules beyond 10.
In the case of BC-3, it is evident that $c_{33,eff}$ is comparatively lower for CS-1 than for CS-1, specifically within the range of 6 microtubules, with regard to the piezoelectric and flexoelectric phenomena. The value of $c_{33,eff}$ is shown to converge with that of CS-1 when considering 6 microtubules, as depicted in \fig\ref{fig:3_43}(c). In addition, when the number of microtubules are 10, it is seen that the effective value of $c_{33,eff}$ for CS-1 starts rapidly increasing.
\subsubsection{Impact of the orientation and shape of the organelles on the effective elastic coefficients}
\noindent The effective elastic coefficients are significantly influenced by the shape and orientation of the organelles. When a compressive displacement is given to the top side of the cell, the effective value of $c_{11,eff}$ is higher for CS-1 compared to CS-2, as depicted in \fig\ref{fig:3_41}(a). Alterations in the cell's shape and orientation of the organelles resulted in different effective coefficients of the cell structures, particularly the microtubules \citep{Kapat2020}. This phenomenon is also observed when there is an increase in the quantity of microtubules, as it leads to a shift in their orientation. 

$c_{13,eff}$ exhibits an initial decrease as the number of microtubules is raised from 2 to 6, as depicted in \fig\ref{fig:3_42}(a). Nevertheless, the orientation of microtubules leads to a significant rise in the effective coefficient, $c_{13,eff}$, when the number of microtubules reaches 10. 
$c_{33}$ shows a positive correlation with the number of microtubules, increasing as the number of microtubules rises from 2 to 4, as depicted in \fig\ref{fig:3_43}(a). Nevertheless, it undergoes a sudden decrease once the quantity of microtubules approaches 6. It is demonstrated that the values of $c_{33,eff}$ rise for CS-1 and CS-2, going beyond the number of microtubules 6.

When applying compressive displacement to all sides of the cell except the bottom, $c_{11,eff}$ shows a declining trend as the number of microtubules increases from 2 to 6 for CS-1, as depicted in \fig\ref{fig:3_41}(b). Nevertheless, it experiences a sharp surge at 10 microtubules, followed by a subsequent decrease. However, unlike CS-1, CS-2 is displaying a distinct pattern. 
$c_{13,eff}$ is greater for CS-1 when compared to CS-2, as illustrated by \fig\ref{fig:3_42}(b). The value of $c_{13,eff}$ remains unchanged as the number of microtubules increases from 2 to 6. However, there is a dramatic decrease in $c_{13,eff}$ thereafter, followed by a period of no apparent variation.
When subjecting the cell to compressive displacement on all sides, it is evident from \fig\ref{fig:3_43}(b) that $c_{33}$ increases as the number of microtubules grows for CS-2. However, the effective $c_{33}$ coefficient exhibits minimal fluctuation for CS-1.

When the cell is subjected to compressive displacement on all sides (BC-3), it is clear that $c_{11,eff}$ is greater for CS-2 compared to CS-1, particularly for up to 4 microtubules, as shown in \fig\ref{fig:3_41}(c). The effective value of $c_{11,eff}$ shows a declining pattern in CS-2. This drop is noticed when the number of microtubules is increased from 4 to 6. Moreover, the effective value of $c_{11,eff}$ converges to a comparable value to that of CS-1 when the number of microtubules surpasses 10.
$c_{13,eff}$ value is lower for CS-2 compared to CS-1, particularly for microtubule numbers up to 10, as seen in \fig\ref{fig:3_42}(c). Afterwards, the value of $c_{13,eff}$ approaches convergence with that of CS-1 when considering the number of microtubules beyond 10.
$c_{33,eff}$ value for CS-2 is lower than that of CS-1, particularly for 6 microtubules, as shown in \fig\ref{fig:3_43}(c). However, it approaches the value of CS-1 when considering 6 microtubules. Furthermore, as the number of microtubules reaches 10, there is a noticeable and quick increase in the effective value of $c_{33,eff}$ for CS-1.

Calculating effective elastic coefficients helps understand biological cell mechanical behavior. For instance, cell organelles have a non-uniform distribution, with osteoblasts having cellular elasticity in the cytoplasmic skirt \citep{Charras2002,Rigato2015,Schillers2019}.
\subsection{Determination of the effective piezoelectric coefficients}
\noindent The presence of piezoelectric inclusions, such as the number of microtubules, significantly impacts the effective elastic moduli of cell structures. The effective piezoelectric coefficients are calculated by dividing the average values of the strain component and the electric flux density vector component, i.e., $e_{31,eff}=\frac{\langle D_{3} \rangle }{\varepsilon_{\text 11}}$, $e_{33,eff}=\frac{\langle D_{3} \rangle }{\varepsilon_{\text 33}}$.
\subsubsection{Consideration of only the piezoelectric effect and the combination of piezoelectric-flexoelectric effects}
\begin{figure}[!t]
	\centering
	\subfloat[BC-1]{\includegraphics[width=0.49\linewidth]{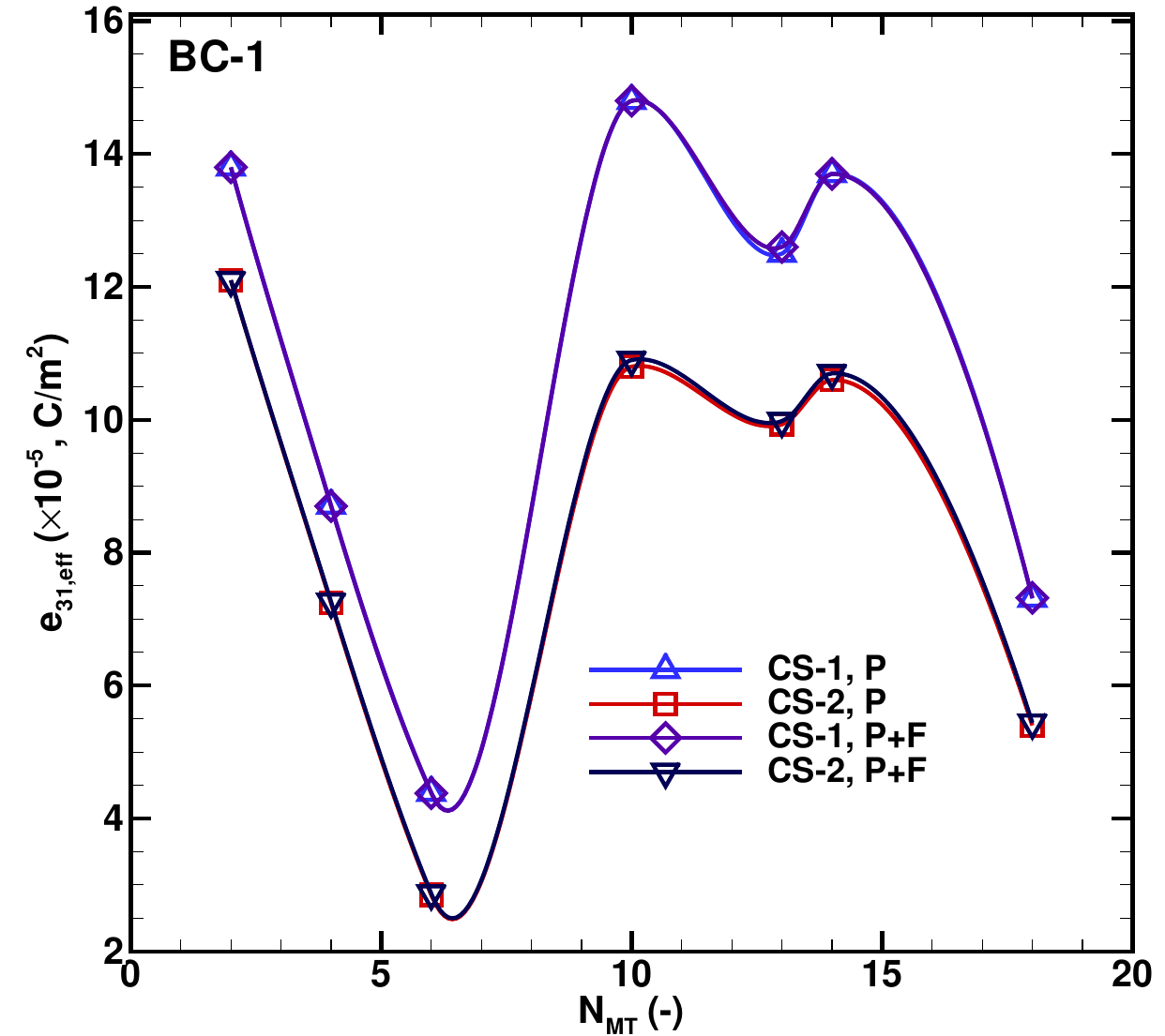}}
	\subfloat[BC-2]{\includegraphics[width=0.49\linewidth]{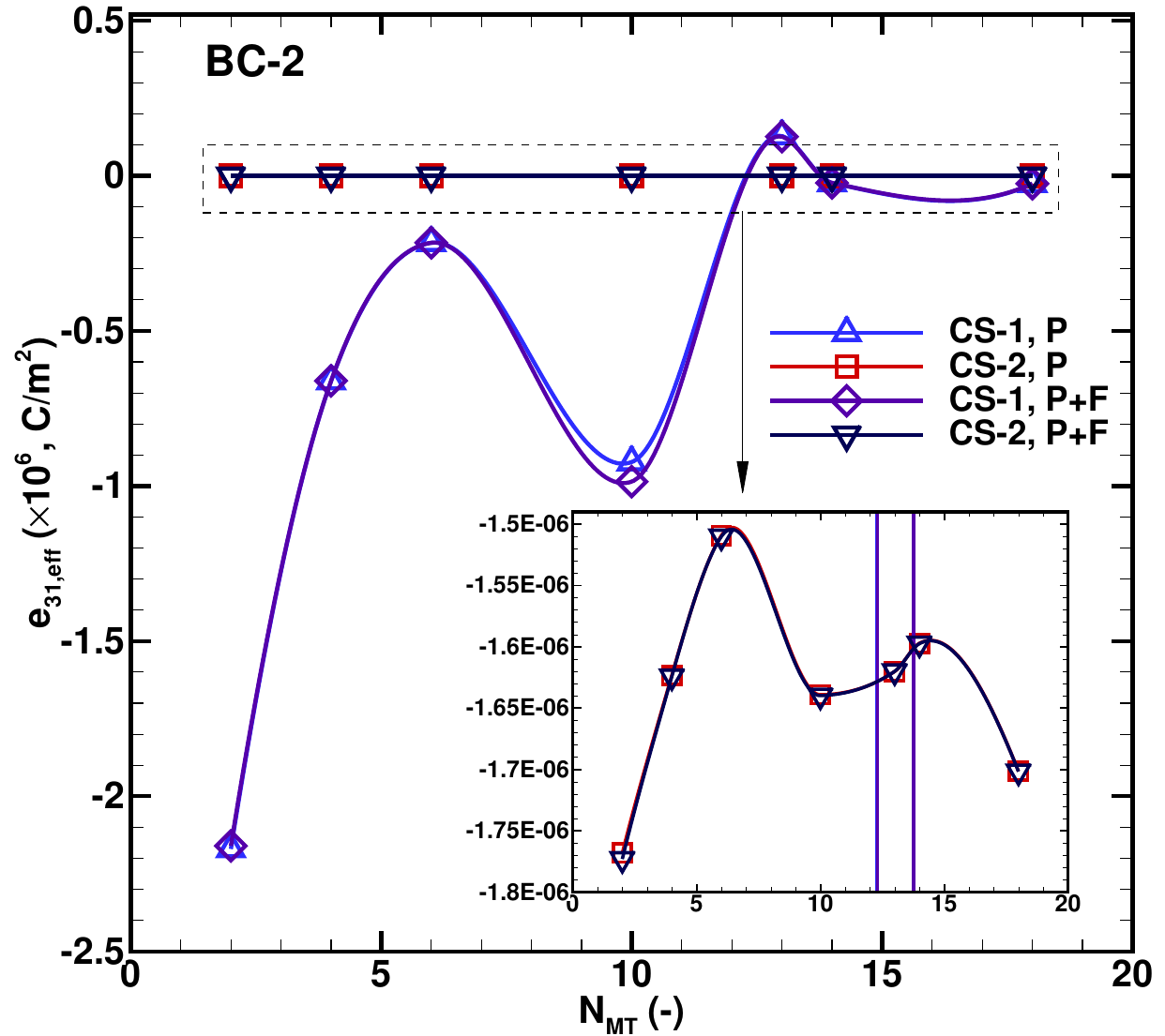}}\\
	\subfloat[BC-3]{\includegraphics[width=0.49\linewidth]{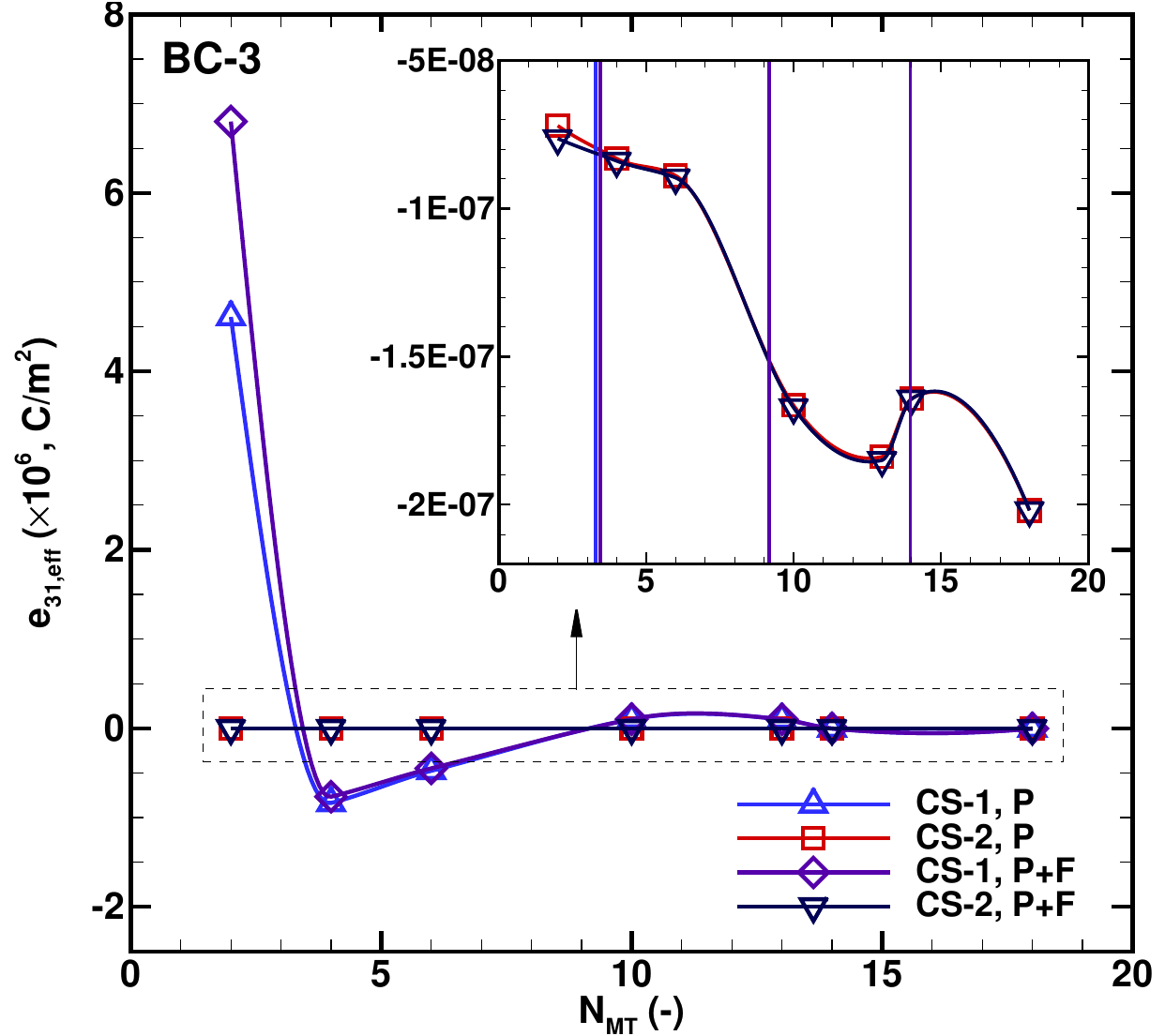}}
	\caption{The effective piezoelectric coefficient ($e_{31,eff} $) of cell structure-1 (CS-1) and cell structure-2 (CS-2) as a function of the number of microtubules considering only the piezoelectric effect (P) and for combined piezoelectric-flexoelectric (P+F) effects are plotted against the compressive displacement at 20 nm for (a) BC-1, (b) BC-2, and (c) BC-3.}
	\label{fig:3_51}
\end{figure}

\noindent \fig\ref{fig:3_51} displays the effective piezoelectric coefficient ($e_{31,eff}$) within the cell for piezoelectrici effect alone and the combination of piezoelectric and flexoelectric effects, i.e., applying mechanical strain in the $x_3$-direction causes a piezoelectric reaction, specifically the production of $e_{31,eff}$ in the $x_1$-direction \citep{Fukada1957}. 

For BC-1, it can be observed that $e_{31,eff}$ displays similar trends for both cell structures (CS-1 and CS-2), encompassing both the piezoelectric and flexoelectric phenomena, as depicted in \fig\ref{fig:3_51}(a). Therefore, the impact of flexoelectricity is insignificant. This phenomenon is caused by the alignment of the microtubules in response to mechanical tension exerted on the top portion of the cell.
When it comes to BC-2, the effective value of $e_{31,eff}$ is higher for CS-2 when compared to CS-1, as depicted in \fig\ref{fig:3_51}(b). The negative values of $e_{31,eff}$ in CS-1 are a result of the compressive stresses in $x_1$ and $x_3$ directions caused by the mechanical stress on three sides of the cell when the bottom side is electrically grounded. Therefore, $e_{31,eff}$ is yielding negative values for CS-1.  
For BC-3, when comparing CS-2 to CS-1, the value of $e_{31,eff}$ shows a trend toward decline with respect to up to four microtubules. This observation relates to the occurrence of piezoelectricity and flexoelectricity. Afterwards, the value of $e_{31,eff}$ shows an increasing pattern and approaches the value of CS-1, as shown in \fig\ref{fig:3_51}(c). 

\begin{figure}[!t]
	\centering
	\subfloat[BC-1]{\includegraphics[width=0.49\linewidth]{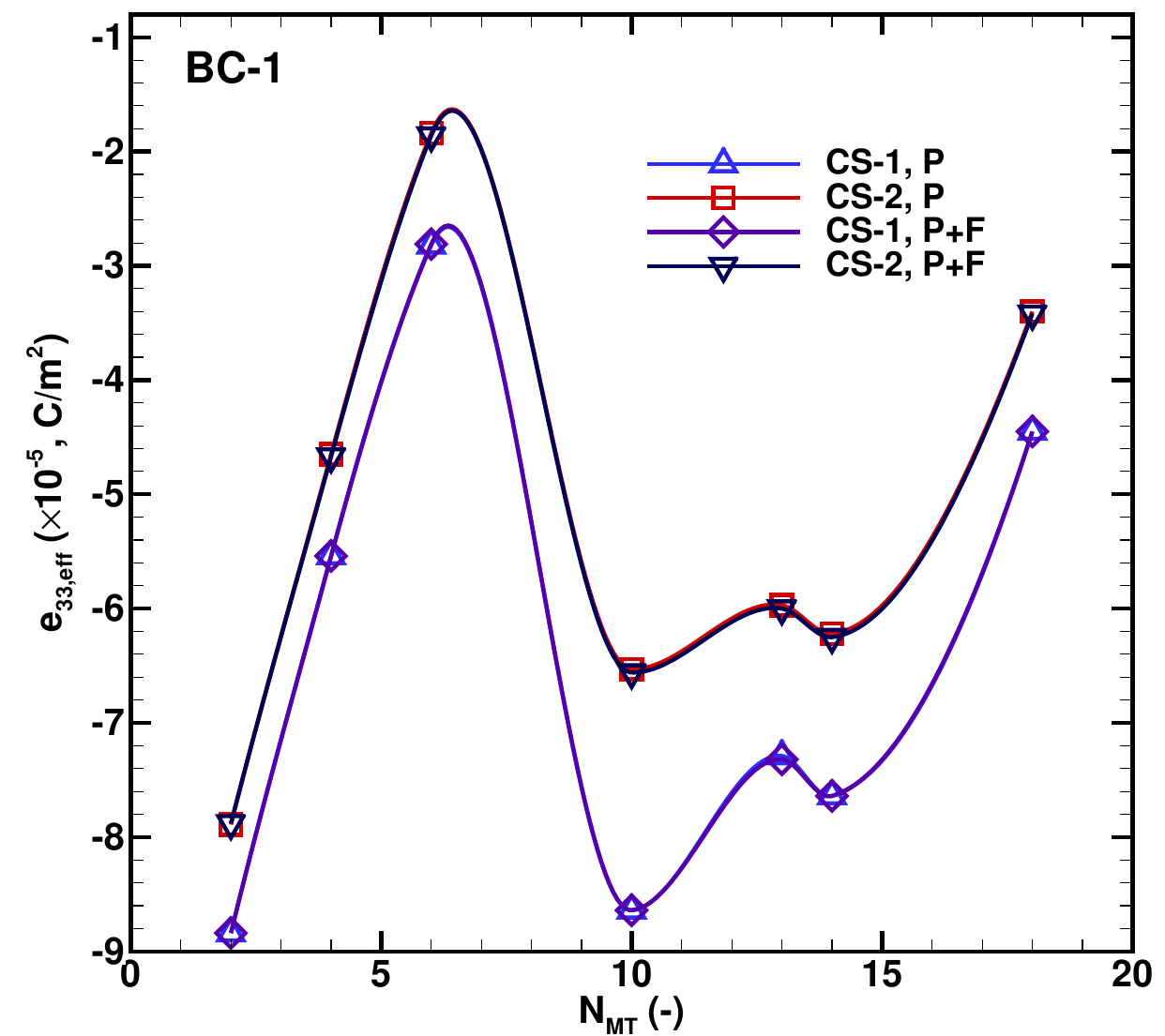}}
	\subfloat[BC-2]{\includegraphics[width=0.49\linewidth]{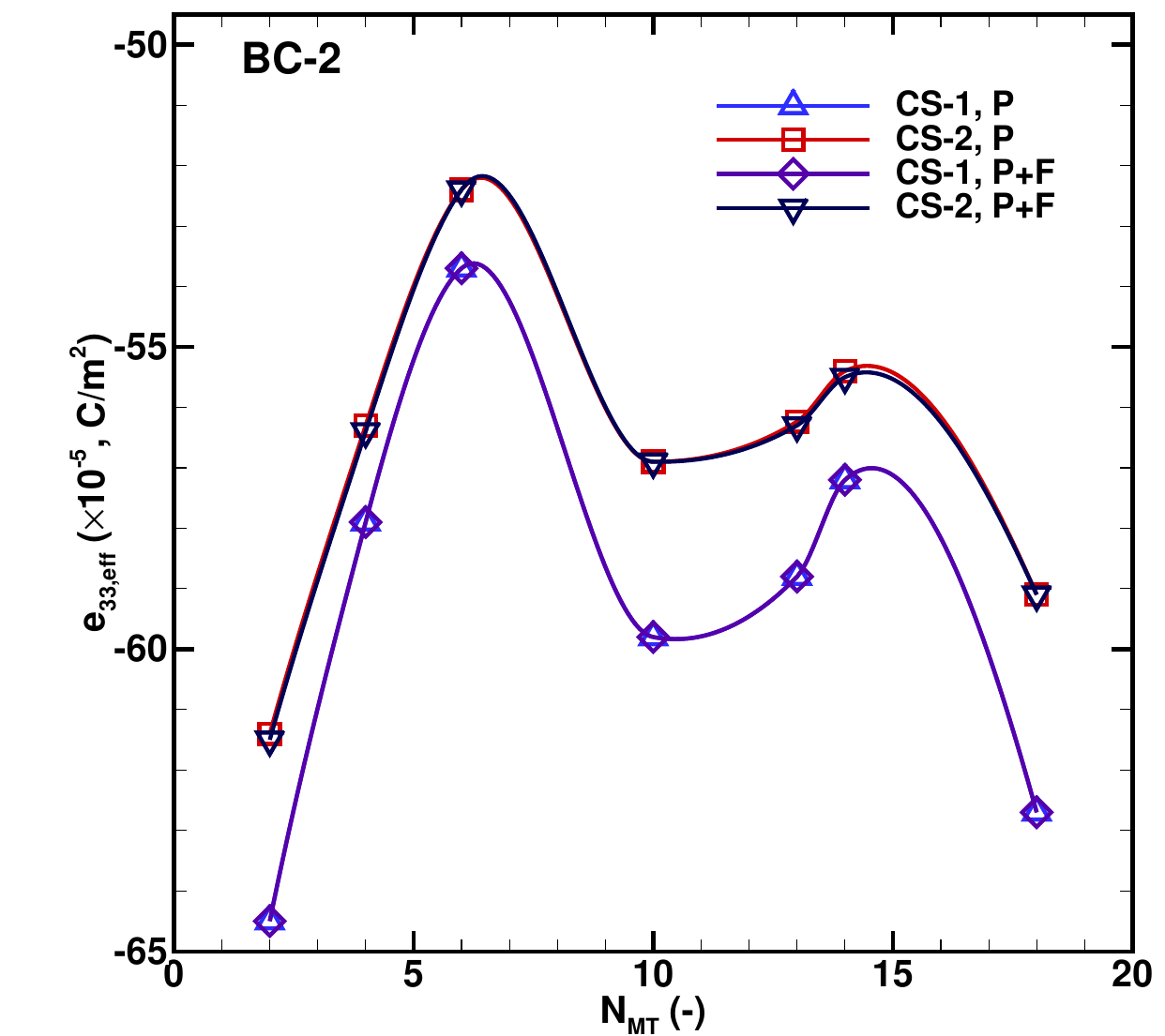}}\\
	\subfloat[BC-3]{\includegraphics[width=0.49\linewidth]{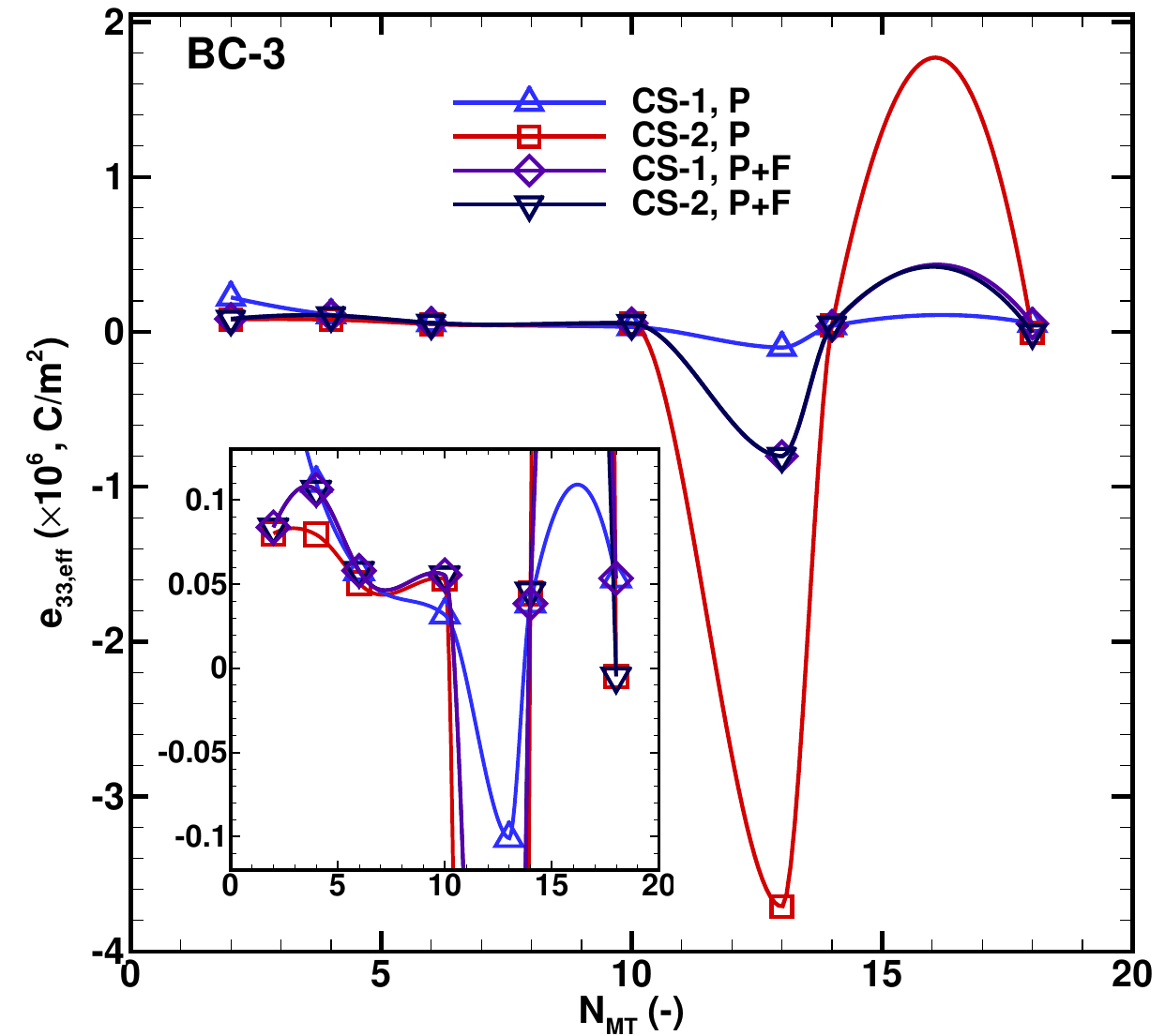}}
	\caption{The effective piezoelectric coefficient ($e_{33,eff} $) of cell structure-1 (CS-1) and cell structure-2 (CS-2) as a function of the number of microtubules considering only the piezoelectric effect (P) and for combined piezoelectric-flexoelectric (P+F) effects are plotted against the compressive displacement at 20 nm for (a) BC-1, (b) BC-2, and (c) BC-3.}
	\label{fig:3_52}
\end{figure}

The effective coefficient $e_{33,eff}$ is calculated for the applied mechanical strain that induces piezoelectric effect in the $x_3$-direction, as shown in \fig\ref{fig:3_52}. It can be observed from \figs\ref{fig:3_52}(a)-(b) that $e_{33,eff}$ is higher for CS-2 than CS-1.
This observation holds true for both the piezoelectric and flexoelectric effects in BC-1 and BC-2. Both CS-1 and CS-2 exhibit similar trends for the values of $e_{33,eff}$. In BC-3, the value of $e_{33,eff}$ remains constant for both the piezoelectric and flexoelectric effects until the number of microtubules reaches 10, as depicted in \fig\ref{fig:3_52}(c), after which it exhibits a significant decrease.
\subsubsection{Influence of the organelle's shapes and orientations on the effective piezoelectric coefficients}
\noindent The effective piezoelectric coefficient, $e_{31,eff}$, has been analyzed under three different applied boundary conditions (BC-1 to BC-3) as a function of the number of microtubules. Despite the symmetry in the shape of the microtubules, the cell structures exhibit different values of $e_{31,eff}$ due to the curved edges at the surfaces.

In the case of BC-1, $e_{31,eff}$ initially decreases as the number of microtubules increases from 2 to 6, as shown in \fig\ref{fig:3_51}(a). However, it subsequently undergoes a significant increase when the number of microtubules reaches 10. This phenomenon is caused by the alignment of the microtubules in response to mechanical displacement exerted on the top portion of the cell. 
From the analysis of \fig\ref{fig:3_52}(a), it is evident that there is an increase in $e_{33,eff}$ for both CS-1 and CS-2 as the number of microtubules is increased from 2 to 6. $e_{33,eff}$ decreases when it transitions from microtubules 6 to 10. Both CS-1 and CS-2 display comparable patterns in terms of the quantity of microtubules for the values of $e_{33,eff}$.

When it comes to BC-2, the value of $e_{31,eff}$ shows no variation regardless of the number of microtubules in cell structure-2, as depicted in \fig\ref{fig:3_51}(b). \fig\ref{fig:3_52}b indicates that there is a rise in $e_{33,eff}$ for both CS-1 and CS-2 when the number of microtubules is raised from 2 to 6 for BC-2. 
The value of $e_{31,eff}$ shows an increasing pattern and approaches the value of CS-1 when the number of microtubules exceeds 10, as shown in \fig\ref{fig:3_51}(c). This behaviour has been caused by the orientation of the microtubules. Moreover, the stress is exerted uniformly on all sides of the cell, leading to no strain gradient. $e_{33,eff}$ consistent for the number of microtubules increases from 2 to 10; beyond that, it experiences a notable reduction, as shown in \fig\ref{fig:3_52}(c). 

In summary, a linearly coupled electromechanical model has been developed to quantify cellular mechanics under external mechanical displacement, considering piezoelectric and nonlocal flexoelectric effects.
The present model assumes that living cells are linear elastic, which provides the first step in the analysis \citep{Kollmannsberger2011,Garcia2018b}. Living cells are visco-elastic, and incorporating dynamic analysis add{into} a three-dimensional study could provide more realistic responses with variable shapes of the organelles and cell structures \citep{Chaudhuri2020}. 
Future extensions could also include cellular tensegrity models, which account for individual components of the cytoskeleton using discrete beam or truss elements \citep{Ingber2014,Sun2023}. Clearly, in realistic scenarios, the electro-elastic properties of biological cells are not uniform and exhibit different qualities in all directions. To capture the intricacies of these phenomena, more sophisticated multiscale modelling methodologies would need to be developed \citep{Ambrosi2019,Schofield2020}. 
Despite limitations, the developed coupled electromechanical model is an important initial model that incorporates piezoelectric and flexoelectric effects within the biological cell. Comparative analysis of the predicted outcomes could help understand the complex electromechanical response of biological cells and assist in the design of electromechanical devices for various medical applications.
\section{Conclusions} \label{Conlusions}
\noindent 
The presented study has explored a two-dimensional conceptual framework of the biological cell for two different structures, encompassing many essential organelles that experience distinct mechanical loading conditions originating from the external environment. The impact of incorporating flexoelectric coupling into the electromechanical model is significantly more prominent than that of solely introducing piezoelectric coupling.
The cell structures used in the present study are classified into two categories: regular and arbitrary shapes. Understanding cell mechanics is highly dependent on these structures and the boundary conditions since they play a significant role in influencing cell behaviour and associated stimulation. 
A parametric analysis revealed that cell structures with regular shapes of the organelles generate a higher maximum electric potential when combined with piezoelectric and flexoelectric effects; this indicates that there are more microtubules in the cellular structure.
The mechanical and piezoelectric responses of the cell under three distinct boundary conditions have been elucidated with the help of the effective elastic and piezoelectric coefficients.  
Cell structures show that the presence of piezoelectric inclusions, specifically microtubules, significantly affects effective elastic moduli, leading to increased effective moduli.
The number of microtubules significantly influences the effective elastic coefficients and piezoelectric coefficients, both with and without considering the piezoelectric properties, due to the dependence of the behaviour of cells on their structure, the orientation of the microtubules, and the direction of mechanical stress.
Microtubule number, orientation, and loading conditions significantly impact cell mechanics, predicting increased electromechanical effects. These findings have practical applications in the development of biocompatible nano-biosensors, drug delivery systems, noninvasive diagnostic techniques, and therapeutic approaches where the properties of biological cells need to be better understood.
\section*{Acknowledgments}
\noindent The authors are grateful to the NSERC and the CRC Program (Canada) for their support. This research was enabled in part by support provided by SHARCNET (www.sharcnet.ca) and Digital Research Alliance of Canada (www.alliancecan.ca).
\appendix
\renewcommand{\thefigure}{A\arabic{figure}}
\renewcommand{\thetable}{A\arabic{table}}
\setcounter{figure}{0}
\setcounter{table}{0}
\section{Mesh Independence analysis}
\noindent The precision of numerical solutions is contingent upon the specific mesh employed to discretize the governing equations and boundary conditions. As the mesh size decreases and tends to zero within its optimal limits, the accuracy of the solutions improves, but the computational cost increases. The degree of freedom needed for model computation significantly impacts time and memory use. It is typically advantageous to estimate the degrees of freedom by considering the number of mesh elements in the model. Hence, a mesh independence analysis has been performed on CS-2 over the entire computational domain using a non-uniform triangular mesh to evaluate the trade-off between the accuracy of the solution and computational cost. 

\begin{figure}[!t]
	\centering
	{\includegraphics[width=0.7\linewidth]{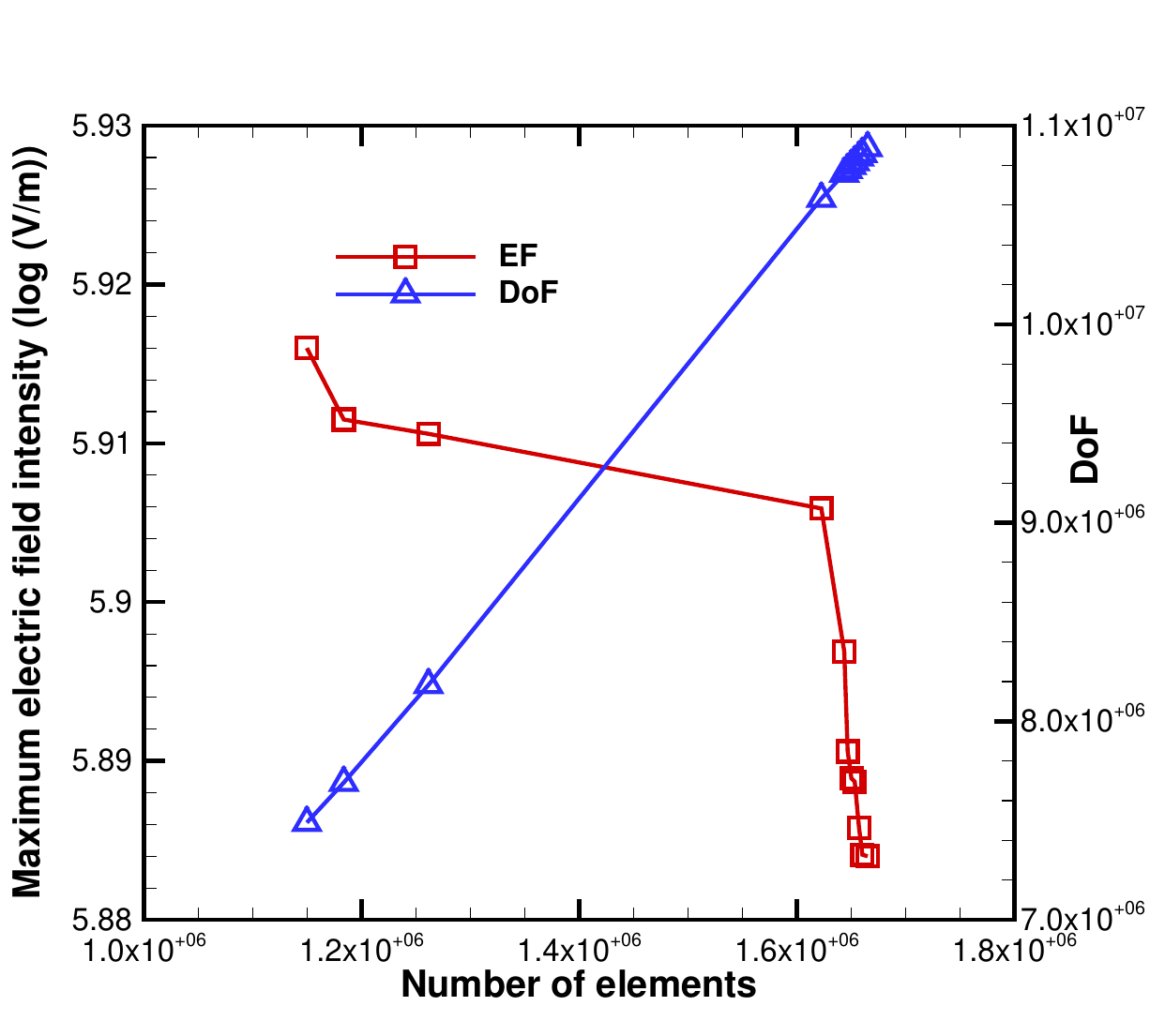}}
	\caption{The relationship between the total number of elements and the electric field intensity, as well as the degrees of freedom (DoF).}
	\label{fig:A1}
\end{figure}

\tab\ref{tab:A1} presents pertinent data regarding the quantity of elements, degrees of freedom, and the maximum electric field intensity within the cell. The present study has noted an increase in the computational cost as a result of the irregular structure of the microtubules and shapes of the mitochondria. This ultimately leads to an increase in both the number of elements and the degrees of freedom, as depicted in \fig\ref{fig:A1}. The trend observed is that the maximum electric field intensity value exhibits an increasing pattern as the number of elements increases up to 1650073. After that, the changes in the solution are negligible in terms of the maximum electric field intensity and relative percentage error. For instance, at 1653103 elements, and the maximum electric field intensity value is 5.887 V/m, with a relative percentage error of 0.0033. Therefore, in order to optimize computational efforts and cost, the total number of elements selected and all subsequent results presented in this work have been given for 1653103 elements and the degrees of freedom, 10817040. The mesh independence analysis for CS-1 has been conducted in a separate study \citep{SUNDEEPSINGH2020}.
\begin{table}[hbpt]
	\begin{center}
		\caption{Details of the mesh independence analysis.}\label{tab:A1}
		\scalebox{0.9}
		{
			\begin{tabular}{|p{0.4in}|p{1.0in}|p{1.0in}|p{1.5in}|p{1.2in}|} \hline
				S.No. &Number of elements &   Degrees of freedom  & Maximum electric field intensity ($\log (V/m)$) &Relative \% error  \\ \hline
				1 & 1149649 & 7489668 & 5.9160 &  -\\
				2 & 1183564 & 7691148 & 5.9115 & 0.0761 \\
				3 & 1261426 & 8183232  & 5.9106 & 0.0152  \\   
				4 & 1622430 & 10632984  & 5.9059 & 0.0795 \\ 
				5 & 1643463 & 10759200 & 5.8969 & 0.1526  \\
				6 & 1646619 & 10778136 & 5.8906 & 0.1069\\ 
				7 & 1650073 & 10798860 & 5.8889 & 0.0288 \\
				\textbf{8} & \textbf{1653103} & \textbf{10817040} & \textbf{5.8887} & \textbf{0.0033} \\
				9 & 1657046 & 10840704 & 5.8858 & 0.0492 \\
				10 & 1660121 & 10859160 & 5.8841 & 0.0288 \\
				11 & 1664968 & 10888236 & 5.8840 & 0.0016 \\	\hline
			\end{tabular}
		}
	\end{center}
\end{table}
\newpage
\bibliography{references}

\begin{thebibliography}{100}
\expandafter\ifx\csname url\endcsname\relax
  \def\url#1{\texttt{#1}}\fi
\expandafter\ifx\csname urlprefix\endcsname\relax\def\urlprefix{URL }\fi
\expandafter\ifx\csname href\endcsname\relax
  \def\href#1#2{#2} \def\path#1{#1}\fi

\bibitem{WANG2015}
Q.~Wang, H.~Cheng, H.~Peng, H.~Zhou, P.~Y. Li, R.~Langer, Non-genetic
  engineering of cells for drug delivery and cell-based therapy, Advanced Drug
  Delivery Reviews 91 (2015) 125--140.

\bibitem{Tang2021}
S.~Tang, Z.~Davoudi, G.~Wang, Z.~Xu, T.~Rehman, A.~Prominski, B.~Tian, K.~M.
  Bratlie, H.~Peng, Q.~Wang, Soft materials as biological and artificial
  membranes, Chemical Society Reviews 50~(22) (2021) 12679--12701.

\bibitem{Holy2003}
J.~Holy, Structure and function of cell organelles, in: S.~A. Krawetz, D.~D.
  Womble (Eds.), Introduction to bioinformatics: A theoretical and practical
  approach, Humana Press, 2003, pp. 25--54.

\bibitem{Link2023}
R.~Link, K.~Weißenbruch, M.~Tanaka, M.~Bastmeyer, U.~S. Schwarz, Cell shape
  and forces in elastic and structured environments: From single cells to
  organoids, Advanced Functional Materials (2023) 2302145.

\bibitem{Basoli2018}
F.~Basoli, S.~M. Giannitelli, M.~Gori, P.~Mozetic, A.~Bonfanti, M.~Trombetta,
  A.~Rainer, Biomechanical characterization at the cell scale: present and
  prospects, Frontiers in Physiology 9 (2018) 1449.

\bibitem{Zhang2007}
{Zhang, Ming and Assouline, Jose G.}, {Cilia containing 9 + 2 structures grown
  from immortalized cells}, Cell Research 17~(6) (2007) 537--545.

\bibitem{Lin2012}
{Lin, Jianfeng and Heuser, Thomas and Song, Kangkang and Fu, Xiaofeng and
  Nicastro, Daniela}, {One of the nine doublet microtubules of eukaryotic
  flagella exhibits unique and partially conserved structures}, PLOS ONE 7~(10)
  (2012) e46494.

\bibitem{HoyerFender2013}
{Hoyer-Fender, Sigrid}, {Primary and motile cilia: Their ultrastructure and
  ciliogenesis}, in: K.~L. Tucker, T.~Caspary (Eds.), {Cilia and nervous system
  development and function}, Springer Netherlands, 2013, pp. 1--53.

\bibitem{Fenton2021}
A.~R. Fenton, T.~A. Jongens, E.~L.~F. Holzbaur, Mitochondrial dynamics: Shaping
  and remodeling an organelle network, Current Opinion in Cell Biology 68
  (2021) 28--36.

\bibitem{Chada2004}
S.~R. Chada, P.~J. Hollenbeck, Nerve growth factor signaling regulates motility
  and docking of axonal mitochondria, Current Biology 14~(14) (2004)
  1272--1276.

\bibitem{Gutnick2019}
A.~Gutnick, M.~R. Banghart, E.~R. West, T.~L. Schwarz, The light-sensitive
  dimerizer zapalog reveals distinct modes of immobilization for axonal
  mitochondria, Nature Cell Biology 21~(6) (2019) 768--777.

\bibitem{Cason2022}
S.~E. Cason, E.~L.~F. Holzbaur, Selective motor activation in organelle
  transport along axons, Nature Reviews Molecular Cell Biology 23~(11) (2022)
  699--714.

\bibitem{RodriguezGarcia2020}
R.~Rodríguez-García, V.~A. Volkov, C.-Y. Chen, E.~A. Katrukha, N.~Olieric,
  A.~Aher, I.~Grigoriev, M.~P. López, M.~O. Steinmetz, L.~C. Kapitein,
  G.~Koenderink, M.~Dogterom, A.~Akhmanova, Mechanisms of motor-independent
  membrane remodeling driven by dynamic microtubules, Current Biology 30~(6)
  (2020) 972--987.

\bibitem{Gudimchuk2021}
N.~B. Gudimchuk, J.~R. McIntosh, Regulation of microtubule dynamics, mechanics
  and function through the growing tip, Nature Reviews Molecular Cell Biology
  22~(12) (2021) 777--795.

\bibitem{Gudimchuk2023}
N.~B. Gudimchuk, V.~V. Alexandrova, Measuring and modeling forces generated by
  microtubules, Biophysical Reviews 15~(5) (2023) 1095--1110.

\bibitem{Tuszynski2020}
J.~A. Tuszynski, D.~Friesen, H.~Freedman, V.~I. Sbitnev, H.~Kim, I.~Santelices,
  A.~P. Kalra, S.~D. Patel, K.~Shankar, L.~O. Chua, Microtubules as
  sub-cellular memristors, Scientific Reports 10~(1) (2020) 2108.

\bibitem{Kummer2021}
E.~Kummer, N.~Ban, Mechanisms and regulation of protein synthesis in
  mitochondria, Nature Reviews Molecular Cell Biology 22~(5) (2021) 307--325.

\bibitem{Li2017}
S.~Li, C.~Wang, P.~Nithiarasu, Three-dimensional transverse vibration of
  microtubules, Journal of Applied Physics 121~(23) (2017).

\bibitem{SETAYANDEH2016}
S.~Setayandeh, A.~Lohrasebi, Multi scale modeling of 2450mhz electric field
  effects on microtubule mechanical properties, Journal of Molecular Graphics
  and Modelling 70 (2016) 122--128.

\bibitem{Ishikawa2017}
T.~Ishikawa, Axoneme structure from motile cilia, Cold Spring Harbor
  Perspectives in Biology 9~(1) (2017).

\bibitem{Wanderley2010}
{de Souza, Wanderley and Attias, Marcia}, {Subpellicular microtubules in
  apicomplexa and trypanosomatids}, in: W.~de~Souza (Ed.), {Structures and
  organelles in pathogenic protists}, Springer Berlin Heidelberg, 2010, pp.
  27--62.

\bibitem{Mollaret1997}
I.~Mollaret, J.-L. Justine, Immunocytochemical study of tubulin in the
  9+‘1’ sperm axoneme of a monogenean (platyhelminthes), pseudodactylogyrus
  sp., Tissue and Cell 29~(6) (1997) 699--706.

\bibitem{Bezler2022}
A.~Bezler, A.~Woglar, F.~Schneider, F.~Douma, L.~Bürgy, C.~Busso, P.~Gönczy,
  Atypical and distinct microtubule radial symmetries in the centriole and the
  axoneme of lecudina tuzetae, Molecular Biology of the Cell 33~(8) (2022)
  ar75.

\bibitem{Roggen1966}
D.~R. Roggen, D.~J. Raski, N.~O. Jones, Cilia in nematode sensory organs,
  Science 152~(3721) (1966) 515--516.

\bibitem{Deurs1973}
B.~van Deurs, Axonemal 12 + 0 pattern in the flagellum of the motile
  spermatozoon of nymphon leptocheles, Journal of Ultrastructure Research
  42~(5) (1973) 594--598.

\bibitem{Dallai1992}
R.~Dallai, L.~Xu{\'e}, W.-Y. Yin, Flagellate spermatozoa of protura (insecta,
  apterygota) are motile, International Journal of Insect Morphology and
  Embryology 21~(2) (1992) 137--148.

\bibitem{Spreng2019}
B.~Spreng, H.~Fleckenstein, P.~Kübler, C.~Di~Biagio, M.~Benz, P.~Patra, U.~S.
  Schwarz, M.~Cyrklaff, F.~Frischknecht, Microtubule number and length
  determine cellular shape and function in plasmodium, The EMBO Journal 38~(15)
  (2019) e100984.

\bibitem{Ferreira2023}
J.~L. Ferreira, V.~Pražák, D.~Vasishtan, M.~Siggel, F.~Hentzschel, A.~M.
  Binder, E.~Pietsch, J.~Kosinski, F.~Frischknecht, T.~W. Gilberger,
  K.~Grünewald, Variable microtubule architecture in the malaria parasite,
  Nature Communications 14~(1) (2023) 1216.

\bibitem{Guolla2012}
L.~Guolla, M.~Bertrand, K.~Haase, A.~E. Pelling, Force transduction and strain
  dynamics in actin stress fibres in response to nanonewton forces, Journal of
  Cell Science 125~(3) (2012) 603--613.

\bibitem{Liu2012}
C.~Liu, S.~Hu, S.~Shen, Effect of flexoelectricity on electrostatic potential
  in a bent piezoelectric nanowire, Smart Materials and Structures 21~(11)
  (2012) 115024.

\bibitem{Yang2015}
W.~Yang, X.~Liang, S.~Shen, Electromechanical responses of piezoelectric
  nanoplates with flexoelectricity, Acta Mechanica 226~(9) (2015) 3097--3110.

\bibitem{Joshan2022}
Y.~S. Joshan, S.~Santapuri, {A gradient electromechanical theory for thin
  dielectric curved beams considering direct and converse flexoelectric
  effects}, Zeitschrift für angewandte Mathematik und Physik 73~(5) (2022)
  178.

\bibitem{Thai2022}
L.~M. Thai, D.~T. Luat, V.~B. Phung, P.~V. Minh, D.~V. Thom, {Finite element
  modeling of mechanical behaviors of piezoelectric nanoplates with
  flexoelectric effects}, Archive of Applied Mechanics 92~(1) (2022) 163--182.

\bibitem{WANG2023}
X.~Wang, Y.~Xue, {Investigation of the electric response of the piezoelectric
  curved beam considering the direct piezoelectric and flexoelectric effects},
  Thin-Walled Structures 188 (2023) 110839.

\bibitem{Iwao1977}
Y.~Iwao, {Electrical callus and callus formation by electret.}, {Clinical
  Orthopaedics and Related Research (1976-2007)} 124 (1977).

\bibitem{Fukada1955}
E.~Fukada, Piezoelectricity of wood, Journal of the Physical Society of Japan
  10~(2) (1955) 149--154.

\bibitem{Fukada1957}
E.~Fukada, I.~Yasuda, On the piezoelectric effect of bone, Journal of the
  physical society of Japan 12~(10) (1957) 1158--1162.

\bibitem{Fukada1964}
E.~Fukada, I.~Yasuda, Piezoelectric effects in collagen, Japanese Journal of
  Applied Physics 3~(2) (1964) 117.

\bibitem{Fukada1968}
E.~Fukada, Piezoelectricity as a fundamental property of wood, Wood Science and
  Technology 2 (1968) 299--307.

\bibitem{Fukada1983}
E.~Fukada, Piezoelectric properties of biological polymers, Quarterly reviews
  of biophysics 16~(1) (1983) 59--87.

\bibitem{Fukada2000}
E.~Fukada, History and recent progress in piezoelectric polymers, IEEE
  Transactions on ultrasonics, ferroelectrics, and frequency control 47~(6)
  (2000) 1277--1290.

\bibitem{Chae2018}
I.~Chae, C.~K. Jeong, Z.~Ounaies, S.~H. Kim, Review on electromechanical
  coupling properties of biomaterials, ACS Applied Bio Materials 1~(4) (2018)
  936--953.

\bibitem{SHAMOS1967}
M.~Shamos, L.~Lavine, Piezoelectricity as a fundamental property of biological
  tissues, Nature 213~(5073) (1967) 267--269.

\bibitem{ANDERSON1970}
J.~C. Anderson, C.~Eriksson, Piezoelectric properties of dry and wet bone,
  Nature 227~(5257) (1970) 491--492.

\bibitem{Fletcher2010}
D.~A. Fletcher, R.~D. Mullins, Cell mechanics and the cytoskeleton, Nature
  463~(7280) (2010) 485--492.

\bibitem{He2017}
L.~He, J.~Lou, A.~Zhang, H.~Wu, J.~Du, J.~Wang, On the coupling effects of
  piezoelectricity and flexoelectricity in piezoelectric nanostructures, AIP
  Advances 7~(10) (2017) 105106.

\bibitem{LUN2020}
Y.~Lun, H.~Zhou, D.~Yao, X.~Wang, J.~Hong, Screening piezoelectricity in
  determination of flexoelectric coefficient at nanoscale, Mechanics of
  Materials 150 (2020) 103591.

\bibitem{ATEF2022}
H.~Atef, A.~El-Dhaba, Modeling the flexoelectric effect via the reduced
  micromorphic model, Composite Structures 290 (2022) 115504.

\bibitem{Rout2023}
S.~K. Rout, S.~Kapuria, A flexoelectric actuator model with shear-lag and peel
  stress effects, Proceedings of the Royal Society A: Mathematical, Physical
  and Engineering Sciences 479~(2273) (2023) 20230099.

\bibitem{DENG2014}
Q.~Deng, L.~Liu, P.~Sharma, Flexoelectricity in soft materials and biological
  membranes, Journal of the Mechanics and Physics of Solids 62 (2014) 209--227.

\bibitem{Deng2017}
F.~Deng, Q.~Deng, W.~Yu, S.~Shen, Mixed finite elements for flexoelectric
  solids, Journal of Applied Mechanics 84~(8) (2017).

\bibitem{Abdollahi2019}
A.~Abdollahi, N.~Domingo, I.~Arias, G.~Catalan, Converse flexoelectricity
  yields large piezoresponse force microscopy signals in non-piezoelectric
  materials, Nature Communications 10~(1) (2019) 1266.

\bibitem{Nguyen2013}
T.~D. Nguyen, S.~Mao, Y.-W. Yeh, P.~K. Purohit, M.~C. McAlpine, Nanoscale
  flexoelectricity, Advanced Materials 25~(7) (2013) 946--974.

\bibitem{Zhang2023}
{Zhang, Rui and Gu, Lusheng and Chen, Wei and Tanaka, Nobutoshi and Zhou,
  Zhengrong and Xu, Honglin and Xu, Tao and Ji, Wei and Liang, Xin and Meng,
  Wenxiang}, {CAMSAP2 and CAMSAP3 localize at microtubule intersections to
  regulate the spatial distribution of microtubules}, Journal of Molecular Cell
  Biology (2023) mjad050.

\bibitem{Kobori2024}
{Kobori, Mako and Abe, Junya and Saito, Reika and Hirai, Yohei}, {CAMSAP3, a
  microtubule orientation regulator, plays a vital role in manifesting
  differentiation-dependent characteristics in keratinocytes}, Experimental
  Cell Research 435~(1) (2024) 113927.

\bibitem{Connie2016b}
R.~Connie, W.~Robert, J.~Vladimir, D.~Jean, C.~Jung, A.~Yael, Biology,
  OpenStax, Houston, Texas, 2016.

\bibitem{Natasha2016a}
R.~S. Natasha, Introduction to biology, Creative Commons Attribution 4.0
  International License, 2016, Ch. 3.2 Comparing prokaryotic and eukaryotic
  cells.

\bibitem{SUNDEEPSINGH2020}
S.~Singh, J.~A. Krishnaswamy, R.~Melnik, {Biological cells and coupled
  electro-mechanical effects: The role of organelles, microtubules, and
  nonlocal contributions}, Journal of the Mechanical Behavior of Biomedical
  Materials 110 (2020) 103859.

\bibitem{Uzman2003}
A.~Uzman, \mbox{Molecular biology of the cell (4th ed.): Alberts, B., Johnson,
  A.,} \mbox{ Lewis, J., Raff, M., Roberts, K., and Walter, P.}, Biochemistry
  and Molecular Biology Education 31~(4) (2003) 212--214.

\bibitem{Parker2017}
N.~Parker, M.~Schneegurt, A.~H.~T. Tu, B.~M. Forster, P.~Lister, Microbiology,
  Open Textbook Library, OpenStax, 2017.

\bibitem{Tange2002}
Y.~Tange, A.~Hirata, O.~Niwa, An evolutionarily conserved fission yeast
  protein, ned1, implicated in normal nuclear morphology and chromosome
  stability, interacts with dis3, pim1/rcc1 and an essential nucleoporin,
  Journal of Cell Science 115~(22) (2002) 4375--4385.

\bibitem{Norppa2003}
H.~Norppa, G.~C.-M. Falck, What do human micronuclei contain?, Mutagenesis
  18~(3) (2003) 221--233.

\bibitem{Robert2004}
R.~D. Goldman, D.~K. Shumaker, M.~R. Erdos, M.~Eriksson, A.~E. Goldman, L.~B.
  Gordon, Y.~Gruenbaum, S.~Khuon, M.~Mendez, R.~Varga, F.~S. Collins,
  Accumulation of mutant lamin a causes progressive changes in nuclear
  architecture in hutchinson–gilford progeria syndrome, Proceedings of the
  National Academy of Sciences 101~(24) (2004) 8963--8968.

\bibitem{Webster2009}
M.~Webster, K.~L. Witkin, O.~Cohen-Fix, Sizing up the nucleus: nuclear shape,
  size and nuclear-envelope assembly, Journal of Cell Science 122~(10) (2009)
  1477--1486.

\bibitem{Khatau2012}
S.~B. Khatau, S.~Kusuma, D.~Hanjaya-Putra, P.~Mali, L.~Cheng, J.~S.~H. Lee,
  S.~Gerecht, D.~Wirtz, The differential formation of the linc-mediated
  perinuclear actin cap in pluripotent and somatic cells, PLOS ONE 7~(5) (2012)
  1--12.

\bibitem{MANNELLA2006}
C.~A. Mannella, The relevance of mitochondrial membrane topology to
  mitochondrial function, Biochimica et Biophysica Acta (BBA) - Molecular Basis
  of Disease 1762~(2) (2006) 140--147.

\bibitem{BOZELLI2020}
J.~C. Bozelli, R.~M. Epand, Membrane shape and the regulation of biological
  processes, Journal of Molecular Biology 432~(18) (2020) 5124--5136.

\bibitem{Kalt1975}
M.~R. Kalt, Mitochondrial pleiomorphism in sustentacular cells of xenopus
  laevis, The Anatomical Record 182~(1) (1975) 53--60.

\bibitem{Paumard2002}
P.~Paumard, J.~Vaillier, B.~Coulary, J.~Schaeffer, V.~Soubannier, D.~M.
  Mueller, D.~Brèthes, J.-P. di~Rago, J.~Velours, The atp synthase is involved
  in generating mitochondrial cristae morphology, The EMBO Journal 21~(3)
  (2002) 221--230.

\bibitem{VELOURS2009}
J.~Velours, A.~Dautant, B.~Salin, I.~Sagot, D.~Brèthes, Mitochondrial f1f0-atp
  synthase and organellar internal architecture, The International Journal of
  Biochemistry \& Cell Biology 41~(10) (2009) 1783--1789.

\bibitem{Vaziri2008}
A.~Vaziri, A.~Gopinath, Cell and biomolecular mechanics in silico, Nature
  Materials 7~(1) (2008) 15--23.

\bibitem{Connie2016a}
R.~Connie, W.~Robert, J.~Vladimir, D.~Jean, C.~Jung, A.~Yael, Biology,
  OpenStax, Houston, Texas, 2016.

\bibitem{Moujaber2020}
O.~Moujaber, U.~Stochaj, The cytoskeleton as regulator of cell signaling
  pathways, Trends in Biochemical Sciences 45~(2) (2020) 96--107.

\bibitem{Ofek2009}
G.~Ofek, D.~C. Wiltz, K.~A. Athanasiou, Contribution of the cytoskeleton to the
  compressive properties and recovery behavior of single cells, Biophysical
  Journal 97~(7) (2009) 1873--1882.

\bibitem{Cooper2000}
G.~M. Cooper, The cell: A molecular approach. 2nd edition, Sinauer Associates,
  2000.

\bibitem{KATTI2017}
D.~R. Katti, K.~S. Katti, {Cancer cell mechanics with altered cytoskeletal
  behavior and substrate effects: A 3D finite element modeling study}, Journal
  of the Mechanical Behavior of Biomedical Materials 76 (2017) 125--134.

\bibitem{GOWRISHANKAR2006}
T.~R. Gowrishankar, A.~T. Esser, Z.~Vasilkoski, K.~C. Smith, J.~C. Weaver,
  {Microdosimetry for conventional and supra-electroporation in cells with
  organelles}, Biochemical and Biophysical Research Communications 341~(4)
  (2006) 1266--1276.

\bibitem{Tiwari2009}
P.~K. Tiwari, S.~K. Kang, G.~J. Kim, J.~Choi, A.-A.~H. Mohamed, J.~K. Lee,
  {Modeling of nanoparticle-mediated electric field enhancement inside
  biological cells exposed to AC electric fields}, Japanese Journal of Applied
  Physics 48 (2009) 087001.

\bibitem{Yongbo2015}
Y.~Li, S.~Honda, K.~Iwami, Y.~Ohta, N.~Umeda, {Analysis of mitochondrial
  mechanical dynamics using a confocal fluorescence microscope with a bent
  optical fibre}, Journal of Microscopy 260~(2) (2015) 140--151.

\bibitem{BARRETO2013}
S.~Barreto, C.~H. Clausen, C.~M. Perrault, D.~A. Fletcher, D.~Lacroix, {A
  multi-structural single cell model of force-induced interactions
  of cytoskeletal components}, Biomaterials 34~(26) (2013) 6119--6126.

\bibitem{PAMPALONI2008}
F.~Pampaloni, E.-L. Florin, {Microtubule architecture: inspiration for novel
  carbon nanotube-based biomimetic materials}, Trends in Biotechnology 26~(6)
  (2008) 302--310.

\bibitem{Ingber2000}
D.~E. Ingber, S.~R. Heidemann, P.~Lamoureux, R.~E. Buxbaum, {Opposing views on
  tensegrity as a structural framework for understanding cell mechanics},
  Journal of Applied Physiology 89~(4) (2000) 1663--1678.

\bibitem{STRICKER2010}
J.~Stricker, T.~Falzone, M.~L. Gardel, {Mechanics of the F-actin cytoskeleton},
  Journal of Biomechanics 43~(1) (2010) 9--14.

\bibitem{Krishnaswamy2019a}
J.~A. Krishnaswamy, F.~C. Buroni, F.~Garcia-Sanchez, R.~Melnik,
  L.~Rodriguez-Tembleque, A.~Saez, {Lead-free piezocomposites with CNT-modified
  matrices: Accounting for agglomerations and molecular defects}, Composite
  Structures 224 (2019) 111033.

\bibitem{Krishnaswamy2019b}
J.~A. Krishnaswamy, F.~C. Buroni, E.~Garc{\'\i}a-Mac{\'\i}as, R.~Melnik,
  L.~Rodriguez-Tembleque, A.~Saez, Design of lead-free pvdf/cnt/batio3
  piezocomposites for sensing and energy harvesting: the role of
  polycrystallinity, nanoadditives, and anisotropy, Smart Materials and
  Structures 29~(1) (2019) 015021.

\bibitem{Krishnaswamy2019c}
J.~A. Krishnaswamy, F.~C. Buroni, F.~Garcia-Sanchez, R.~Melnik,
  L.~Rodriguez-Tembleque, A.~Saez, Improving the performance of lead-free
  piezoelectric composites by using polycrystalline inclusions and tuning the
  dielectric matrix environment, Smart Materials and Structures 28~(7) (2019)
  075032.

\bibitem{Garcini1990}
{de Garcini, E. Montejo and Carrascosa, J. L. and Nieto, A. and Avila, J.},
  {Collagenous structures present in brain contain epitopes shared by collagen
  and microtubule-associated protein tau}, {Journal of Structural Biology}
  103~(1) (1990) 34--39.

\bibitem{Kushagra2015}
A.~Kushagra, {Thermal fluctuation induced piezoelectric effect in cytoskeletal
  microtubules: Model for energy harvesting and their intracellular
  communication}, {Journal of Biomedical Science and Engineering} 8 (2015)
  511--519.

\bibitem{Denning2017}
D.~Denning, J.~I. Kilpatrick, E.~Fukada, N.~Zhang, S.~Habelitz, A.~Fertala,
  M.~D. Gilchrist, Y.~Zhang, S.~A.~M. Tofail, B.~J. Rodriguez, {Piezoelectric
  tensor of collagen fibrils determined at the nanoscale}, ACS Biomaterials
  Science \& Engineering 3~(6) (2017) 929--935.

\bibitem{GARCIA2018}
P.~D. Garcia, R.~Garcia, {Determination of the elastic moduli of a single cell
  cultured on a rigid support by force microscopy}, Biophysical Journal
  114~(12) (2018) 2923--2932.

\bibitem{Garcia2018b}
P.~D. Garcia, R.~Garcia, {Determination of the viscoelastic properties of a
  single cell cultured on a rigid support by force microscopy}, Nanoscale
  10~(42) (2018) 19799--19809.

\bibitem{Li2018}
M.~Li, L.~Liu, X.~Xu, X.~Xing, D.~Dang, N.~Xi, Y.~Wang, {Nanoscale
  characterization of dynamic cellular viscoelasticity by atomic force
  microscopy with varying measurement parameters}, Journal of the Mechanical
  Behavior of Biomedical Materials 82 (2018) 193--201.

\bibitem{Marcotti2019}
S.~Marcotti, G.~C. Reilly, D.~Lacroix, Effect of cell sample size in atomic
  force microscopy nanoindentation, Journal of the Mechanical Behavior of
  Biomedical Materials 94 (2019) 259--266.

\bibitem{Nakamura2021}
{Nakamura, Tsukasa and Takata, Mineaki and Michimoto, Itsuki and Koyama,
  Daisuke and Matsukawa, Mami}, {Site dependence of ultrasonically induced
  electrical potentials in bone}, JASA Express Letters 1~(1) (2021) 012002.

\bibitem{Wang2007}
Z.~L. Wang, Nanopiezotronics, Advanced Materials 19~(6) (2007) 889--892.

\bibitem{Tichy2010}
J.~Tich{\`y}, J.~Erhart, E.~Kittinger, J.~Privratska, Fundamentals of
  piezoelectric sensorics: mechanical, dielectric, and thermodynamical
  properties of piezoelectric materials, Springer Science \& Business Media,
  2010.

\bibitem{Zhang2021}
C.~Zhang, W.~Wang, X.~Hao, Y.~Peng, Y.~Zheng, J.~Liu, Y.~Kang, F.~Zhao, Z.~Luo,
  J.~Guo, B.~Xu, L.~Shao, G.~Li, A novel approach to enhance bone regeneration
  by controlling the polarity of gan/algan heterostructures, Advanced
  Functional Materials 31~(5) (2021) 2007487.

\bibitem{Chunyu2022}
C.~Yang, J.~Ji, Y.~Lv, Z.~Li, D.~Luo, Application of piezoelectric material and
  devices in bone regeneration, Nanomaterials 12~(24) (2022).

\bibitem{Berger2005}
{Berger, Harald and Kari, Sreedhar and Gabbert, Ulrich and Rodriguez-Ramos,
  Reinaldo and Guinovart, Raul and Otero, Jose A. and Bravo-Castillero,
  Julian}, {An analytical and numerical approach for calculating effective
  material coefficients of piezoelectric fiber composites}, {International
  Journal of Solids and Structures} 42~(21) (2005) 5692--5714.

\bibitem{Dascalescu2014}
{Dascalescu, Lucian and Qin, Ri-Song and Xiao, Yi and Lan, Haitian}, {Numerical
  simulation of effective properties of 3D piezoelectric composites}, Journal
  of Engineering 2014 (2014) 824806.

\bibitem{Saputra2017}
{Saputra, Albert Artha and Sladek, Vladimir and Sladek, Jan and Song,
  Chongmin}, {Micromechanics determination of effective material coefficients
  of cement-based piezoelectric ceramic composites}, {Journal of Intelligent
  Material Systems and Structures} 29~(5) (2017) 845--862.

\bibitem{Hamdia2024}
{Hamdia, Khader M.}, {A representative volume element model to evaluate the
  effective properties of flexoelectric nanocomposite}, {European Journal of
  Mechanics - A/Solids} 103 (2024) 105149.

\bibitem{Titushkin2009}
I.~Titushkin, M.~Cho, Regulation of cell cytoskeleton and membrane mechanics by
  electric field: Role of linker proteins, Biophysical Journal 96~(2) (2009)
  717--728.

\bibitem{Schofield2020}
Z.~Schofield, G.~N. Meloni, P.~Tran, C.~Zerfass, G.~Sena, Y.~Hayashi, M.~Grant,
  S.~A. Contera, S.~D. Minteer, M.~Kim, A.~Prindle, P.~Rocha, M.~B.~A. Djamgoz,
  T.~Pilizota, P.~R. Unwin, M.~Asally, O.~S. Soyer, Bioelectrical understanding
  and engineering of cell biology, Journal of The Royal Society Interface
  17~(166) (2020) 20200013.

\bibitem{Markx2008}
G.~H. Markx, The use of electric fields in tissue engineering, Organogenesis
  4~(1) (2008) 11--17.

\bibitem{Chen2019}
{Chen, Cen and Bai, Xue and Ding, Yahui and Lee, In-Seop}, {Electrical
  stimulation as a novel tool for regulating cell behavior in tissue
  engineering}, Biomaterials Research 23~(1) (2019) 25.

\bibitem{Boehm2005}
K.~J. Böhm, N.~E. Mavromatos, A.~Michette, R.~Stracke, E.~Unger, Movement and
  alignment of microtubules in electric fields and electric-dipole-moment
  estimates, Electromagnetic Biology and Medicine 24~(3) (2005) 319--330.

\bibitem{Antebi2012}
{Antebi, Ben and Cheng, Xingguo and Harris, Jeffrey N. and Gower, Laurie B. and
  Chen, Xiao-Dong and Ling, Jian}, {Biomimetic collagen-hydroxyapatite
  composite fabricated via a novel perfusion-flow mineralization technique},
  Tissue Engineering Part C: Methods 19~(7) (2012) 487--496.

\bibitem{Peppas2016}
N.~A. Peppas, J.~R. Clegg, The challenge to improve the response of
  biomaterials to the physiological environment, Regenerative Biomaterials
  3~(2) (2016) 67--71.

\bibitem{Ashammakhi2018}
N.~Ashammakhi, S.~Ahadian, F.~Zengjie, K.~Suthiwanich, F.~Lorestani, G.~Orive,
  S.~Ostrovidov, A.~Khademhosseini, Advances and future perspectives in 4d
  bioprinting, Biotechnology Journal 13~(12) (2018) 1800148.

\bibitem{Kamel2022}
N.~A. Kamel, Bio-piezoelectricity: fundamentals and applications in tissue
  engineering and regenerative medicine, Biophysical Reviews 14~(3) (2022)
  717--733.

\bibitem{Tsui2022}
C.~T. Tsui, P.~Lal, K.~V.~R. Fox, M.~A. Churchward, K.~G. Todd, The effects of
  electrical stimulation on glial cell behaviour, BMC Biomedical Engineering
  4~(1) (2022) 7.

\bibitem{Yaojin2012}
{Wang, Yaojin and Hasanyan, Davresh and Li, Menghui and Gao, Junqi and Li,
  Jiefang and Viehland, D. and Luo, Haosu}, {Theoretical model for
  geometry-dependent magnetoelectric effect in magnetostrictive/piezoelectric
  composites}, {Journal of Applied Physics} 111~(12) (2012) 124513.

\bibitem{Kapat2020}
{Kapat, Kausik and Shubhra, Quazi T. H. and Zhou, Miao and Leeuwenburgh,
  Sander}, {Piezoelectric nano-biomaterials for biomedicine and tissue
  regeneration}, {Advanced Functional Materials} 30~(44) (2020) 1909045.

\bibitem{Charras2002}
G.~T. Charras, M.~A. Horton, Determination of cellular strains by combined
  atomic force microscopy and finite element modeling, Biophysical Journal
  83~(2) (2002) 858--879.

\bibitem{Rigato2015}
A.~Rigato, F.~Rico, F.~Eghiaian, M.~Piel, S.~Scheuring, Atomic force microscopy
  mechanical mapping of micropatterned cells shows adhesion geometry-dependent
  mechanical response on local and global scales, ACS Nano 9~(6) (2015)
  5846--5856.

\bibitem{Schillers2019}
H.~Schillers, Measuring the elastic properties of living cells, in: N.~C.
  Santos, F.~A. Carvalho (Eds.), Atomic Force Microscopy: Methods and
  Protocols, Springer New York, 2019, pp. 291--313.

\bibitem{Kollmannsberger2011}
P.~Kollmannsberger, B.~Fabry, Linear and nonlinear rheology of living cells,
  Annu. Rev. Mater. Res. 41~(1) (2011) 75--97.

\bibitem{Chaudhuri2020}
O.~Chaudhuri, J.~Cooper-White, P.~A. Janmey, D.~J. Mooney, V.~B. Shenoy,
  Effects of extracellular matrix viscoelasticity on cellular behaviour, Nature
  584~(7822) (2020) 535--546.

\bibitem{Ingber2014}
D.~E. Ingber, N.~Wang, D.~Stamenović, Tensegrity, cellular biophysics, and the
  mechanics of living systems, Reports on Progress in Physics 77~(4) (2014)
  046603.

\bibitem{Sun2023}
S.-Y. Sun, L.-Y. Zhang, X.~Chen, X.-Q. Feng, Biochemomechanical tensegrity
  model of cytoskeletons, Journal of the Mechanics and Physics of Solids 175
  (2023) 105288.

\bibitem{Ambrosi2019}
D.~Ambrosi, M.~Ben~Amar, C.~J. Cyron, A.~DeSimone, A.~Goriely, J.~D. Humphrey,
  E.~Kuhl, Growth and remodelling of living tissues: perspectives, challenges
  and opportunities, Journal of The Royal Society Interface 16~(157) (2019)
  20190233.

\end{thebibliography}
\end{document}